\documentclass[lettersize,journal]{IEEEtran}
\usepackage{amsmath,amsfonts}
\usepackage{algorithmic}
\usepackage{algorithm}
\usepackage{array}
\usepackage[caption=false,font=normalsize,labelfont=sf,textfont=sf]{subfig}
\usepackage{textcomp}
\usepackage{stfloats}
\usepackage{url}
\usepackage{verbatim}
\usepackage{graphicx}
\usepackage{cite}

\usepackage{makecell}

\usepackage[dvipsnames]{xcolor}
\definecolor{darkgreen}{rgb}{0.0, 0.5, 0.0}

\usepackage{textcomp}
\usepackage[normalem]{ulem}
\usepackage{multirow}

\usepackage{booktabs}

\newcommand{\mysubfigure}[3]{%
    \subfloat[#2]{\includegraphics[width=#1]{#3}}%
}
\graphicspath{{./figures/}{./results/}{./bios/}}
\DeclareGraphicsExtensions{.pdf,.jpeg,.png}

\usepackage{hyperref}
\usepackage{hyperxmp}
\hypersetup{
bookmarks=true,
bookmarksnumbered=true,            
hypertexnames=false,               
breaklinks=true,                   
linktoc=all,
colorlinks=true,
linkcolor=blue,
citecolor=blue,
urlcolor=blue,                     
pdfborder={0 0 112.0},              
hyperfootnotes=false
}                        

\usepackage{colortbl}
\usepackage{rotating}

\usepackage{hyphenat}

\usepackage{siunitx}

\usepackage{tikz}
\usetikzlibrary{positioning}
\usetikzlibrary{arrows.meta,positioning}

\usepackage{todonotes}

\newcommand{\FOneScore}{F_{1}}
\newcommand{\FOneScoreBold}{\textbf{\textit{F}}_{\textit{1}}}
\newcommand{\FOneScoreG}{\FOneScore^{G}}
\newcommand{\FOneScoreCZero}{\FOneScore^{C0}}
\newcommand{\FOneScoreCOne}{\FOneScore^{C1}}

\newcommand{\distanceThr}{\delta_{thr}}
\newcommand{\distanceThrSet}{D_{thr}}

\newcommand{\NumDataFrames}{N_F}

\newcommand{\MarlinksDataset}{\texttt{Marlinks-NS DAS}}

\begin{document}

\title{Vessel Detection and Localization Using Distributed Acoustic Sensing in Submarine Optical Fiber Cables}

\author{Erick Eduardo Ramirez-Torres,
  Javier Macias-Guarasa~\IEEEmembership{Member,~IEEE},
  Daniel Pizarro,
  Javier Tejedor,
  Sira Elena Palazuelos-Cagigas~\IEEEmembership{Senior Member,~IEEE},
  Pedro J. Vidal-Moreno,
  Sonia Martin-Lopez,
  Miguel Gonzalez-Herraez,
  Roel Vanthillo
  \thanks{Author-accepted manuscript. Accepted for publication in the \textit{IEEE Journal of Selected Topics in Applied Earth Observations and Remote Sensing}. Digital Object Identifier: \href{https://doi.org/10.1109/JSTARS.2026.3716768}{10.1109/JSTARS.2026.3716768}. \textcopyright~2026 The Authors. This work is licensed under a Creative Commons Attribution 4.0 License; see \url{https://creativecommons.org/licenses/by/4.0/}.}
  \thanks{This work has been partially supported by the Spanish Ministry of Science and Innovation MCIN/AEI/10.13039/501100011033 and by the European Union NextGenerationEU/PRTR program under grants PSI (PLEC2021-007875), REMO (CPP2021-008869), EYEFUL-UAH (PID2020-113118RB-C31) and NeurEYE (PID2024-156576OB-C31); by FEDER Una manera de hacer Europa under grant PRECISION (PID2021-128000OBC21); by the European Innovation Council under grant SAFE (101098992)  and Horizon Europe under grants SUBMERSE (101095055) and ECSTATIC (101189595).}
\thanks{P.J.V-M was supported by FPI-2021 Grant from the University of Alcalá Research Program.}
  \thanks{We gratefully acknowledge the computer resources at Artemisa, funded by the European Union ERDF and Comunitat Valenciana as well as the technical support provided by the Instituto de Fisica Corpuscular, IFIC (CSIC-UV).}
  \thanks{E.E. Ramirez-Torres, J. Macias-Guarasa, D. Pizarro, P.J. Vidal-Moreno, S.E. Palazuelos-Cagigas, S. Martin-Lopez, and M. Gonzalez-Herraez are with Universidad de Alcalá, Department of Electronics, Alcalá de Henares, Madrid, Spain. email: javier.maciasguarasa@uah.es. S. Martin-Lopez, and M. Gonzalez-Herraez are also with Daza de Valdés Institute of Optics (IO-CSIC), Madrid, Spain. J. Tejedor is with Institute of Technology, Universidad San Pablo-CEU, CEU Universities, Urbanización Montepríncipe, 28668 Boadilla del Monte, Spain. Roel Vanthillo is with Marlinks, Sint-Maartenstraat 5, 3000 Leuven, Belgium. }
}

\markboth{Accepted for Publication in IEEE J-STARS}%
{Ramirez-Torres \MakeLowercase{\textit{et al.}}: Vessel Detection and Localization Using Distributed Acoustic Sensing}


\maketitle

\begin{abstract}
  Submarine cables play a critical role in global internet connectivity, energy transmission, and communication but remain vulnerable to accidental damage and sabotage. Recent incidents in the Baltic Sea highlighted the need for enhanced monitoring to protect this vital infrastructure. Traditional vessel detection methods, such as synthetic aperture radar, video surveillance, and multispectral satellite imagery, face limitations in real-time processing, adverse weather conditions, and coverage range. This paper explores Distributed Acoustic Sensing (DAS) as an alternative by repurposing submarine telecommunication cables as large-scale acoustic sensor arrays. DAS offers continuous real-time monitoring, operates independently of cooperative systems like the \emph{Automatic Identification System} (AIS), being largely unaffected by lighting or weather conditions. However, existing research on DAS for vessel tracking is limited in scale and lacks validation under real-world conditions. To address these gaps, a general and systematic methodology is presented for vessel detection and distance estimation using DAS. Advanced machine learning models are applied to improve detection and localization accuracy in dynamic maritime environments. The approach is evaluated on a real pre-existing $\mathbf{28\,km}$ submarine fiber-optic cable deployed in the North Sea over a continuous ten-day period, covering diverse ship and operational conditions, representing one of the largest-scale DAS-based vessel monitoring studies to date, and for which we release the full evaluation dataset. Results demonstrate DAS as a practical tool for maritime surveillance, with an overall $\FOneScoreBold$-score of over $\mathbf{90\%}$ in vessel detection at the selected $\mathbf{1000\,m}$ operating threshold, and a mean absolute error of $\mathbf{141\textit{m}}$ for vessel distance estimation, towards bridging the gap between experimental research and real-world deployment.
\end{abstract}


\section{Introduction}
\label{sec:intro}

\IEEEPARstart{S}{ubmarine} cables, essential for global internet connectivity, energy transmission, and communication, are particularly vulnerable to accidental damage (e.g., fishing nets, ship anchors) and sabotage linked to geopolitical tensions~\cite{euromaidan_baltic_2025}. Recent incidents highlight this risk: in November 2024, damage to two Baltic Sea cables disrupted internet services between Finland, Germany, and Lithuania, with suspicions of foreign interference~\cite{cbsnews_baltic_cables_2024}. In December 2024 the EstLink2 power cable between Finland and Estonia was severed, prompting an investigation~\cite{navalnews_estlink2_2024}. In February 2025, the C-Lion1 cable between Finland and Germany suffered damage near Gotland Island~\cite{euromaidan_baltic_2025}. The European Commission recently addressed the need for stronger monitoring and security measures in its Joint Communication on submarine cable resilience~\cite{eu_submarine_cables2025}. 

Detecting and localizing vessels near submarine cables is thus crucial to prevent disruptions in international data communications and energy transmission. In this study, vessel detection is addressed as a cable-protection capability, since nearby ships may threaten submarine communication and power infrastructure through anchoring, fishing activity, or other operations that can damage the cable.

Existing vessel localization methods use various sensing techniques, each with strengths and limitations. Yu et al.~\cite{Yu_2021} demonstrated vessel detection with synthetic aperture radar (SAR), which enables wide-area surveillance in all weather conditions. However, SAR struggles with real-time processing and false alarms from sea clutter. Wawrzyniak et al.~\cite{Wawrzyniak_2019} used video surveillance in ports, applying vessel detection algorithms to track ships, but these systems are limited by lighting conditions and range. Similarly, Xie et al.~\cite{Xie_2020} improved ship detection in multispectral satellite imagery with machine learning (ML), enhancing performance under mist and clouds, though severe weather and nighttime conditions significantly reduce accuracy.

Distributed Acoustic Sensing (DAS) has emerged as a transformative technology enabling dense spatial sampling of acoustic signals along fiber-optic cables, and has seen increasing applications across geophysics, infrastructure monitoring, and environmental sensing. Recent reviews such as~\cite{cheng2024photonic} provide an overview of the evolution, challenges, and potentials of DAS across different domains. In the paper context, DAS
overcomes many limitations of traditional methods by repurposing submarine telecommunication cables as virtual sensor arrays~\cite{Landro_2022}. Unlike SAR and optical imagery, DAS enables continuous real-time monitoring with high spatial and temporal resolution, independently of lighting and weather conditions~\cite{Thiem_2023}. It also operates without cooperative systems like the \emph{Automatic Identification System} (AIS), allowing the detection of ``dark'' ships involved in unauthorized activities~\cite{Malaprade2019}. By making use of existing infrastructure, DAS provides wide-area coverage at lower costs without requiring dedicated hardware like hydrophones~\cite{Malaprade2019, Landro_2022}. DAS detects ships by analyzing their acoustic signatures, such as broadband noise from machinery and tonal signals from propellers. These characteristics have been studied for vessel classification, speed estimation, and maritime traffic monitoring~\cite{Malinowski_2002}. However, existing methods must be adapted to the specific signal-processing requirements of DAS. Despite advancements, significant gaps remain in the DAS literature for vessel detection and localization, specially those oriented to submarine cable protection. Many studies use limited datasets, lack diversity in vessel types and environmental conditions, or focus on proof-of-concept setups without real-world validation.

The main contributions of this work are:

\begin{itemize}
    \item A general and systematic methodology for \emph{joint vessel detection and distance estimation} using DAS on submarine cables, explicitly oriented to \emph{cable-protection applications} and early-warning monitoring scenarios rather than generic maritime surveillance.

    \item The application of advanced ML models, including XGBoost and neural networks, to enhance vessel detection and localization accuracy in dynamic maritime environments.

    \item A \emph{large-scale and rigorous experimental validation} under real-world conditions, based on a continuous ten-day monitoring period with diverse vessel types, sizes, and operational scenarios, substantially exceeding the experimental scale typically reported in previous DAS-based vessel-monitoring studies.

    \item The \emph{release of a fully annotated evaluation dataset} and \emph{associated source code} in public repositories, enabling reproducibility and future comparative research.

    \item Evidence that DAS can operate as a practical tool for protecting submarine communication and power-cable infrastructure, helping bridge the gap between experimental research and real-world deployment.
\end{itemize}

To the best of our knowledge, this is the first openly documented scientific study to systematically address  vessel detection and continuous cable-relative distance estimation using machine learning on submarine-cable DAS data, specifically focused on submarine cable protection and at larger scale than any other previous work. This single-deployment validation is not intended to demonstrate universal cross-cable generalization of the trained models, but to provide a realistic and reproducible field evaluation of the proposed methodology under operational submarine-cable conditions.

The remainder of this paper is structured as follows. 
Section~\ref{sec:sota} introduces the previous work on vessel detection, classification and localization using DAS. Section~\ref{sec:das-submarine} provides a general overview of DAS technology in submarine cables, with Section~\ref{sec:database-description} describing the data used in this study. Section~\ref{sec:proposed-system} provides details on the proposed system architecture and methodological aspects, while Section~\ref{sec:exper-work-results} presents the experimental work and results. Finally, Section~\ref{sec:discussion} provides additional discussion details, and Section~\ref{sec:concl-future-work} concludes the paper and outlines future research directions.

\section{Previous Work}
\label{sec:sota}

DAS data present significant challenges due to its high-dimensionality, noise levels, and large volume, often surpassing the capabilities of traditional data storage and signal processing capabilities~\cite{Nur_DAS_storage2024,dong2020dassa,Rivet_2021}. On the other hand, advanced techniques for data interpretation such as artificial intelligence and ML methods have demonstrated potential in overcoming these challenges for vessel detection~\cite{Malaprade2019, Rivet_2021}, seismic monitoring~\cite{Stork_2020}, and acoustic recognition~\cite{Drylerakis2024source, zhan2024application}. Malaprade et al.\cite{Malaprade2019} illustrated the feasibility of ML-based ship detection using frequency band energy plots but were limited by a small dataset (only seven crossings) and sparse experimental details. Thiem et al.\cite{Thiem_2023} localized a single passenger ship in a Norwegian fjord using noise-reduction filtering and traveltime inversion, showing good agreement with AIS but focusing on just one vessel type and not assessing other factors like tides or weather. Rivet et al.\cite{Rivet_2021} estimated tanker trajectories off Toulon, France, leveraging Doppler effects and simulation models; while results aligned with observations, the study considered only one vessel and lacked broader environmental variability. Chen et al.\cite{Chen_2022} deployed a photonic integrated sensing and communication system in the Pearl River estuary, effectively tracking a ferry route, though only one vessel type and route were examined. Wienecke et al.\cite{wienecke2023new} detected bottom trawlers via seabed vibrations up to $2.5\,km$ from the cable, but tested no multi-vessel scenarios. Similarly, Drylerakis et al.\cite{Drylerakis2024source} used PCA-based denoising to improve signals from a single passing vessel.

Other works emphasize shorter cable deployments or more constrained tests. Dias et al.\cite{dias2014} used a $500\,m$ cable in shallow waters to detect artificial ship noise, while Douglass et al.\cite{Douglass_2023} captured broadband acoustic signals along a $3.5\,km$ long cable and compared DAS responses to hydrophones. Paap et al.\cite{paap2025leveragin} demonstrated automatic detection and localization for three specific vessels, achieving good performance but not exploring more diverse vessel types. Martins et al.\cite{Martins2026ModelingSurface} also explored physics-based strategies to develop a mathematical framework to derive the hyperbolic vessel moveout equations, yet to be validated as their experimental work is limited to two vessels. Shao et al.\cite{shao2025tracking} used Doppler shifts to track ships in a short $180\,m$ river cable deployment; the method aligned well with GPS but only covered three vessel passages.

Very recently, Huang et al.\cite{Huang_2025} proposed DASHip, a large-scale labeled dataset for ship detection based on AIS-guided annotation of 55,875 events. They reported strong detection performance (73.6\% joint detection ratio) using an object visual detection algorithm (YOLO), yet important higher-frequency acoustic cues ($10--200\,Hz$) were sacrificed by downsampling to $10\,Hz$, and their wake-centric approach could not identify vessels before cable crossings. This domain gap between clean labels and real-world conditions potentially reduces generalization~\cite{Karasalo_2017, Landro_2022}. Shan et al.\cite{shan2026distributed} extended this image-based DAS monitoring paradigm, proposing a multi-task learning framework for vessel detection, type classification, and heading classification from enhanced DAS spatio-temporal maps; however, the study is focused again on image-based vessel-attribute extraction and have the same dataset and wake-centric limitations than~\cite{Huang_2025}, as they used a subset of the DASHip data. Van Oers et al.\cite{Oers2025EnabledDetection} also explored YOLO-based vessel detection using public submarine DAS data and AIS-derived cable-crossing labels, showing that Fourier-spectrum representations improved detection with respect to seismic-power images; however, their evaluation was limited to 26 labeled cable crossings and did not address continuous cable-relative distance estimation. Similarly, Anhaus et al.\cite{Anhaus2025TowardsShip} investigated signal-based features for DAS ship detection but, again, evaluating a single ship passage over the explored cable.

Beyond direct ship detection, Landr{\o} et al.\cite{Landro_2022} analyzed various acoustic phenomena, including seismic events and whale calls, using an Arctic cable. They tracked an 86-meter cargo ship via hyperbolic wavefront analysis and near-field beamforming. Stork et al.\cite{Stork_2020} likewise demonstrated microseismic and whale vocalization detection on a $2\,km$ cable, highlighting the versatility of DAS for multi-purpose acoustic monitoring. While such studies do not center on vessel detection, they confirm DAS’s broad applicability in ocean acoustic sensing. Pedersen et al.~\cite{pedersen2025feasibility} recently proposed an automated DAS signal-detection and classification framework using KL-divergence-based change detection, spectral features, PCA, and HDBSCAN, showing effective discrimination among ships, vehicles, earthquakes, and cable-related signals; however, the work addresses general signal classification rather than cable-centered vessel detection and distance estimation, and their vessel detection task is limited to 41 AIS-labeled ship signals. Similarly, Zhang et al.~\cite{zhang2026deeplearningframeworkmarine} very recently proposed DASNet, a Mask-RCNN-based deep-learning framework for multi-class marine DAS monitoring, but their experimental work on vessel localization only considers a two-hour continuous recording of one vessel passing near the cable. The locations inferred by their network formed a coherent trajectory that closely matched the vessel's actual movement recorded by the AIS, but the very limited data does not allow for performance extrapolation.

From the above discussion, despite demonstrating the feasibility of DAS-based vessel detection, previous works tend to rely on very small or narrowly focused datasets, single vessels, or limited scenarios. Many studies lack broad environmental variability, often considering short cable segments, and do not systematically address cable protection goals such as quantifying vessel proximity or implementing detection prior to a direct cable crossing. In contrast, our proposal is explicitly oriented to submarine cable protection, jointly addressing vessel detection and distance estimation in support of early-warning monitoring, and validating the approach on a substantially larger and more diverse real-world dataset than typically reported in the literature.

\begin{figure*}
  \centering
  \mysubfigure{0.296\textwidth}{Regional map.}{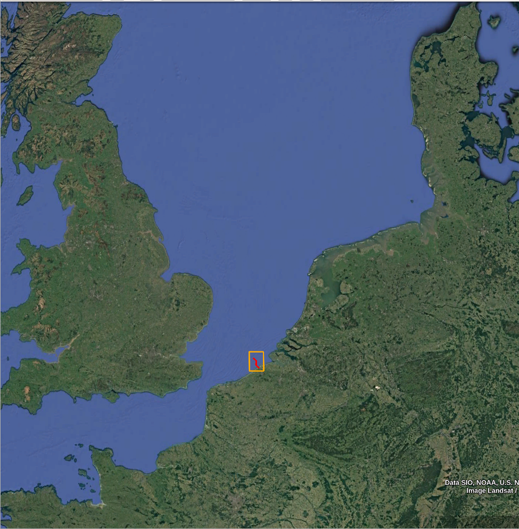}
  \label{fig:regional-map}
  ~
  \mysubfigure{0.294\textwidth}{Local map showing cable (red line).}{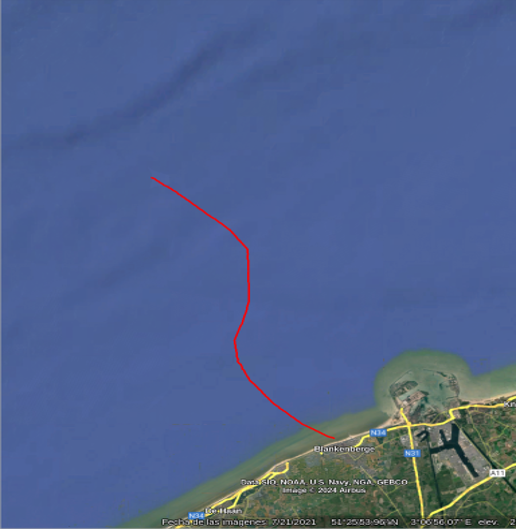}
  \label{fig:local-map}
  ~
  \mysubfigure{0.38\textwidth}{Local map showing cable (red line) and  bathymetry.}{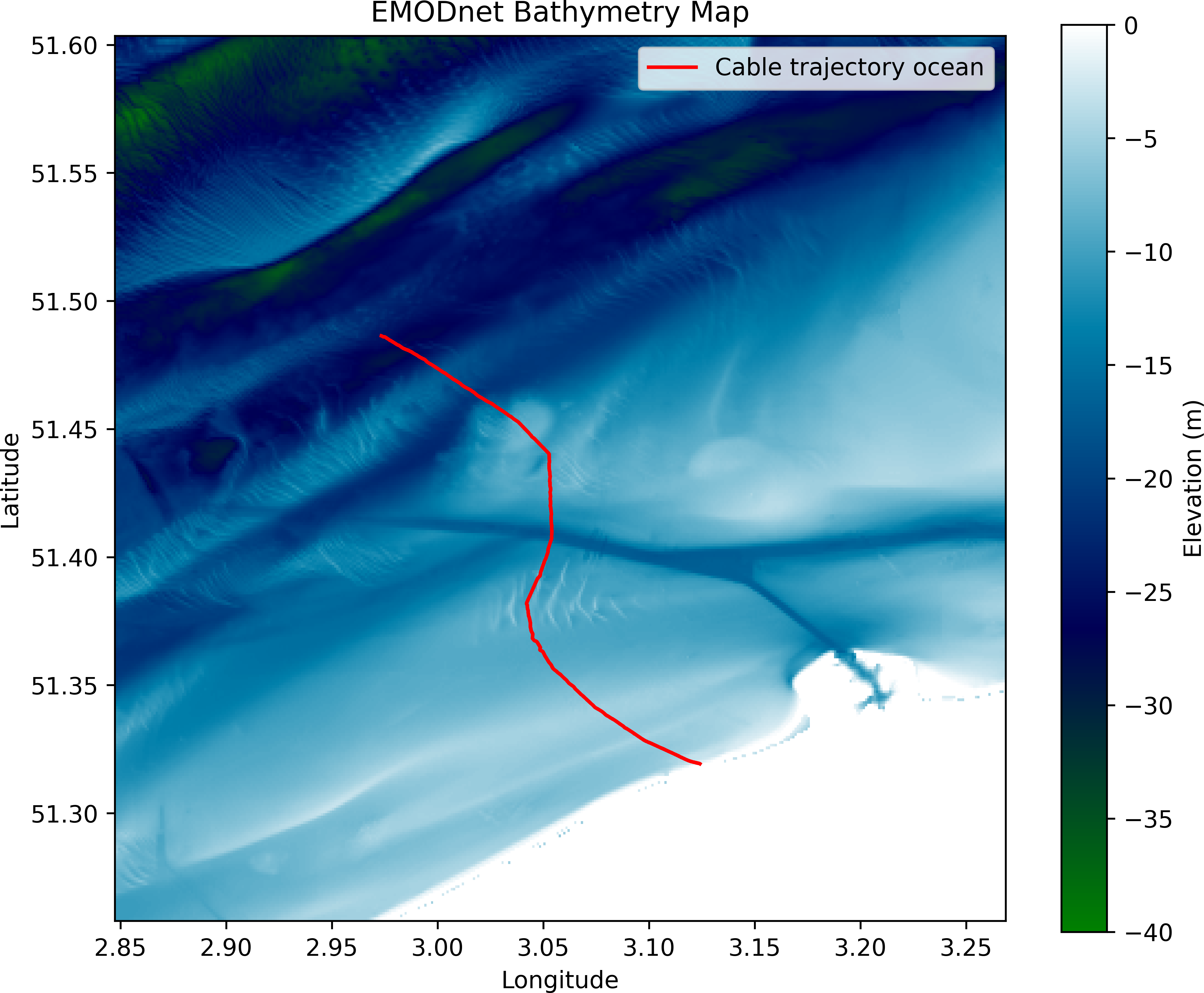}
  \label{fig:bathymetry-map}
  \caption{General location and bathymetry (cable location has been displaced for security considerations).}
  \label{fig:geographical-location}
\end{figure*}

\section{Distributed Acoustic Sensing in Submarine Cables}
\label{sec:das-submarine}

Optical fiber-based distributed acoustic sensors are advanced sensing systems that use optical fibers as the primary medium for detecting and monitoring dynamic acoustic signals along their entire length. Unlike traditional sensor networks, DAS systems convert a conventional optical fiber into a vast array of virtual sensors that can measure dynamic strain changes along its entire length~\cite{Zhan_2020}. Conventional DAS technology probes the fiber-optic cable with a coherent laser pulse, and measures changes in the phase of the returning Rayleigh back-scatter signal time-series. Optical phase shifts in the back-scattered signal are proportional to longitudinal deformation in the fiber and can be translated into local strain changes across a fiber segment (termed gauge length $L$) by integration. The primary issue with the conventional phase-measuring approach is that a substantial fraction of positions along the fiber exhibit zero or near-zero intensity. 
These low-intensity points (typically called “fading points”) in the back-scattered signal lead to a defective determination of the phase of the local back-scattered field, implying inhomogeneous sensitivity variations across the DAS array~\cite{gabai2016sensitivity}. Phase-measuring DAS systems generally available in the market achieve a fully distributed measurement of vibrations of up to several kHz, with a resolution of a few meters and measurement ranges of a few tens of kilometers. Optical coding methods are one of the preferred approaches for extending the measurement range of DAS systems. 
Among these, optical pulse compression reflectometry~\cite{Zou_2015} has been widely developed, allowing measurement ranges well exceeding $100\,km$~\cite{Waagaard_2021}.

Regardless of the interrogation method, DAS sensitivity depends on the optical fiber type, installation conditions, and local fiber-ground coupling. In existing installations, these factors are essentially not controllable; thus, effective data processing algorithms must account for them. Recent advances in signal processing and machine learning techniques have enabled improved handling of coupling variations, significantly enhancing DAS performance and making it a powerful tool for complex, large-scale monitoring tasks~\cite{munoz2022enhancing}.

One of the mainstream applications of DAS is the monitoring of submarine cables for various purposes. In particular, DAS technology has proven effective in the surveillance of both communication and energy transmission submarine cables~\cite{Brenne2019, Malaprade2019}. Traditional offshore monitoring methods typically rely on fixed platforms, buoys, or ocean-bottom point sensors, offering limited spatial coverage. This constraint hinders the detection of dynamic and spatially diverse events, such as maritime activity involving vessels, large marine mammals, or other wildlife, occurring at the sea surface or seabed.

\section{Database Description}
\label{sec:database-description}

\subsection{Geographical and Instrumentation Details}
\label{sec:geographical-details}
In our study, we used DAS and AIS data collected over a 10-day period between June $16^{th}$ and $25^{th}$ 2023, to develop models for vessel detection and localization.

The dataset was recorded using a $28\,km$ long pre-existing ocean-bottom fiber-optic cable in the Southern Bight of the North Sea offshore Zeebrugge, Belgium~(see Fig.~\ref{fig:geographical-location}). The fiber-optic cable was originally installed to monitor a power cable from an offshore wind farm facility
. The cable is buried between $1.4\,m$ and $7.2\,m$ below the seafloor, with an average burial depth of $4.7\,m$, which provides a good acoustic coupling between the seabed vibrations and the fiber~\cite{paap2025leveragin}. The Alcatel OptoDAS interrogator~\cite{OptoDAS} was used for strain data acquisition. It is a phase-measuring DAS system that uses optical pulse compression reflectometry to extend the measurement range. The gauge length in this case is $L\approx10\,m$, which is significantly larger than the nominal optical resolution of the system ($1\,m$), related to the bandwidth of the optical probing signal used. This significantly reduces the sensitivity issues related to fading points.

The interrogator operates with a spatial resolution of $10.21\,m$ (channel spacing), creating $2774$ simultaneously recording strain sensors, and generating differential phase data signals at a $f_{s}=3125\,Hz$ sampling frequency. These raw signals are further preprocessed following the procedure described in Section~\ref{sec:signal-preprocessing}.

\subsection{Metadata Processing}
\label{sec:metadata-processing}

Effective analysis of the DAS data required careful processing and integration of several crucial auxiliary datasets (metadata). This section details the acquisition, characteristics, and processing steps applied to the vessel tracking data (AIS), the fiber's geographical and bathymetric information, and addresses key considerations regarding data balance and spatio-temporal calibration necessary for the subsequent analysis. Fig.~\ref{fig:system-architecture} illustrates how these metadata processing steps are integrated within the overall system architecture presented in section~\ref{sec:proposed-system}.

\subsubsection{AIS Data}
\label{sec:ais-data}

In this study, AIS data were employed primarily for geographically labeling the DAS data, providing ground-truth vessel positions, which are  essential for both training and evaluation purposes~\cite{emmens2021promisesAIS}. The raw AIS data, provided by the cable owner, included Maritime Mobile Service Identity (MMSI) numbers, UTC timestamps, vessel positions (longitude and latitude), course, speed, and heading for 745 unique vessels, totaling 64,417 reported positions. We further augmented this dataset by web-scraping static vessel information (name, IMO number, type, dimensions, tonnage), although dynamic details like navigational status and draught were unavailable. Analysis of vessel types revealed 45 distinct categories, with cargo ships being the most prevalent in the monitored area, alongside notable presences of fishing vessels and tugs (see Table~\ref{tab:vessel-stats-number-per-type} for detailed statistics, extracted from the cable owner raw AIS data).
\begin{table}[htbp]
  \centering
  \caption{Vessel type distribution in AIS data (sorted by occurrences).}
  \label{tab:vessel-stats-number-per-type}
  \resizebox{\columnwidth}{!}{
    \begin{tabular}{|c|c|c|c|}
      \cline{2-4}    \multicolumn{1}{c|}{} & \# vessels & \% total & Avg length \\
      \cline{2-4}\hline
      Container Ship & 134   & 18\%  & 263 \\
      \hline
      Chemical/Oil Products Tanker & 124   & 17\%  & 151 \\
      \hline
      General Cargo Ship & 96    & 13\%  & 114 \\
      \hline
      Vehicles Carrier & 62    & 8\%   & 182 \\
      \hline
      Bulk Carrier & 61    & 8\%   & 194 \\
      \hline
      Unknown & 41    & 6\%   &  N/A \\
      \hline
      Ro-Ro Cargo Ship & 39    & 5\%   & 201 \\
      \hline
      Lpg Tanker & 29    & 4\%   & 128 \\
      \hline
      Hopper Dredger & 19    & 3\%   & 106 \\
      \hline
      Fishing Vessel & 19    & 3\%   & 22 \\
      \hline
      Tug   & 16    & 2\%   & 31 \\
      \hline
      Crude Oil Tanker & 10    & 1\%   & 248 \\
      \hline
    \end{tabular}%
  }
  \label{tab:addlabel}%
\end{table}%

\begin{figure}[!t]
	\centering
  \includegraphics[width=.8\columnwidth]{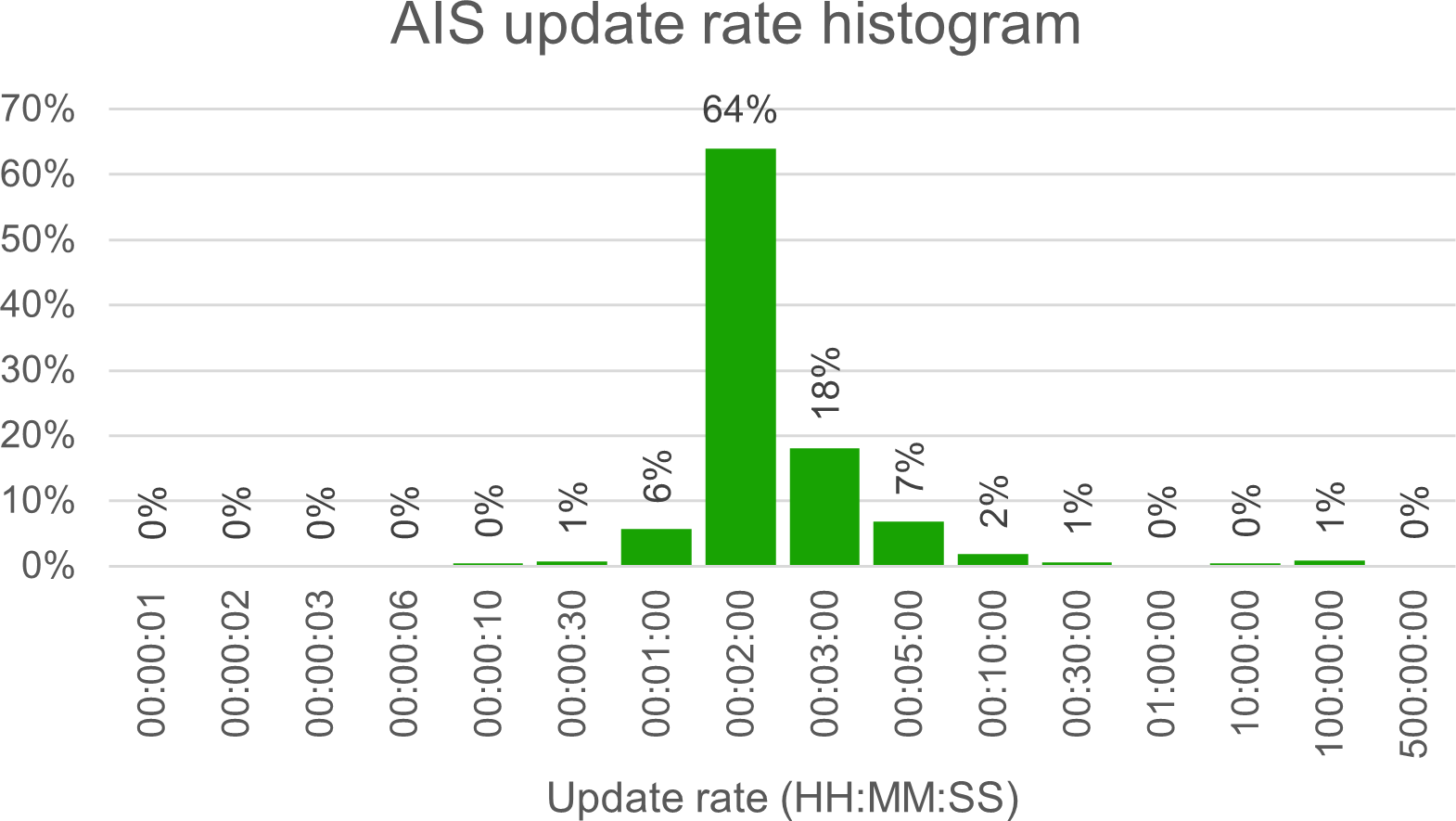}
	\caption{AIS data update rate histogram for the used dataset.}
	\label{fig:ais-update-rate-histogram}
\end{figure}

\begin{figure}[!t]
	\centering
	\includegraphics[width=.8\columnwidth]{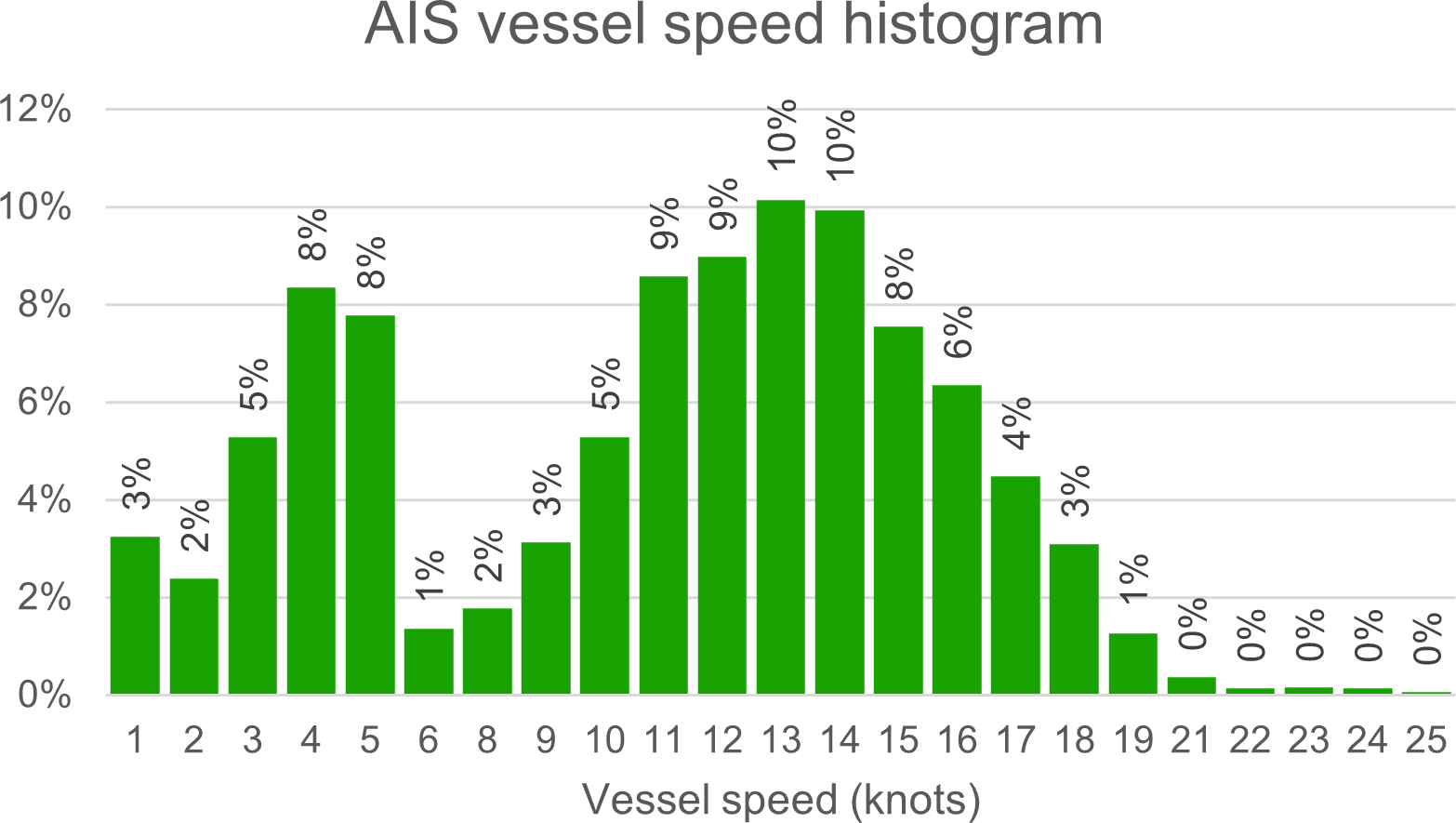}
	\caption{AIS vessel speed histogram for the used dataset.}
	\label{fig:ais-speed-histogram}
\end{figure}

A significant challenge encountered when utilizing AIS data for machine learning applications, including ours, is the typically low and variable update rate~\cite{Harati-Mokhtari_Wall_Brooks_Wang_2007AIS, emmens2021promisesAIS}. While ITU recommendations suggest reporting intervals under 30 seconds for most moving vessels~\cite{ITU_R_M1371_5AIS}, our dataset exhibited much lower frequencies, with $88\%$ of vessels reporting only every 1 to 3 minutes (see Fig.~\ref{fig:ais-update-rate-histogram} for details). Considering the average vessel speed of 10.2 knots observed in the dataset (Fig.~\ref{fig:ais-speed-histogram}), such infrequent updates can lead to significant positional uncertainty (approximately 315 meters per minute in average), preventing accurate labeling. To address this limitation, we applied linear interpolation between reported positions with a 1-second resolution, excluding intervals with vessels not reporting during 60 minutes, and incorporating additional heuristic rules to minimize the generation of erroneous trajectories. Given the update rate distribution and the good results achieved, we believe the linear interpolation is good enough for the explored scenario, although in more demanding situations, alternative interpolation methods should be adopted~\cite{guo2021improved} (more details in the \href{https://geintra-uah.org/psi/index.html#interp}{supplementary material Web page}). This interpolation process resulted in a significantly denser dataset of 1,207,446 position entries used for subsequent analysis.
Therefore, AIS-derived labels should be interpreted as interpolated ground-truth estimates for cooperative vessels, rather than as direct continuous measurements of vessel position. In addition, AIS does not provide complete information on the vessel acoustic state, such as propulsion regime, loading condition, draught, or radiated source level.

\subsubsection{Geometrical + bathymetry data acquisition}
\label{sec:bathymetry-information}

Accurate knowledge of the fiber-optic cable's geographical position and the surrounding bathymetry is critical for DAS-based location-dependent monitoring. Variations in seafloor morphology and cable burial conditions directly impact the coupling between the cable and the seabed, significantly influencing the quality and characteristics of the recorded acoustic signals~\cite{sladen2019distributed, Mata_Flores_2023}. Furthermore, precise fiber localization is essential for accurately positioning detected events, such as vessel passages. While cable deployment aims to follow a specific trajectory, the final resting position on the seafloor can deviate, and its burial status may change over time~\cite{sladen2019distributed}.

\begin{figure*}[!t]
	\centering
	\mysubfigure{0.49\textwidth}{\scriptsize Full fiber-optic cable.\label{fig:bathymetry-cmp-emodnet-owner}
	}{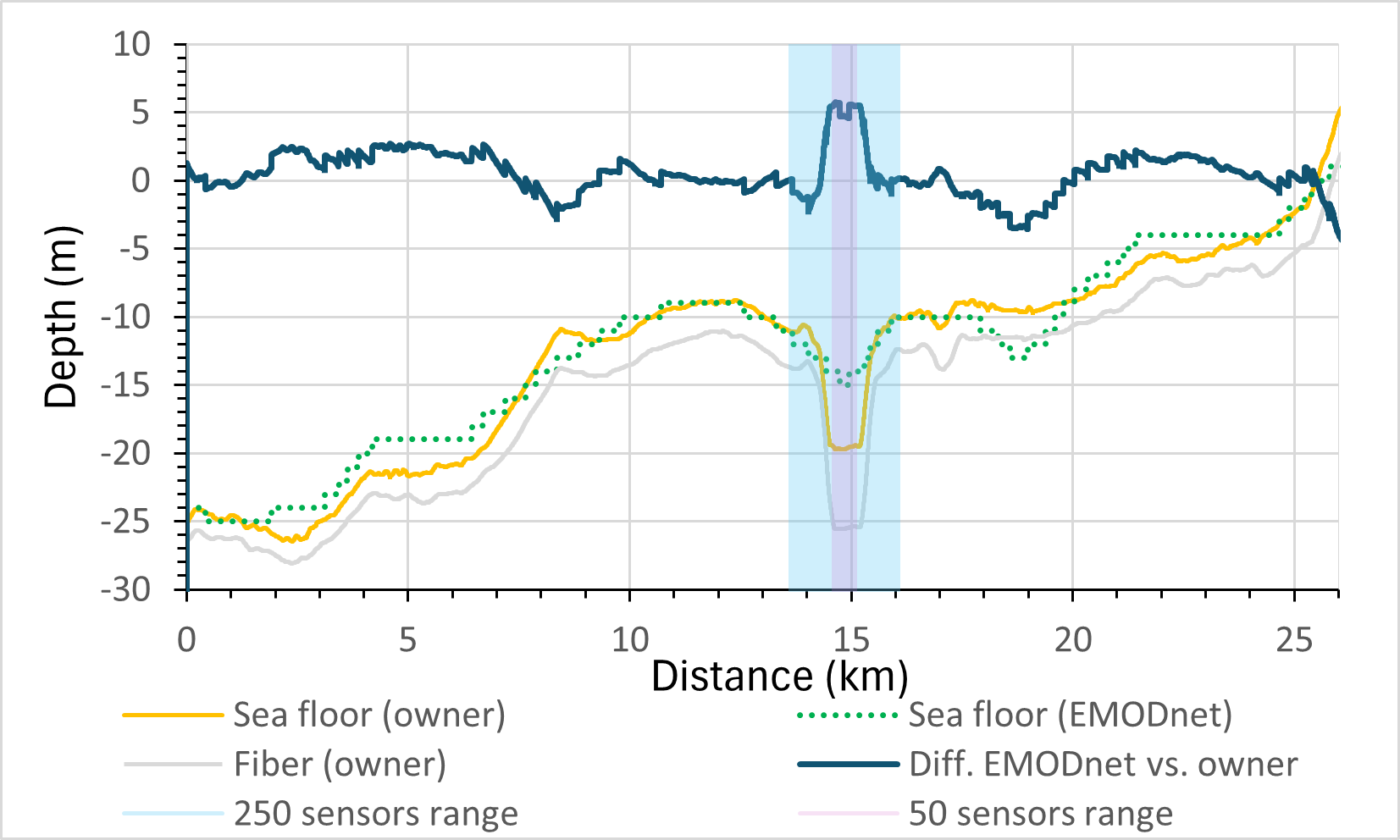}
	~
	\mysubfigure{0.49\textwidth}{\scriptsize Selected fiber-optic cable region.\label{fig:bathymetry-cmp-emodnet-owner-zoom}}{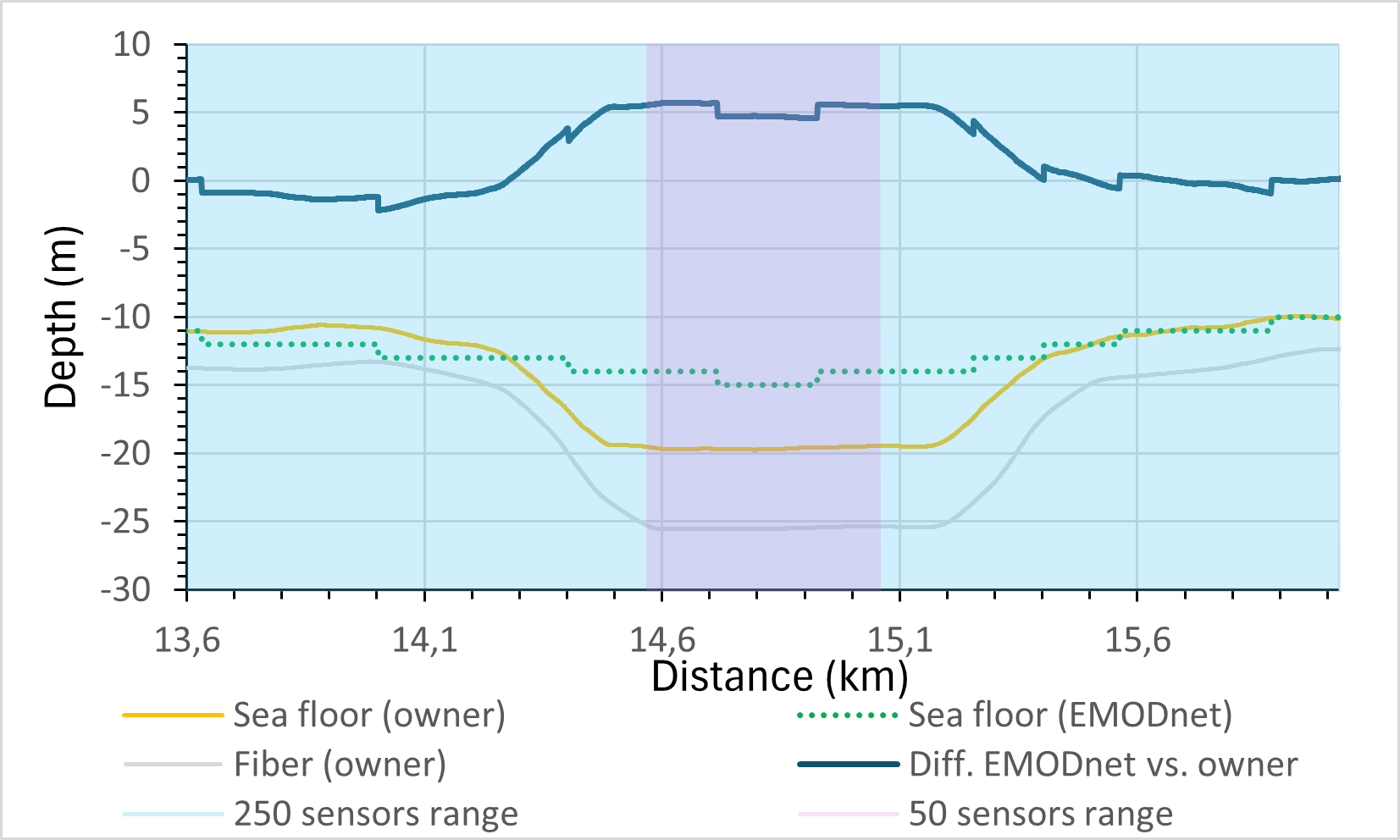} 
	\caption{Comparison of bathymetry sources in the cable under study. \textit{Sea floor} depth provided by the cable owner (orange trace) and EMODnet (green dotted line). The \textit{Fiber} position (buried cable) is also shown (gray trace). \textit{Diff. EMODnet vs. owner} measures depth differences between EMODnet and owner data (dark blue trace). Selected fiber-optic cable regions are shown in blue background (250 sensors range) and pink background (50 sensors range).}
	\label{fig:bathymetry-cmp}
\end{figure*}

For this study, we utilized precise geometrical and bathymetry measurements for the full $28\,km$ cable length, provided directly by the cable owner. This dataset offers significantly higher precision than publicly available sources like EMODNet~\cite{emodnet2024} (which has a resolution of 115 meters). A comparison, shown in Fig.~\ref{fig:bathymetry-cmp}, reveals depth discrepancies of up to 6 meters between the owner's data and EMODNet along the cable path used in this work.

Fig.~\ref{fig:bathymetry-cmp}.a illustrates the bathymetry along the entire cable, while Fig.~\ref{fig:bathymetry-cmp}.b provides a detailed view of the specific segment used for our experiments. This region corresponds to a section where the cable crosses a dredged artificial channel (approximately $20\,m$ deep) providing harbor access. Here, the cable itself is buried at depths reaching $25\,m$. Our experiments analyzed data from segments spanning $102.1\,m$ to $2552.5\,m$ (corresponding to $10$ and $250$ sensed positions), as indicated by the highlighted blue and pink regions in Fig.~\ref{fig:bathymetry-cmp}.a and Fig.~\ref{fig:bathymetry-cmp}.b, respectively. The use of this high-resolution, owner-provided data ensures accurate geographical referencing for our analysis.

\begin{figure}[!b]
  \centering
  \includegraphics[width=\columnwidth]{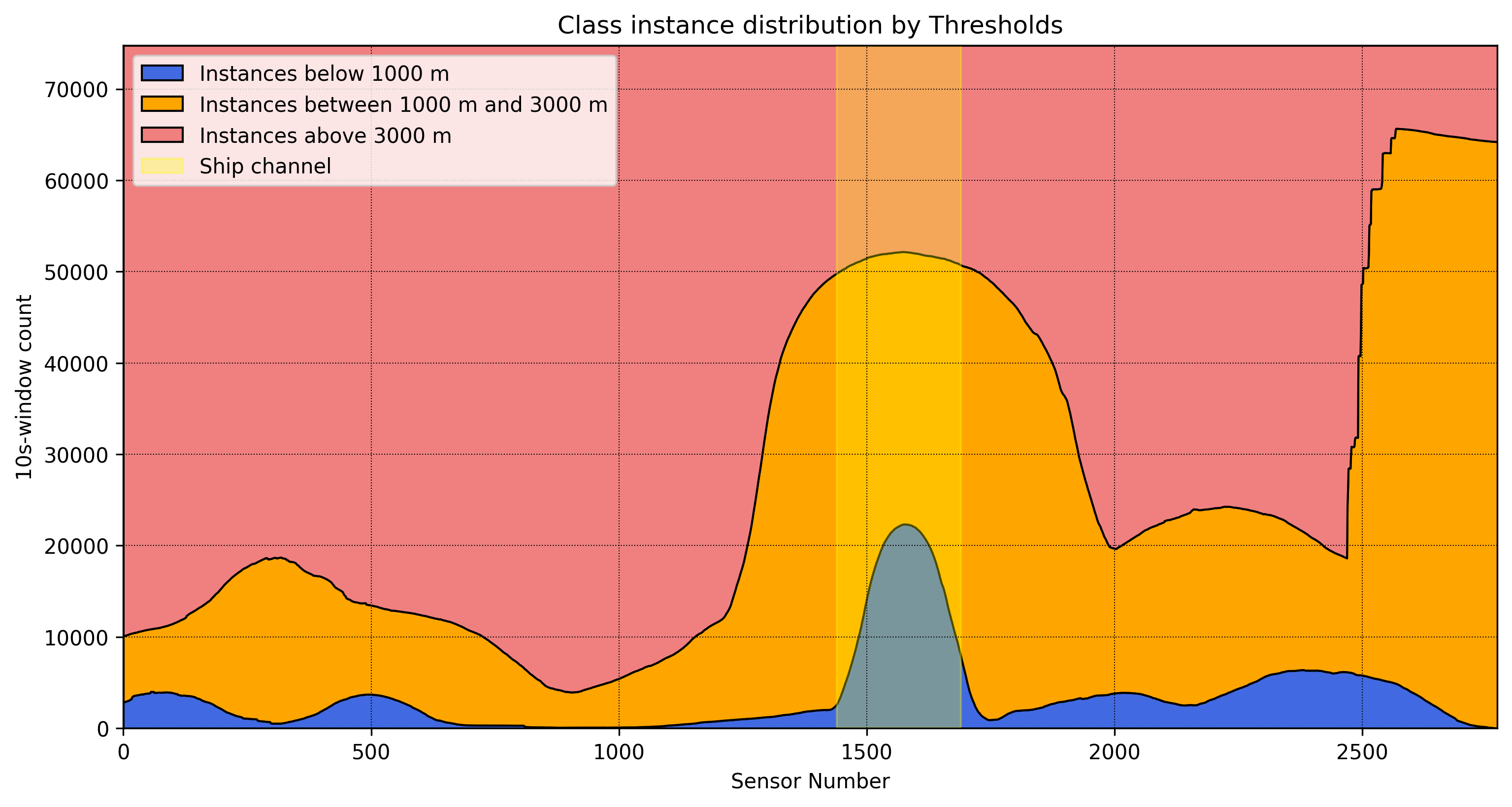}
  \caption{Class instance distribution for different distance thresholds.}
  \label{fig:data-balance}
\end{figure}
\subsubsection{Data calibration and synchronization}
\label{sec:data-proc-calibr}

Accurate geographical localization is essential, requiring each DAS sensed data point to be mapped first to a specific location along the fiber, and then to its corresponding geographical coordinates (longitude, latitude, and depth). This geographical calibration process typically relies on identifying anchor points (locations with known coordinates) along the fiber. Establishing the mapping between fiber distance and geographical position, whether automatically or semi-automatically, can be challenging, particularly for sections far from the sensing equipment.

The best approach to fulfill the geographical calibration task would be direct measurements of the cable position (longitude, latitude and depth) combined with knowledge on anchor points of well-known fiber positions. This scenario is not typically realistic, so that alternative methods based on general bathymetry sources combined with cable deployment information and interpolation strategies can also be used.

Finally, the availability of UTC timestamps in AIS data allows time synchronization with the acquired DAS data, which is also accurately timed.

\subsection{Data Balance Considerations}
\label{sec:data-balance-cons}

The classification task categorizes data frames based on vessel proximity relative to a distance threshold. Our dataset exhibits significant class imbalance, meaning the number of frames with vessels closer versus further than the threshold is often unequal. This imbalance varies considerably both along the fiber's length and depending on the specific distance threshold chosen. As an example, Fig.~\ref{fig:data-balance} shows the distribution along the fiber length (horizontal axis) of the number of available data frames with a vessel closer than $1000\,m$ (blue area), data frames with vessels further than $3000\,m$ (reddish area), and data frames with vessels at distances between them (orange region). Consequently, these data balance characteristics were crucial considerations in our experimental design and required careful interpretation of results, particularly favouring performance scores that properly assess the results in unbalanced scenarios.

\subsection{Data Availability}
\label{sec:data-availability}

Due to data-owner restrictions, the original raw differential-strain recordings, the precise geographical localization data, and the full AIS records cannot be made publicly available. Instead, we have released the \MarlinksDataset{} dataset in Zenodo~\cite{ramirez2024dasvesseldataset}, that includes the complete set of processed data for the experiments in this work. It comprises the feature vectors and their corresponding ground-truth labels (timestamp, closest-vessel distance information, and vessel related metadata that includes vessel type and size) in a structured \texttt{hdf5} archive. To ease data distribution and reproducibility, we have also generated a GitHub repository (available at \url{https://github.com/UAH-PSI/das-vessel-detection}) that provides the data description, initial source code for loading and processing the data (to be completed with the evaluated machine learning approaches), as well as a small showcase data subset. The public release of the processed dataset and source code will also enable future fair comparisons with alternative methods, under a common benchmark.


\section{DAS+ML system for vessel detection and localization}
\label{sec:proposed-system}

\begin{figure*}[!t]
  \centering
  \includegraphics[width=\textwidth]{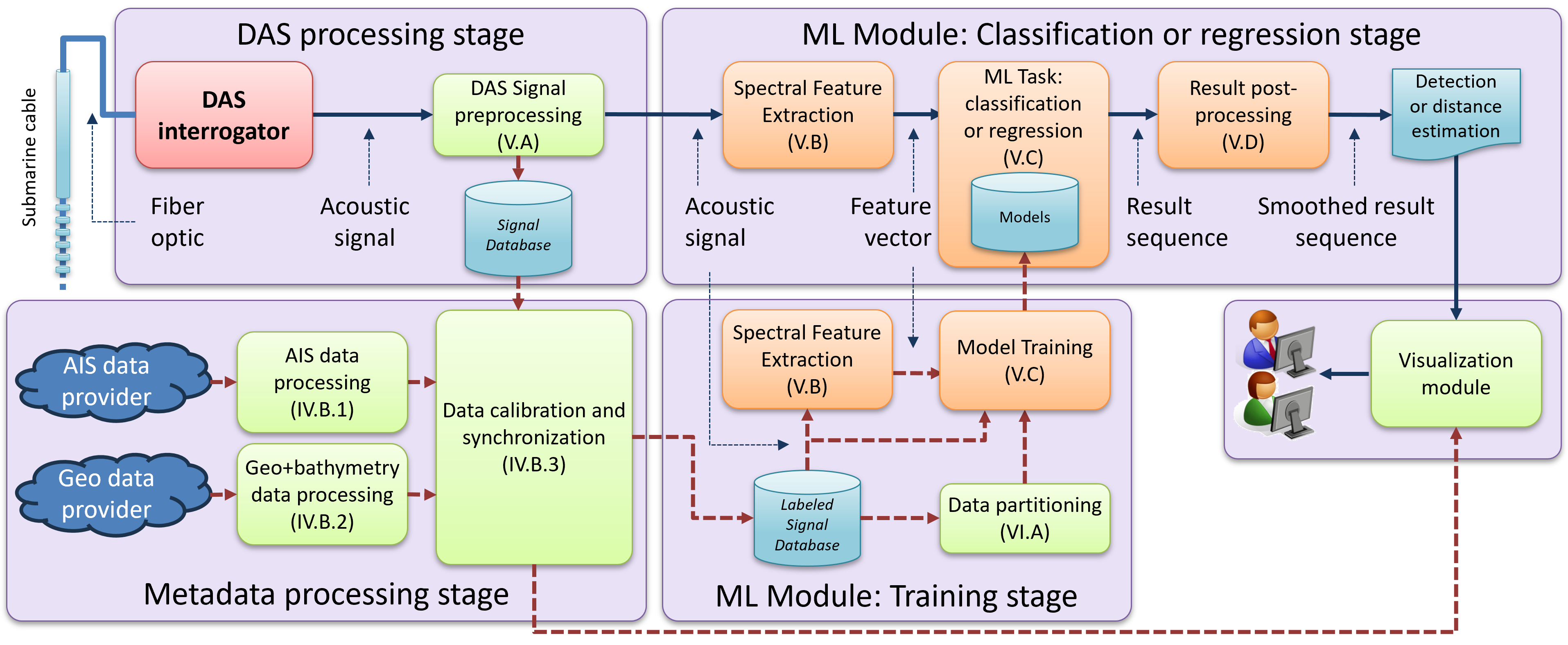}
  \caption{Detailed architecture of the proposed system, with references to the corresponding paper sections for the relevant modules.}
  \label{fig:system-architecture}
\end{figure*}

The detailed system architecture proposed in this work is shown in Fig.~\ref{fig:system-architecture}, which we detail next.

\subsection{DAS Signal Preprocessing}
\label{sec:signal-preprocessing}

The DAS interrogator used in this work is a phase DAS, and the generated raw signal is differential phase (proportional to the average axial strain variation over the DAS gauge length). To convert this into a direct strain measurement, the following signal preprocessing operations are carried out (additional implementation details are provided in the \href{https://geintra-uah.org/psi/index.html#spike}{supplementary material Web page}):
\begin{itemize}
    \item {Data scaling:} Adjusts the raw differential phase measurements based on system gain and calibration coefficients, ensuring subsequent steps operate on signals with consistent amplitude levels.
    \item {Spike detection and removal:} Removes impulsive spikes using an adaptive thresholding method, where samples exceeding a sliding-window mean plus an offset are set to zero before integration.
    \item {Phase unwrapping:} Converts the wrapped phase (limited to $[-\pi, \pi)$) into
    a continuous time series by correcting abrupt $2\pi$ jumps, thus preserving the true optical path length changes.
    \item {Time integration:} Integrates the unwrapped phase over time and scale it to recover the cumulative strain in the fiber
    . This step yields a direct strain signal rather than a differential measurement. Fig.~\ref{fig:energy-dt-map}.a shows an example of a strain signal acquired during a vessel crossing above the cable.
\end{itemize}
Strain-rate and strain are linearly related in the frequency domain (essentially involving a multiplication of the strain-rate spectrum by $1/j\omega)$, 
so that the use of either magnitude would simply lead to changes in model tuning, and being related by a linear operation, it should definitely not affect the system performance.

\subsection{Spectral Feature Extraction}
\label{sec:spectral-analysis}

Extracting robust features from DAS data for vessel detection and localization is complicated due to the intricate physical processes converting ship noise into measured strain. Accurately modeling this conversion is challenging due to several factors: the complex and variable nature of noise radiated by different vessels~\cite{Rivet_2021}; the complexities of underwater acoustic propagation, including attenuation, reflection, and Doppler effects~\cite{Karasalo_2017, vadov2001long, li2024precise}; variations in the coupling between the fiber cable and seabed, which strongly influence strain measurements~\cite{Lior_2021, Mata_Flores_2023}; and potential artifacts introduced by the DAS system itself (due to electronic and optical design, and the conditions in the interrogator deployment environment).

Given that acquiring the precise knowledge needed for a full physics-based model is often impractical due to environmental variability and incomplete deployment information, our proposal is adopting a purely data-driven strategy. A frequent technique involves analyzing the energy within specific frequency bands, which has proven effective for interpreting DAS signals and providing features for associated ML tasks~\cite{tejedor2016toward,hartog2017introduction,Malaprade2019}. Adopting this principle, our work utilizes a data-driven spectral analysis, informed by available AIS data, to generate feature vectors suitable for guiding the ML process. Further details are given in Section~\ref{sec:feature-extraction-analysis}, as it implies experimental work for feature vector design.
\subsection{Machine Learning Tasks and Training Modules}
\label{sec:machine-learning-strategy}

The experimental work can be oriented to different ML tasks. In this work, we focus on vessel detection and localization, as follows:

\begin{itemize}
  \item Vessel detection: This addresses a binary classification task, and aims to detect the presence of any vessel at a distance closer than a given threshold. This threshold must be designed considering  criteria related to security, fiber response capabilities, and data balance. Regarding security considerations, in a real world deployment, an alarm should be triggered if the vessel is a potential threat, and with enough time in advance to be able to take the necessary measures. As an example, a Spanish company managing deployed submarine cables being monitored with AIS data to prevent vessel threats to the cable, sets a distance threshold of $1000\,m$ to consider further detailed vessel tracking. Also, a suitable distance threshold should be chosen to allow a detectable response in the fiber (which may be an issue in case buried cables are used), with a reasonable data availability balance between the two classes (threat and non-threat).
  \item Vessel localization: This addresses a regression task, aimed to estimate the distance of the closest vessel to the cable. In this case, distance thresholds need to be set to deal with distance estimation at different distance ranges, assuming that different models will be built for each range.
\end{itemize}

To tackle these detection and localization tasks, we evaluated two well-established machine learning approaches:
\begin{itemize}
	\item {XGBoost (eXtreme Gradient Boosting):} An optimized gradient boosting framework that learns an ensemble of decision trees~\cite{Chen:2016:XST:2939672.2939785}. XGBoost was selected for its proven effectiveness, scalability to large datasets, and inherent mechanisms to mitigate overfitting, making it suitable for handling potentially high-dimensional feature spaces derived from DAS data.
	\item {Neural Network (NN):} A simple feed-forward neural network architecture was also implemented, providing a contrast based on NN learning principles.
\end{itemize}

The objective of this work is not to exhaustively optimize machine-learning architectures, but to validate a complete DAS-based vessel-detection and distance-estimation framework under realistic submarine-cable conditions. Therefore, we used fixed, representative configurations for XGBoost and the NN across all folds and experimental variants. 

In both classification and regression scenarios, an associated training process is required to generate the corresponding models. The availability of sufficient and adequately balanced data among the considered classes or distance ranges is crucial. This issue can be particularly significant in marine environments around deployed submarine cables, as maritime traffic varies heavily. Therefore, a detailed analysis of the available data is necessary to assess its suitability and guide the data acquisition and partitioning tasks.

Another key element that impacts the reliability and robustness of a system and its real generalization capabilities is the rigor of the experimental approach~\cite{tejedor2017machine}. In time-series data, such as DAS recordings, temporal dependencies can introduce biases if not properly handled. The typically used random partitioning strategy may result in training and validation sets containing temporally adjacent data, leading to information leakage and overestimation of model performance. Special care must be taken in this stage, adopting specific cross-validation strategies that include restrictions on the selection of temporal ranges for training and testing subsets.

Further details are given in Section~\ref{sec:experimental-setup}.

\subsection{Result Post-Processing}
\label{sec:result-smoothing-or}

In ML tasks dealing with time series, we can further improve the classification results by applying post-processing methods. Typical strategies are smoothing via sliding window averaging, median or mode filtering, majority voting (using ensembles or temporal ranges), multiple intervals, etc.~\cite{bagnall2017great}.

Finally, one key benefit of DAS is the extensive range of sensed positions, which allows the systems to exploit the spatial and temporal redundancy of the acquired data. This should be explicitly included in either the feature extraction or ML stages. Suitable alternatives include exploiting array signal processing~\cite{munoz2022enhancing}, or data integration across temporal and spatial ranges~\cite{tejedor2019contextual,huynh2025real}, which will be detailed in Section~\ref{sec:experimental-setup}.
\begin{figure}[!h]
  \centering
  \includegraphics[width=.85\columnwidth]{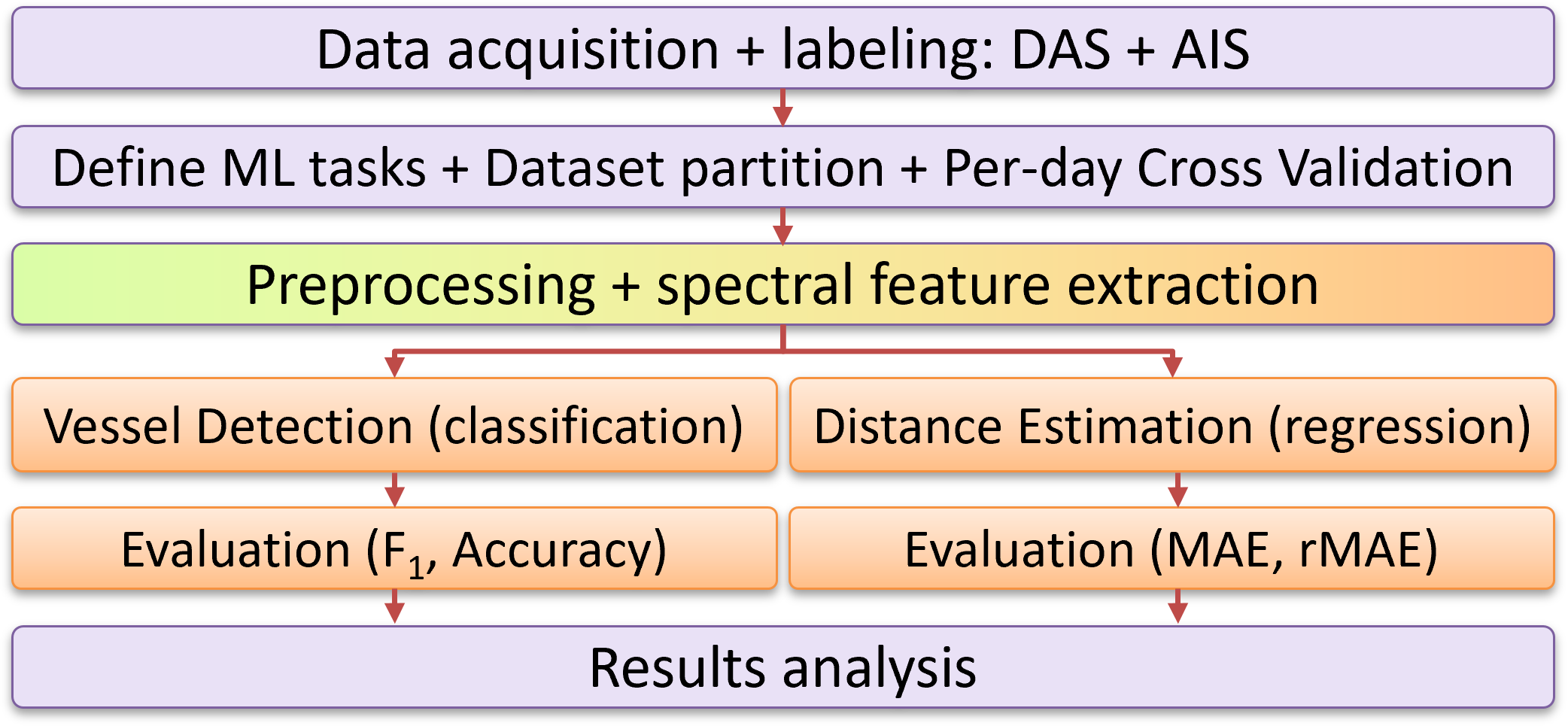}
  \caption{Methodological flowchart illustrating the sequential stages of the experimental study.}
 \label{fig:high_level_workflow}
\end{figure}
\section{Experimental Work and Results}
\label{sec:exper-work-results}

Fig.~\ref{fig:high_level_workflow} shows the simplified methodological flowchart of this study, 
complementing the architectural description of Fig.~\ref{fig:system-architecture}.
\subsection{Database Partitioning}
\label{sec:datab-part-exper}

Given the characteristics of our task, we adopted the $k$-fold cross-validation strategy, a widely used technique for assessing model generalization capability~\cite{Wong2019-kfold}. This involves dividing the whole dataset into $k$ equal subsets (folds). The model is trained on $k-1$ folds and validated (tested) on the remaining fold. This process repeats $k$ times, with each fold serving once as the test set. The results from all test sets are then averaged to estimate the overall performance. Therefore, this method ensures that every data point is used for both training and testing, providing a comprehensive evaluation of the model's performance while maintaining a strict separation between training and testing data.

As described in Section~\ref{sec:database-description}, the dataset was collected over a 10-day period to develop models for vessel detection and distance estimation. Therefore, to ensure a rigorous and robust evaluation and mitigate potential biases, we adopted a 10-fold cross-validation strategy, assigning each day's data to a distinct fold. That way, we maintain the temporal continuity and integrity within each fold and also ensure that the testing set comprises data points that are not temporally interleaved with those of the training set. Additionally, this data partition strategy ensures that higher variability is introduced in the model generation, promoting a better generalization on unseen data. Additional daily-fold statistics are provided in the \href{https://geintra-uah.org/psi/index.html#daily-folds}{supplementary material Web page}. Considering all $k$-folds, it is worth mentioning than only 11 vessels (out of $565$ in the vicinity of the explored sensor range) appear in more than 5 folds along the recording period, which indicates the difficulty of the task.

\subsection{Performance Metrics}
\label{sec:performance-metrics}
The performance metrics depend on the addressed task (vessel detection or vessel localization). For vessel detection, we define $Class\,0$ as the condition in which a vessel is closer than a predefined threshold distance, and $Class\,1$ when there is no vessel closer than the predefined threshold distance. The following classification performance metrics are used:

\begin{itemize}
	\item {Accuracy}: This is defined as the percentage of correctly classified instances for all classes (the higher, the better).

	\item {$\FOneScore{}$-score}: The harmonic mean of $Precision$ and $Recall$ (the higher, the better), computed as:
        $\FOneScore{} \nobreak = \nobreak 2 \cdot\nobreak \frac{Precision \cdot Recall}{Precision + Recall}$, where $Precision$ quantifies the number of correctly detected positive samples out of all the samples classified as positive, while $Recall$ quantifies the percentage of positive samples that are correctly detected.

        Because the class distribution depends on the selected distance threshold and can be imbalanced, we report both the global $\FOneScore{}$-score (referred to from now on as $\FOneScoreG{}$) and class-wise $\FOneScore{}$-scores (referred to as $\FOneScoreCZero{}$ and $\FOneScoreCOne{}$ for class 0 and class 1 $\FOneScore{}$-scores, respectively). 
        That way, class-wise $\FOneScore$-scores will be the main criterion for interpreting detection performance, as they allow us to assess the precision-recall balance of each class independently, avoiding conclusions dominated by the majority class. We will also report the precision and recall values for the selected final configuration.

  \item {Area Under the Curve (AUC)}: The area under the Receiver Operating Characteristic curve, which evaluates the trade-off between true and false positive rates.
\end{itemize}

On the other hand, for vessel localization, the regression model is evaluated using the {Mean Absolute Error (MAE)}. This measures the average absolute deviation between predictions and actual values (the lower, the better):
\begin{equation}
  MAE = \frac{1}{\NumDataFrames{}} \sum_{i=1}^{\NumDataFrames{}} |y_i - \hat{y}_i|,
\end{equation}
where $\NumDataFrames$ is the number of data frames for which we generate a distance estimation value $\hat{y}_{i}$, and $y_{i}$ is the real distance.

In the vessel distance-estimation task, where the MAE will be calculated on data subsets with different distance thresholds, we will also normalize the MAE by these distance thresholds ($\distanceThr \in \distanceThrSet$), to get a relative estimation of the error, as compared with this distance threshold: $rMAE = \nobreak \frac{MAE}{\distanceThr}$.

To compare the results between the two ML approaches, we will quantify the uncertainty of our performance metrics calculating confidence intervals using a non-parametric bootstrap procedure~\cite{efron1994introduction} (adapted to our $k$-fold cross-validation strategy) that has been further validated by subsequent works on bootstrap methods in ML~\cite{goo2024confidence, shabbir2024estimation}. Briefly, confidence intervals were obtained by resampling held-out daily confusion matrices, whereas regression intervals were obtained by resampling pooled held-out residuals/predictions. Additional details are provided in the \href{https://geintra-uah.org/psi/index.html#bootstrap}{supplementary material Web page}.

\subsection{Feature Extraction Analysis}
\label{sec:feature-extraction-analysis}

The initial spectral analysis we carried out to guide our feature extraction design (as discussed in Section~\ref{sec:spectral-analysis}) is based on the calculation of long term spectral averages for different environmental conditions related to the presence or not of nearby vessels. We defined \textit{nearby} in relation to different distance thresholds with respect to fiber sensing positions. Specifically, calculations were computed using data frames corresponding to fiber positions:
\begin{itemize}
  \item That had no nearby vessels, with distance thresholds of $3000\,m$ and $5000\,m$ (condition referred to as \textit{noise}).
  \item That had nearby vessels, with distance thresholds of $1000\,m$ and $2000\,m$ (condition referred to as \textit{vessels}).
\end{itemize}

The computation of the spectrum average for each recording is carried out from $10$-second windows. We also limited the displayed bandwidth to $250\,Hz$ as we found no significant information to appear in higher frequencies.

\begin{figure*}[ht]
  \centering
  \mysubfigure{0.33\textwidth}{\scriptsize Spectral content in the $(0.1$-$250)\,Hz$ range.\label{fig:spectra-comparison-fullBW-250Hz}}{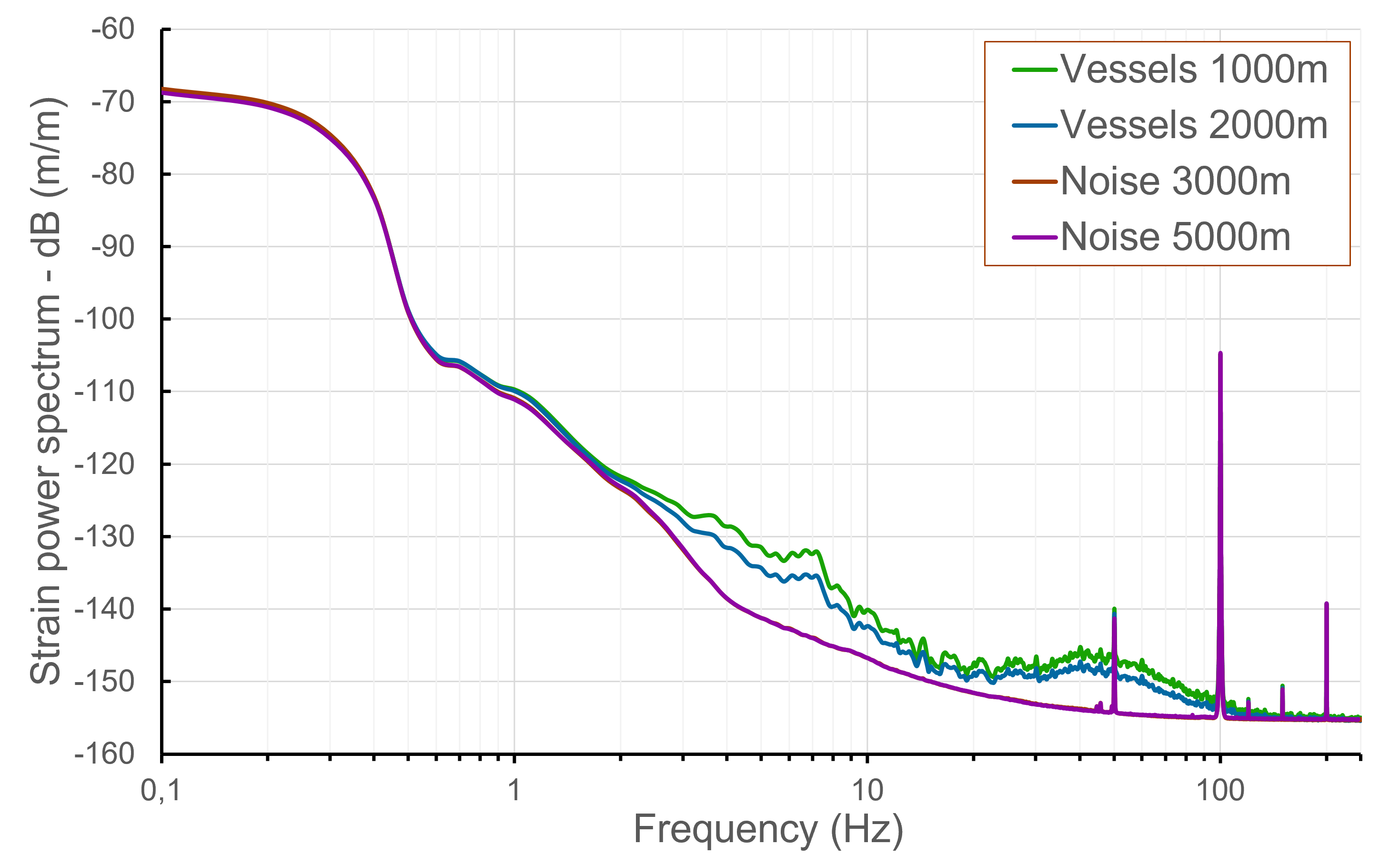}
    ~
  \mysubfigure{0.33\textwidth}{\scriptsize Zoomed spectral content (vessels only).\label{fig:spectra-comparison-4-250Hz}}{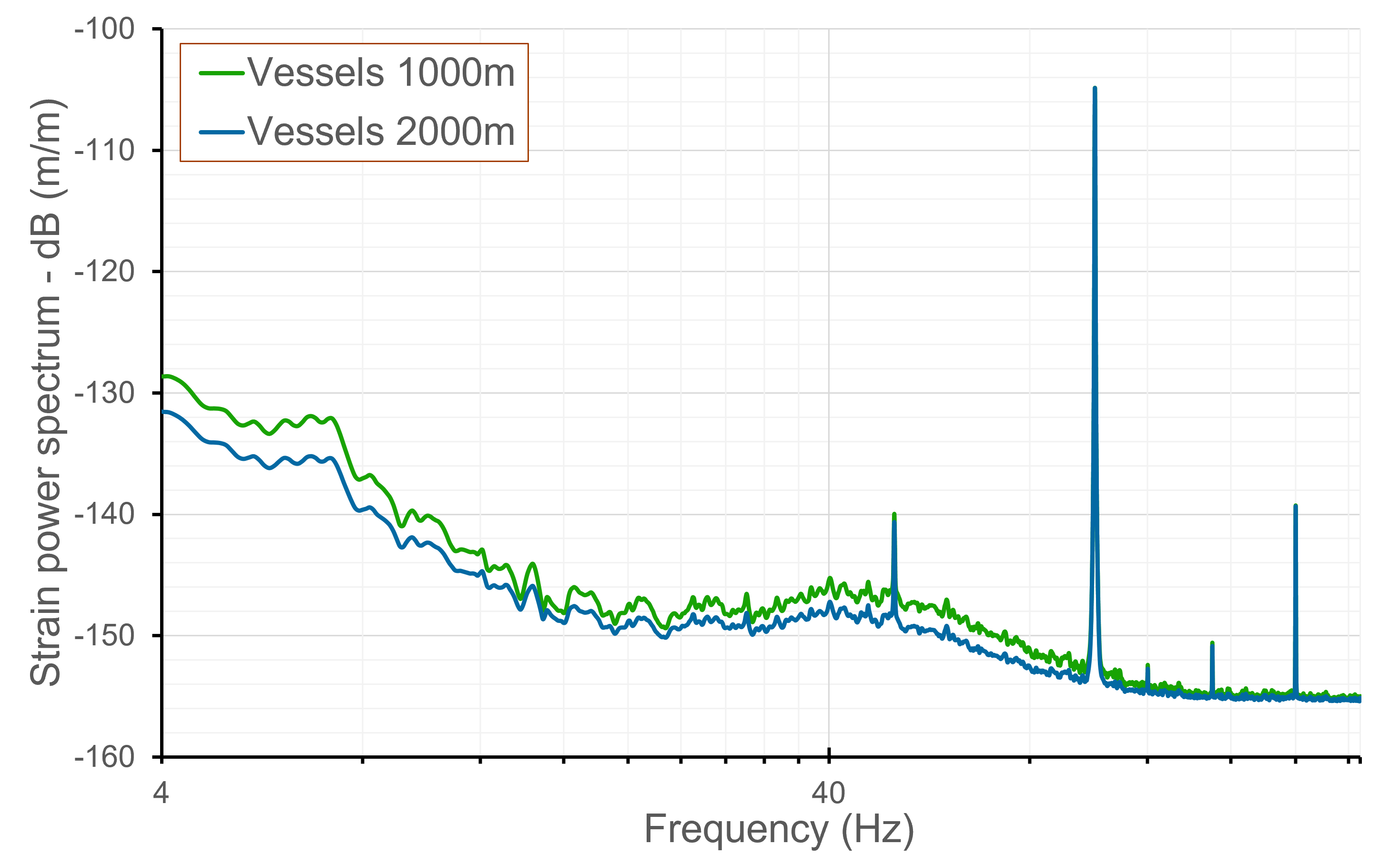}
    ~
  \mysubfigure{0.33\textwidth}{\scriptsize Spectral content for feature vector design.\label{fig:spectra-comparison-4-250Hz-featureVector}
}{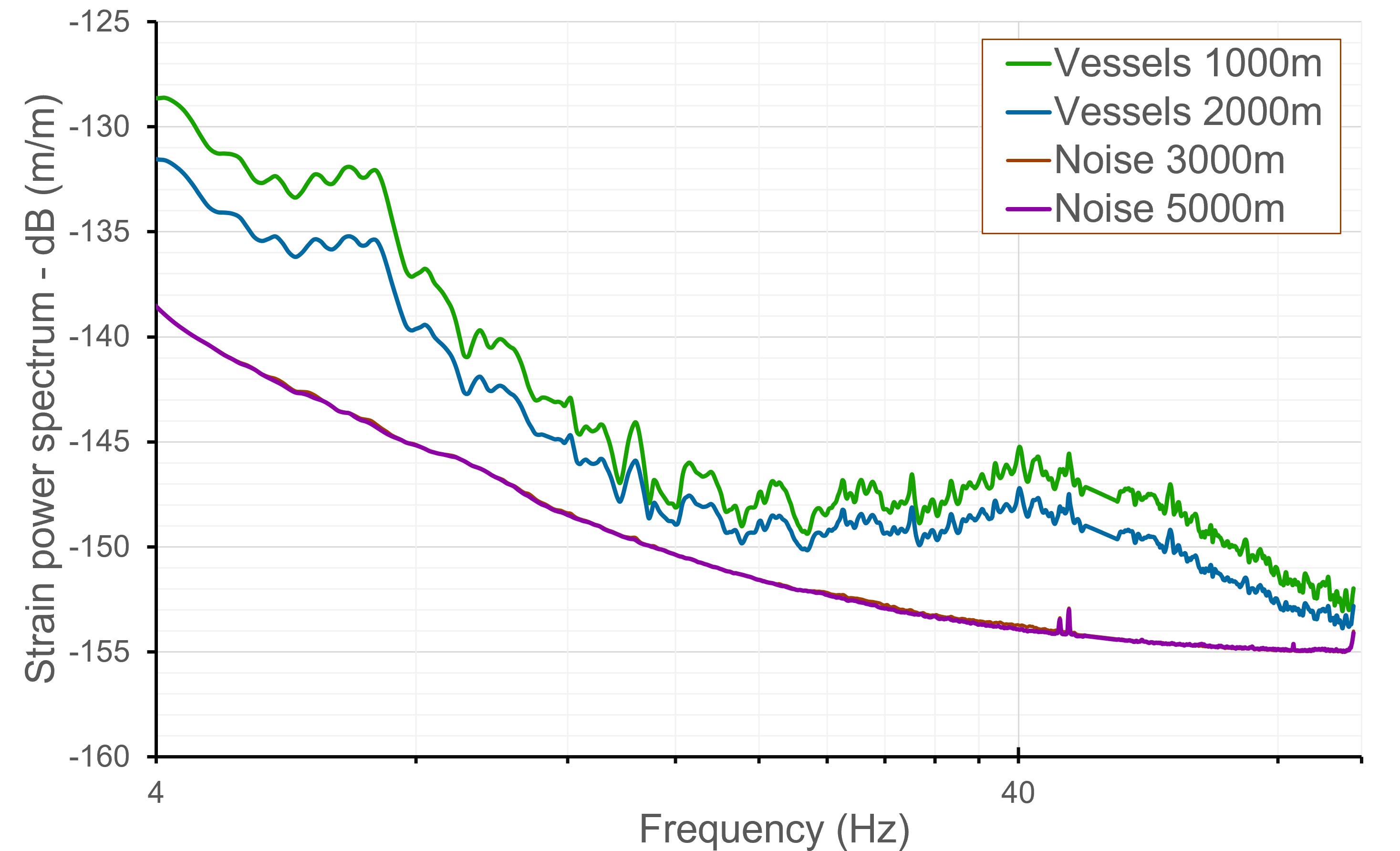}
    \caption{Long term averaged strain power spectrum plots for strain measurements under different environmental conditions (expressed in dBs).}
  \label{fig:avg_noise_spectrum}
\end{figure*}
\begin{figure*}[!t]
  \centering
  \includegraphics[width=1\textwidth]{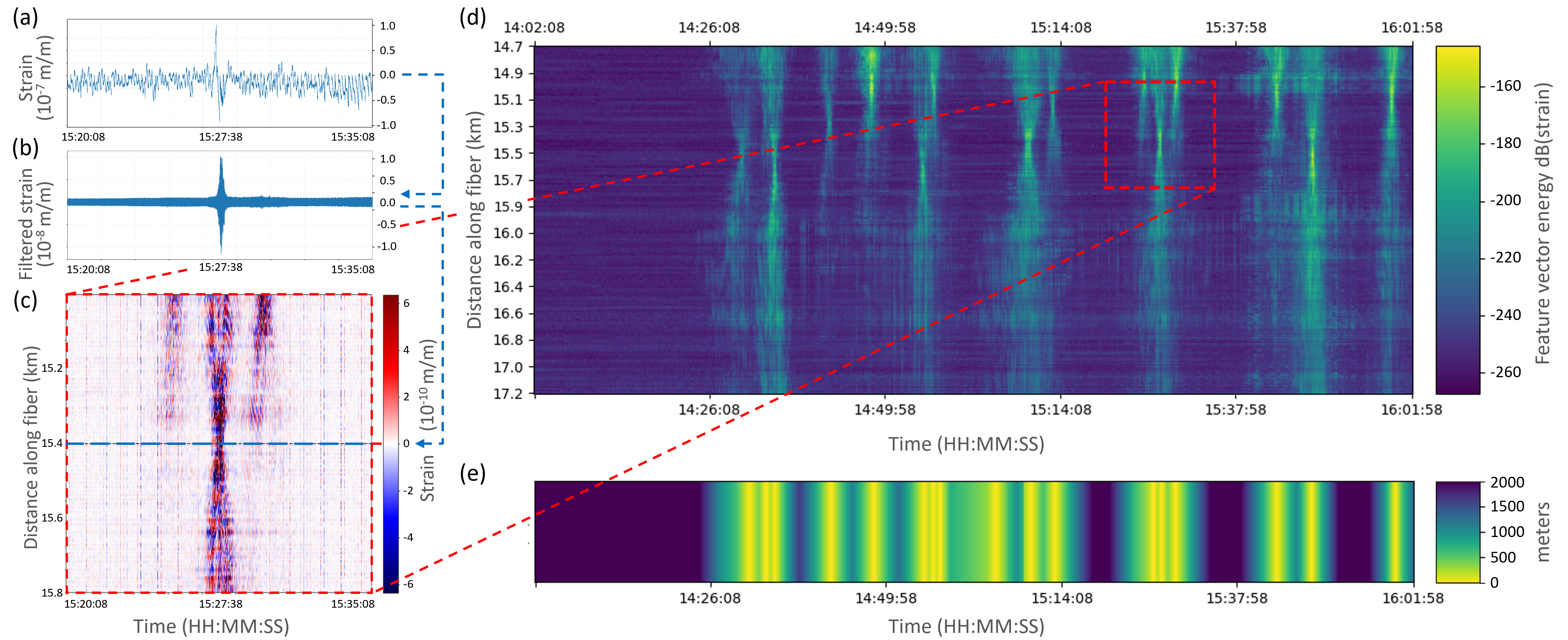}
  \caption{(a) Raw strain signal sensed at $15.4\,km$ in the time interval of the red dashed square in Fig.~\ref{fig:energy-dt-map}.d. (b) Filtered strain signal ($4\,Hz$ to $98\,Hz$, excluding the $(49-51)\,Hz$) corresponding to the raw signal in Fig.~\ref{fig:energy-dt-map}.a and time-synchronized with it. (c) Filtered strain data as a function of time and distance along the cable corresponding to the red dashed square in Fig.~\ref{fig:energy-dt-map}.d, time synchronized with Figs.~\ref{fig:energy-dt-map}.a and~\ref{fig:energy-dt-map}.b. (d) Feature vector energy (in dBs) as a function of time and distance along the cable. (e) Distance from cable to nearest vessel (in meters), time-synchronized with Fig.~\ref{fig:energy-dt-map}.d.}
  \label{fig:energy-dt-map}
\end{figure*}

Fig.~\ref{fig:avg_noise_spectrum}.a shows the long term averaged strain power spectrum (expressed in dBs) for all environmental conditions in the $(0.1 - 250)\,Hz$ range. From the calculated averaged spectra, it can be seen that:

\begin{itemize}
  \item There is a constant strong peak located at $100\,Hz$, with clear harmonics and subharmonics at $50\,Hz$, $200\,Hz$ (and $300\,Hz$, $400\,Hz$ and $500\,Hz$ not displayed), with smaller but clear peaks between $100\,Hz$ and $200\,Hz$ (and between $300\,Hz$ and $400\,Hz$ not displayed). The fact that these components appear for both noise and vessel spectra, suggest that they are probably caused by physical or mechanical systems (motors, rotating shaft, vibrating structures, etc.) in the interrogator environment (Fig.~\ref{fig:avg_noise_spectrum}.b shows vessel spectra only, to allow comparison of the harmonic peaks).
  \item There is a clear distinction between the average spectrum for the noise and vessel conditions in the $(2-100)\,Hz$, opening the path to derive discriminative features for ML tasks (see zoomed spectra in Fig.~\ref{fig:avg_noise_spectrum}.b).
  \item There seems to be no differences between the considered \textit{noise} conditions (for $3000\,m$ and $5000\,m$ distance thresholds). 
  \item The impact of the distance threshold in the vessel average spectrum is consistent: the closer the distance, the stronger the amplitudes. It is also important to note that the spectral shape is maintained for the considered thresholds, so that the generated ML models will face similar spectral conditions.
\end{itemize}

These spectral observations directly guided our feature vector design, which will be composed of energy values in logarithmically spaced frequency bands (as shown in the logarithmic frequency axis in Fig.~\ref{fig:avg_noise_spectrum}), to adequately represent lower frequency content. Each vector component is calculated by summing the FFT spectral coefficients corresponding to each defined frequency band. The considered bandwidth ranges from $4\,Hz$ to $98\,Hz$, excluding the $(49-51)\,Hz$ interval to remove the $50\,Hz$ and $100\,Hz$ peaks shown in the spectral analysis (see Fig.~\ref{fig:avg_noise_spectrum}.c), so that they will not mask the events captured by the DAS system. Fig.~\ref{fig:energy-dt-map}.b shows an example of the filtered strain signal corresponding to the raw strain signal in Fig.~\ref{fig:energy-dt-map}.a, while Fig.~\ref{fig:energy-dt-map}.c illustrates the evolution of the filtered strain over time and along the cable.

These results are consistent with the findings in previous literature~\cite{Rivet_2021}, but we were not able to exploit frequency content that is known to be relevant in other studies, such as the $50\,Hz$ frequency band (in which we find a strong noise related harmonic) or that above $1\,kHz$ mentioned in~\cite{Rivet_2021}, neither on the $100\,Hz-200\,Hz$ frequency band in~\cite{Thiem_2023}, with a very small spectral response except for the harmonics described above. This behavior is in fact a direct consequence of the $L\approx 10\,m$ gauge length spatial sampling used in the interrogator setup.  By the spatial Nyquist criterion an acoustic wavelength must be at least $2\times$gauge length (i.e.~$\geq 20\,m$) to be resolved, and since the sound speed in the main propagation path was found to be $\approx 1750\,m/s$ (through the top sediment layer, see Section~\ref{sec:beamforming_comparison} for additional details), the upper resolvable frequency without aliasing is $f_\mathrm{max} \,=\,\frac{c}{2L}\approx 88\,Hz$. Due to the averaging effect of the gauge length~\cite{nasholm2022array}, the DAS system's frequency response rolls off until reaching the first zero (notch) at a frequency of $\approx \nobreak 175\,Hz$ (cf. eq. (9) in~\cite{nasholm2022array}). Any content above $88–100\,Hz$ is therefore attenuated, which explains the reduced response above $100\,Hz$, and, considering Figure~\ref{fig:avg_noise_spectrum}.c, suggests that there is very relevant information in the $40–100\,Hz$ frequency range (which is consistent with~\cite{paap2025leveragin} where they also observed vessel induced response above their nominal cutoff).

As a prospective study of the feasibility of the proposed ML tasks, we generated a number of visualizations to relate the extracted features with the target elements, from which Figs.~\ref{fig:energy-dt-map}.d+e show an example. Fig.~\ref{fig:energy-dt-map}.d displays the temporal variation (horizontal axis, spanning for two hours) of the feature vector energy measured at each sensor (vertical axis, corresponding to $250$ sensors, spanning for $2553\,m$, from $14702$ to $17255$ meters in fiber distance from the interrogator). The feature vector energy is calculated as the sum of the energy components in all the selected frequency bands, and getting its value in db. Fig.~\ref{fig:energy-dt-map}.e is time synchronized with Fig.~\ref{fig:energy-dt-map}.d and represents, at each time instant, the distance from the sensor range to the closest vessel, between $0$ and $2\,km$ (higher distances are saturated at the dark blue constant color to resemble lack of detected activity in the fiber). From Figs.~\ref{fig:energy-dt-map}.d+e, we can clearly see a very good match between the fiber measurements and the presence of vessels as they approach the fiber, cross over it, and progressively move away. We can also see different magnitudes of the vessel effect on the fiber (with longer or shorter spatial and temporal impacts), and with a clear relationship with the vessel presence in all the cases. Visible fiber response is detected along the full displayed fiber distance range, and up to roughly 10 minutes before and after the vessel crossing instant, which at an average speed of $10.2\,knots$ implies a distance above $3\,km$.

\subsection{Experimental Setup}
\label{sec:experimental-setup}

The feature vector used in our experiments follows the considerations outlined in Section~\ref{sec:spectral-analysis}, and the results discussed in Section~\ref{sec:feature-extraction-analysis}. We computed $N_{P}=100$ logarithmically distributed energy band values spanning the $4\,Hz$ to $98\,Hz$ bandwidth, excluding the $(49-51)\,Hz$ interval (for full transparency and reproducibility, details on the frequency bands calculation are included in the \href{https://geintra-uah.org/psi/index.html\#fbands}{supplementary material Web page}, and the frequency-band limits are provided in \href{https://github.com/UAH-PSI/das-vessel-detection}{the project GitHub repository}). The window length for feature vector computation is 10 seconds. Thus, the total number of feature vectors (corresponding to processed data frames) is $74771$, which will be distributed in training and testing subsets using the $k$-fold cross-validation approach described in Section~\ref{sec:datab-part-exper}.

We evaluated experimental conditions based on specific distance thresholds separating the vessel presence (Class 0) and absence (Class 1) for the classification task. The set of thresholds used, measured in meters, was $D_{thr}\nobreak=\nobreak\lbrace 500, 1000, 1500, 2000, 3000, 5000\rbrace$. For the vessel localization (regression) task, separate models were trained to estimate distances within ranges derived from these thresholds.

\begin{figure*}[!t]
	\centering
	\includegraphics[width=.9\textwidth]{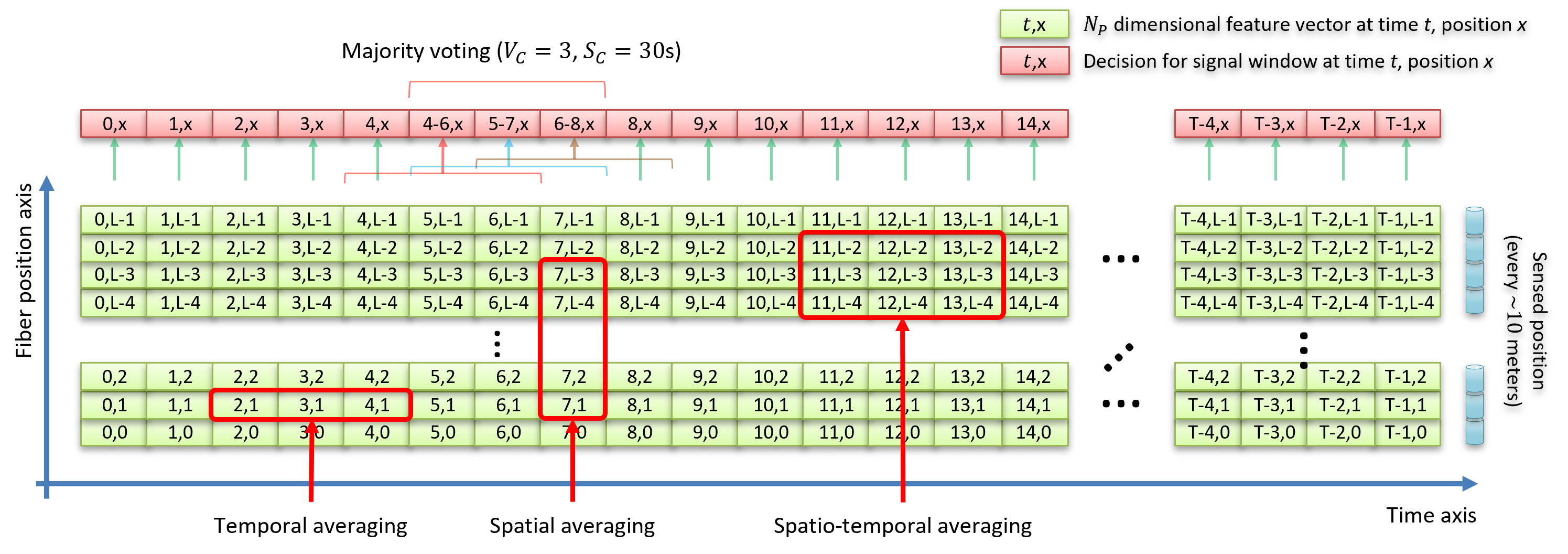}
	\caption{Summary of frame processing strategies to exploit spatial and temporal redundancy in the feature space, and majority voting in the output results.}
	\label{fig:frame-processing}
\end{figure*}

We also explored algorithmic strategies exploiting spatial and temporal redundancy (graphically summarized in Fig.~\ref{fig:frame-processing}):
\begin{itemize}
	\item Spatial redundancy: With configurations using $N_{S}=\nobreak\lbrace 10, 25, 50, 100, 250\rbrace$ sensors, approximately spanning $100$, $250$, $500$, $1000$, and $2500$ meters, respectively.
	\item Temporal redundancy: With temporal contexts of $S_{C} =\nobreak \lbrace 10, 30, 50\rbrace$ seconds, using feature vectors from 1, 3, or 5 consecutive 10-second windows.
	\item Feature averaging: Averaging features across the spatial or temporal dimensions, or both. Spatial averaging is always carried out over the selected sensor range.
	\item Majority voting: Applied majority voting schemes with window lengths $V_{C}= \lbrace 1, 3, 5\rbrace$ (where 1 means no voting). In our implementation, we consider $\texttt{v}$ windows with the evaluated $\texttt{ss}$ window length using a temporal shift of 10 seconds. For example, applying majority voting with $\texttt{v}=5$ windows of $\texttt{ss}=50\,s$, the considered temporal span is $90$ seconds. We did not evaluate longer temporal spans nor majority voting windows to avoid excessive waiting time in a realistic deployment.
\end{itemize}

The specific configurations for the chosen machine learning models were as follows:

\begin{itemize}
	\item {XGBoost:} The baseline model definition was used with the following key parameters:
	\begin{itemize}
		\item Objective function: Binary cross-entropy for classification and mean squared error for regression.
		\item Booster type: Gradient boosted trees (\texttt{gbtree}).
		\item Learning rate (\texttt{eta}): $0.05$.
		\item Maximum tree depth (\texttt{max\_depth}): $10$.
		\item Number of boosting rounds (\texttt{n\_estimators}): $500$.
	\end{itemize}
	\item {Neural Network (NN):} A simple architecture comprising:
	\begin{itemize}
		\item Three fully connected (dense) layers.
		\item ReLU (Rectified Linear Unit) activation functions.
		\item Dropout rate of $50\%$.
		\item Input layer size matching the feature vector dimension (dependent on averaging strategy).
		\item Output layer size of 2 (for classification probabilities) or 1 (for regression distance).
		\item Training batch size of 32.
	\end{itemize}
\end{itemize}
\begin{figure*}[!ht]
  \centering
  \includegraphics[width=\textwidth]{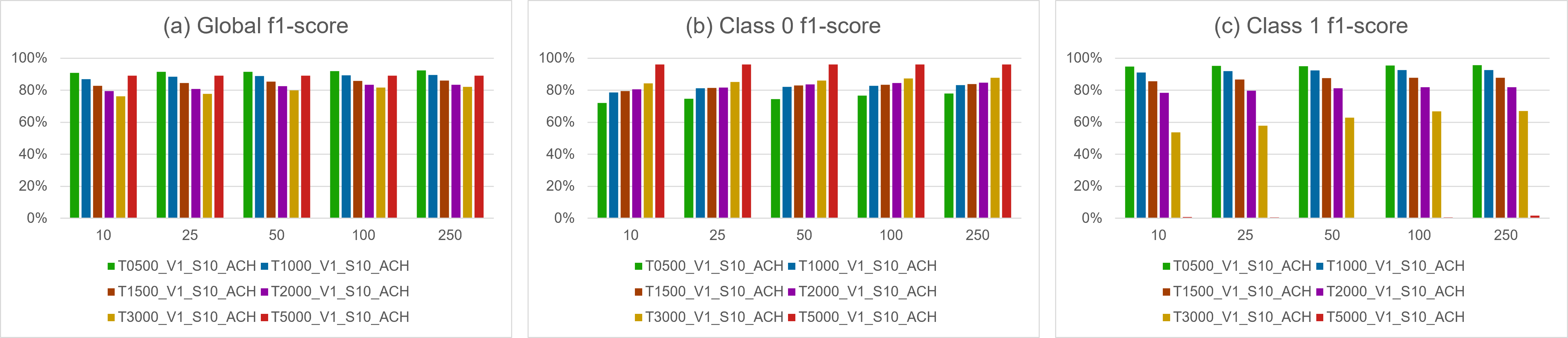}
  \caption{Baseline $\FOneScore{}$-scores as a function of distance threshold (\texttt{T[500,1000,1500,2000,3000,5000]\_V1\_S10\_ACH}), by varying the number of sensors used ($N_{S}=\nobreak\lbrace 10, 25, 50, 100, 250\rbrace$): (a) $\FOneScoreG$, (b) $\FOneScoreCZero$, (c) $\FOneScoreCOne$.}
  \label{fig:f1-vs-distance-V1_S10_ACH}
\end{figure*}
\begin{figure*}[!b]
  \centering
  \includegraphics[width=\textwidth]{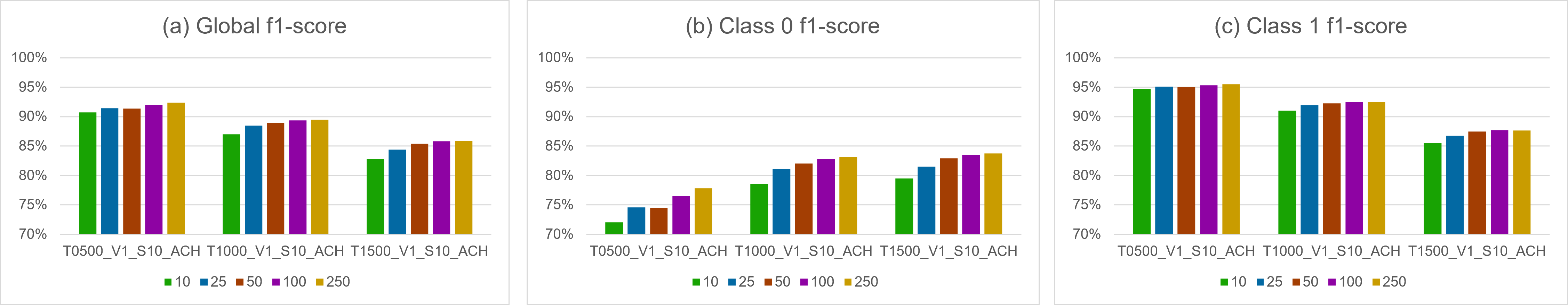}
  \caption{Baseline $\FOneScore{}$-scores as a function of distance threshold (\texttt{T[500,1000,1500]\_V1\_S1\_ACH}) by varying the number of sensors used (10, 25, 50, 100 and 250): (a) $\FOneScoreG$, (b) $\FOneScoreCZero$, (c) $\FOneScoreCOne$.}
  \label{fig:f1-vs-distance-500-1000-1500}
\end{figure*}

To mitigate overfitting and improve generalization, early stopping was applied during model training for both XGBoost and the NN. A validation set was systematically extracted from each training fold and used to monitor performance. Training was stopped if the relevant evaluation metric (e.g., validation loss or $\FOneScore{}$-score) did not improve for 50 consecutive iterations.

To provide a clear view on the algorithmic variations in the included figures, we will refer to the different variants with the following notation: \texttt{Ttttt\_Vv\_Sss\_Aaa}, where \texttt{tttt} refers to the distance threshold ($\texttt{tttt} \in D_{thr}$), \texttt{v} will be the number of temporal windows for majority voting ($\texttt{v} \in V_{C}$), \texttt{ss} will be the considered window length in seconds ($\texttt{ss} \in S_{C}$), and \texttt{aa} states if the feature vector averaging is done on temporal and/or spatial dimensions: $\texttt{aa} = \texttt{CH}$ for spatial averaging, and $\texttt{aa}=\texttt{TI}$ for combined spatial and temporal averaging. Note that spatial averaging is always carried out, with different number of sensors.

\subsection{Vessel Detection Experiments}
\label{sec:vess-detect-exper}

We will describe here the experiments using the XGBoost system, to select the best algorithmic variants. Then, we will compare it with the NN approach.

For the classification task, we first evaluated the baseline strategy with no majority voting, and $10$ seconds temporal context, only exploiting spatial redundancy using the considered number of sensors $N_{S}$.

Fig.~\ref{fig:f1-vs-distance-V1_S10_ACH} shows the $\FOneScore{}$-scores for different number of sensors (horizontal axis), and different distance thresholds. The first observation is the relatively high performance rates, most of them being above $80\%$, with peaks above $90\%$. Considering the $\FOneScoreG{}$-score (Fig.~\ref{fig:f1-vs-distance-V1_S10_ACH}.a), the classification performance decreases as the distance threshold increases, as expected, since the further the vessels, the lower the quality of the spectral information. This is true except for the $5000\,m$ threshold, which suddenly increases. This effect is due to the extreme data unbalance towards class 0 in that case
, which leads to a close to zero $\FOneScoreCOne{}$ as can be seen in Fig.~\ref{fig:f1-vs-distance-V1_S10_ACH}.c.

Regarding the impact of the spatial redundancy exploitation, we can observe a consistent tendency to get better results by using a higher number of sensors, which can be more clearly seen  in Fig.~\ref{fig:f1-vs-distance-500-1000-1500}, for which the data in Fig.~\ref{fig:f1-vs-distance-V1_S10_ACH} has been restricted to $500\,m$, $1000\,m$ and $1500\,m$ thresholds. In the same way, the vertical axis range has been modified to facilitate the visualization.

Considering the results shown so far, it would seem that the classification task for the $500\,m$ distance threshold gets the highest performance, but it should be noted that it suffers from severe data unbalance (see Fig.~\ref{fig:data-balance}), and its superiority is only true if we evaluate the $\FOneScoreG{}$ or $\FOneScoreCOne$ scores.
In fact, the best result selection should also consider the behavior of the class wise $\FOneScore{}$-scores. In particular, as also shown in Fig.~\ref{fig:f1-vs-distance-500-1000-1500}, the $500\,m$ threshold experiment $\FOneScoreCZero{}$s presents the lowest values. This confirms the importance of considering the class-wise $\FOneScore$-scores, since the global score alone may be misleading when the distance threshold generates unbalanced class distributions.  On the contrary, the behavior of the $1000\,m$ threshold task exhibits a better balance between the $\FOneScoreCZero{}$ and $\FOneScoreCOne{}$ scores. So, from now on, we will focus on the $1000\,m$ threshold task spatially integrating the feature vectors from $250$ sensors.
\begin{figure*}[!t]
  \centering
  \includegraphics[width=\textwidth]{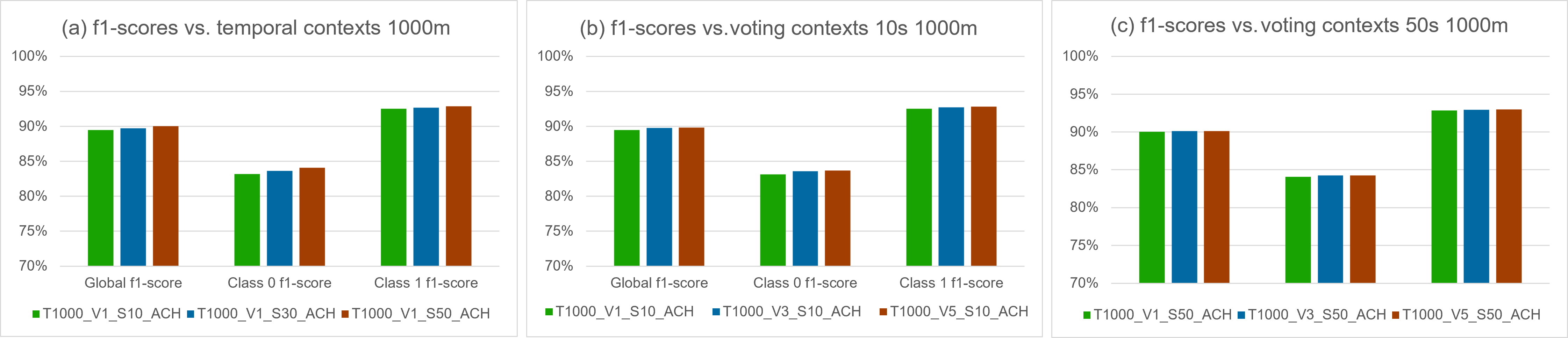}
\caption{$\FOneScore{}$-scores for $1000\,m$ distance threshold as a function of: (a) temporal context duration ($S_{C} = \lbrace 10, 30, 50\rbrace$ seconds long, \texttt{T1000\_V1\_S[10,30,50]\_ACH}), (b) $10\,s$ temporal context with majority window length ($V_{C}= \lbrace 1, 3, 5\rbrace$ windows, \texttt{T1000\_V[1,3,5]\_S10\_ACH}), (c) $50\,s$ temporal context with majority window length ($V_{C}= \lbrace 1, 3, 5\rbrace$ windows, \texttt{T1000\_V[1,3,5]\_S50\_ACH}).}
  \label{fig:f1-vs-tctx+voting-1000}
\end{figure*}
\begin{figure}[!t]
  \centering
  \includegraphics[width=.8\columnwidth]{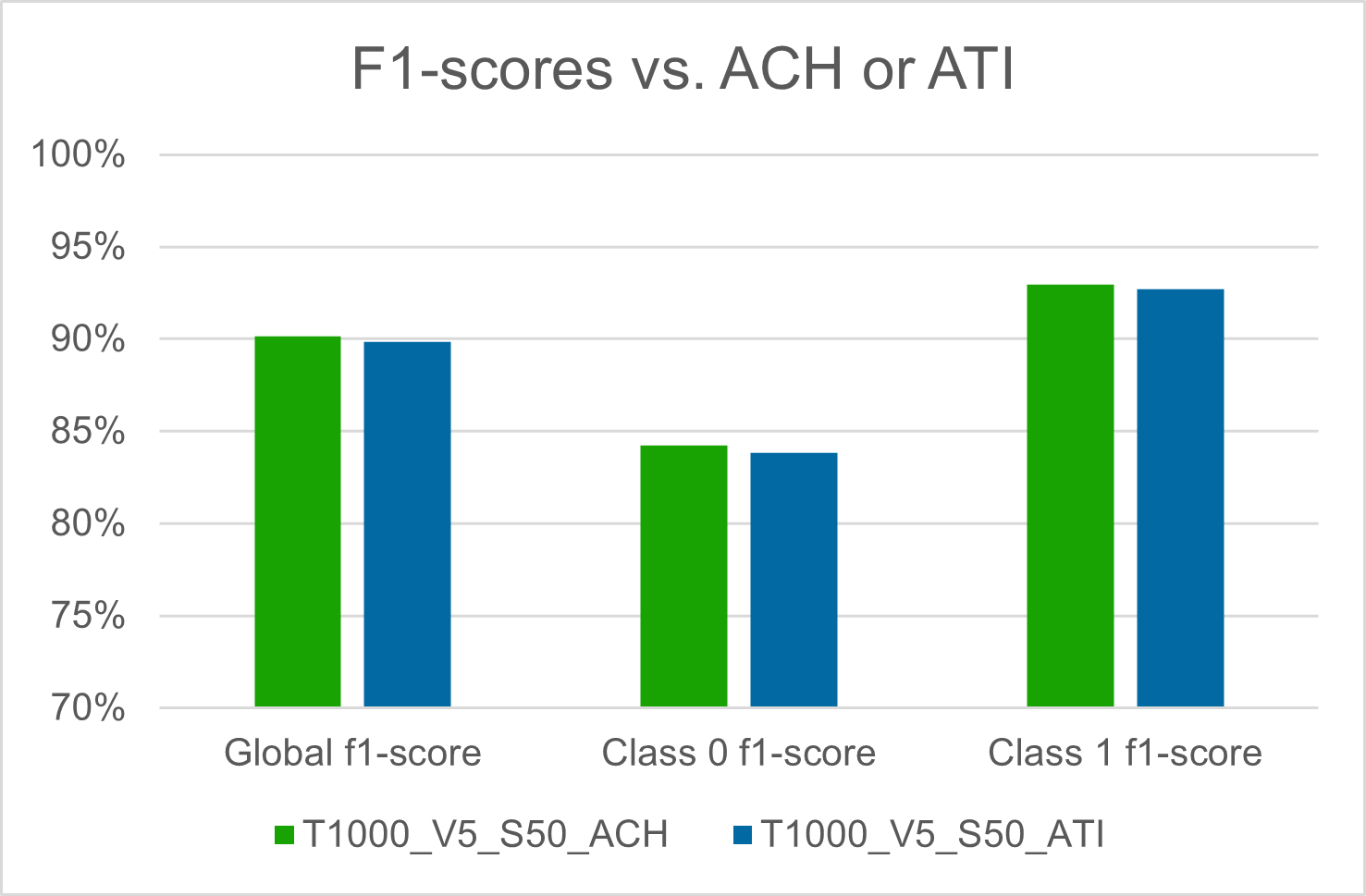}
  \caption{Global $\FOneScore{}$-score as a function of the averaging type.}
  \label{fig:f1-1000-V5_S50_ACH+TI-simple}
\end{figure}

Fig.~\ref{fig:f1-vs-tctx+voting-1000}.a shows the $\FOneScore{}$-scores for the $1000\,m$ threshold with $10$, $30$ and $50$ second-long windows, with improvements for longer windows, specially in the $\FOneScoreCZero$ metric. Similarly, Figures~\ref{fig:f1-vs-tctx+voting-1000}.b and Fig.~\ref{fig:f1-vs-tctx+voting-1000}.c show the $\FOneScore{}$-scores for the $1000\,m$ threshold when using different majority voting window lengths ($1$, $3$ or $5$) for temporal contexts of $10$ and $50$ seconds, respectively. From the results, when increasing the number of windows for majority voting, there are improvements for the smallest time span ($10\,s$), but the differences are small when the considered temporal context is increased ($50\,s$). This can be explained by the fact that the increased temporal context considers the same long temporal range than the increased voting scheme window length.

\begin{figure}[!t]
  \centering
  \includegraphics[width=.8\columnwidth]{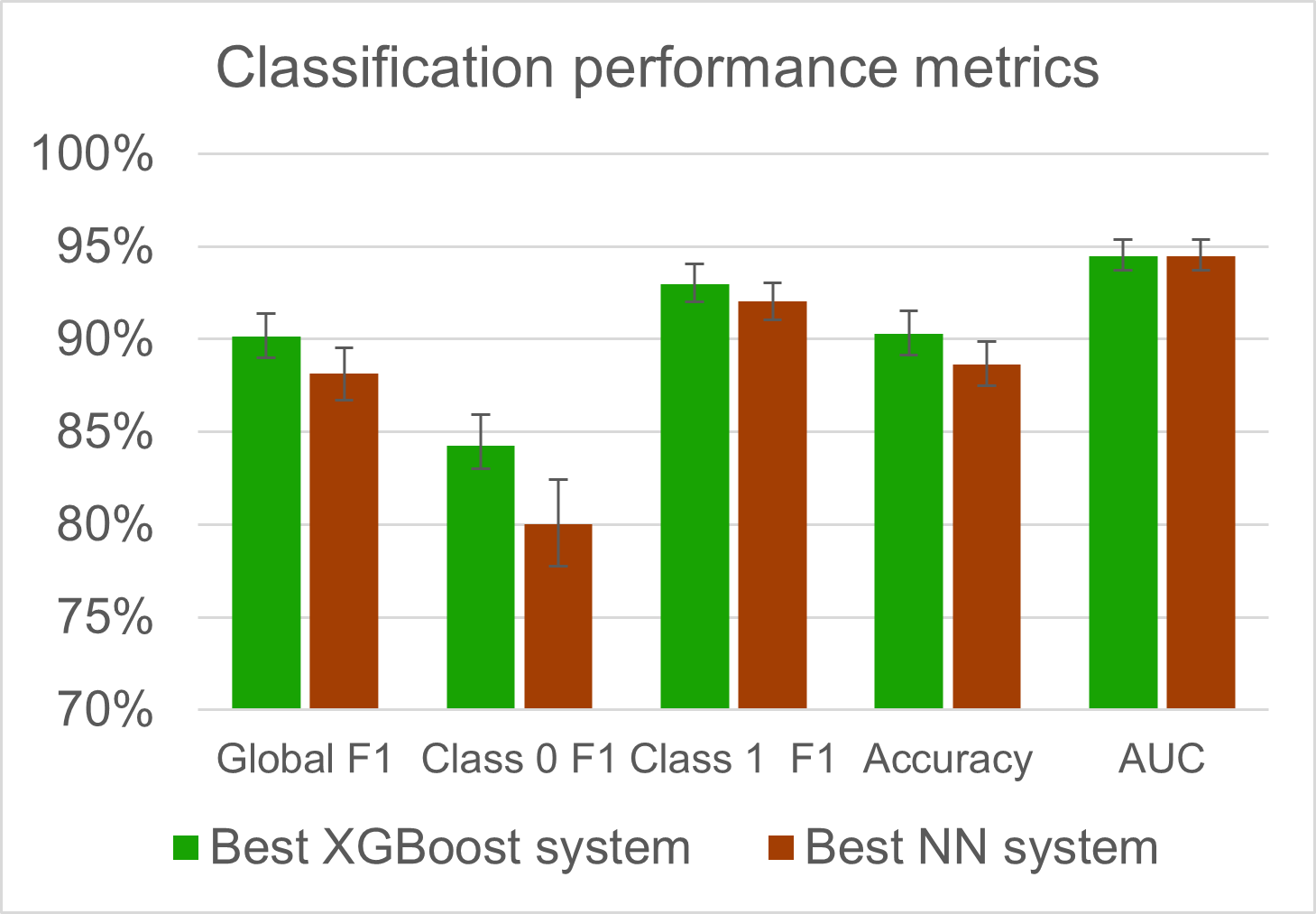}
  \caption{Classification task: Performance comparison between the best XGBoost and NN systems.}
  \label{fig:cmp-classification-best-nn-xgboost}
\end{figure}

Finally, Fig.~\ref{fig:f1-1000-V5_S50_ACH+TI-simple} shows the effect of spatial (\texttt{ACH}) vs. spatial+temporal averaging (\texttt{ATI}) for the $1000\,m$ distance threshold, $50\,s$ temporal context, and $5$ windows of majority voting. The results show that the inclusion of temporal averaging is not useful when applied to strategies that already make use of the redundancy in temporal information (long temporal contexts and majority voting).

Therefore, considering the contributions of the algorithmic variations and aiming for a balanced performance across the $\FOneScore{}$-scores, the final system we select for the production vessel-detection task is the one using a distance threshold of $1000\,m$ (consistent with known criteria adopted by cable companies, as discussed in Section~\ref{sec:machine-learning-strategy}),
data from $250$ sensors, a $50$-second-length temporal context, majority voting with $5$ windows, and spatial averaging only.

To complement the class-wise $\FOneScore$ analysis, Table~\ref{tab:precision-recall-best-classifier} reports the corresponding precision and recall values for the selected classifier at the 1000~m operating threshold. These results make explicit the precision-recall trade-off summarized by the class-wise $\FOneScore$-scores.

We also ran the experiments using the simple NN-based system described in Section~\ref{sec:machine-learning-strategy}. The observations regarding the effect of the algorithmic modifications are very similar to those previously discussed. However, the results are significantly worse than those obtained with the XGBoost system, as shown in Fig.~\ref{fig:cmp-classification-best-nn-xgboost}. Future work will be devoted to a more exhaustive evaluation of variants and parameter tuning of both approaches.

\begin{table}[!t]
  \centering
\caption{Class-wise precision, recall, and $\FOneScore$-scores for the selected 1000~m vessel-detection XGBoost configuration.}
\label{tab:precision-recall-best-classifier}
\setlength{\tabcolsep}{4pt}
  \begin{tabular}{lccc}
\toprule
Class & Precision & Recall & $\FOneScore$-score \\
\hline
Class 0 &
$88.84\%$ &
$80.11\%$ &
$84.25\%$ \\
Class 1 &
$90.88\%$ &
$95.16\%$ &
$92.97\%$ \\
\bottomrule
    \end{tabular}
\end{table}

\subsection{Vessel Localization Experiments}
\label{sec:vess-local-exper}

As in the previous section, we will first describe the experiments using the XGBoost system, and then the comparison with the NN-based approach.

\begin{figure*}[!t]
  \centering
  \includegraphics[width=\textwidth]{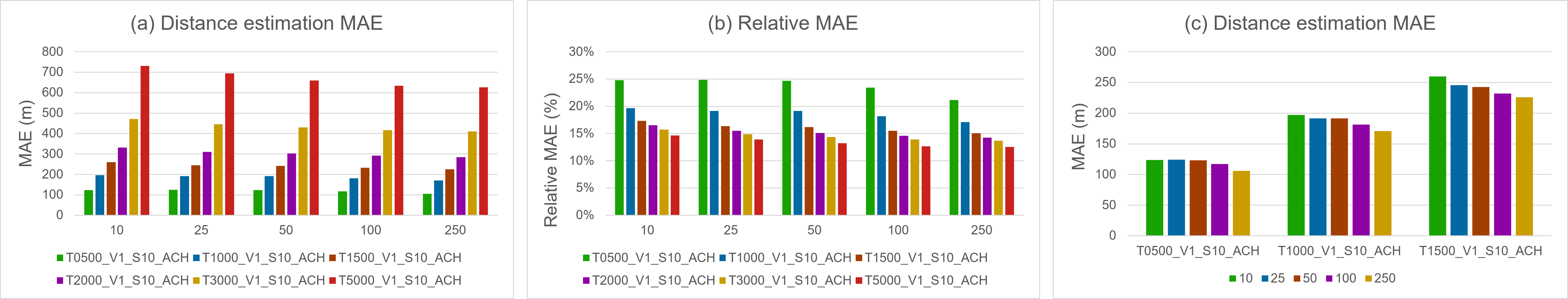}
  \caption{Baseline MAE and rMAE scores (\texttt{T[500,1000,1500,2000,3000,5000]\_V1\_S10\_ACH}): (a) Distance estimation MAE as a function of distance thresholds, (b) Relative MAE  as a function of distance thresholds, (c) Distance estimation MAE as a function of the number of sensors used.}
  \label{fig:dmae+rdmae-vsdistance+vsnumsensors-5000-1000-1500}
\end{figure*}

For the vessel localization task, we adopted the same experimental structure applied for vessel detection. The results for the baseline experiments are shown in Fig.~\ref{fig:dmae+rdmae-vsdistance+vsnumsensors-5000-1000-1500}.a. As expected, the distance-estimation MAE increases as the distance threshold increases. To better compare the impact of distance thresholds, Fig.~\ref{fig:dmae+rdmae-vsdistance+vsnumsensors-5000-1000-1500}.b provides the relative distance MAE, showing that the MAE ranges from approximately $12\%$ to $25\%$ of the distance threshold, decreasing as the threshold increases. This indicates that the system exhibits greater ambiguity when estimating closer vessel positions. A possible explanation is that closer vessels increase the likelihood of multiple vessels falling within the fiber's detection range, thus interfering with the regression process.

As in the classification task, there is a consistent tendency to get better results when using a higher number of sensors, as shown in Fig.~\ref{fig:dmae+rdmae-vsdistance+vsnumsensors-5000-1000-1500}.c (again only using $500\,m$, $1000\,m$ and $1500\,m$ thresholds).

From now on, we will focus on the $1000\,m$ distance threshold condition using $250$ sensors, as it was the one selected in the classification task.

\begin{figure*}[!t]
  \centering
  \includegraphics[width=\textwidth]{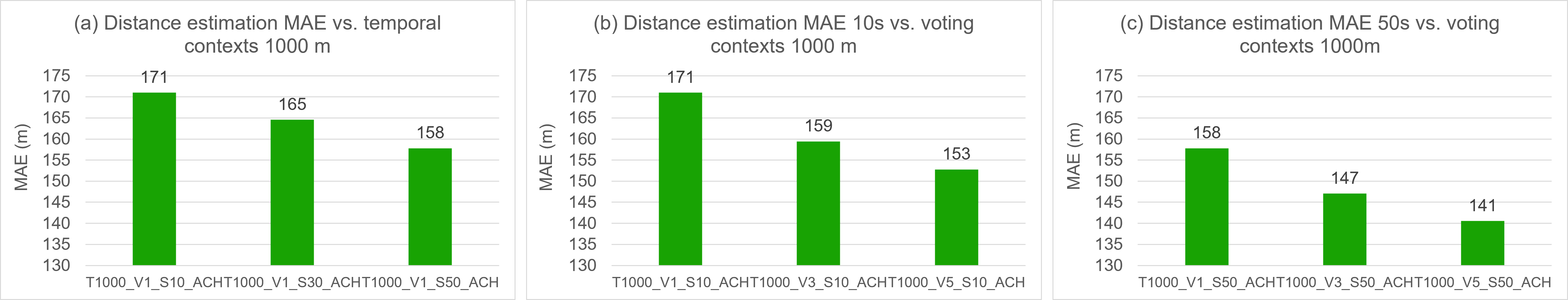}
    \caption{Distance estimation MAE for the $1000\,m$ distance threshold as a function of (a) temporal context duration ($S_{C} = \lbrace 10, 30, 50\rbrace$ seconds long, \texttt{T1000\_V1\_S[10,30,50]\_ACH}), (b) $10\,s$ temporal context with majority window length ($V_{C}= \lbrace 1, 3, 5\rbrace$ windows, \texttt{T1000\_V[1,3,5]\_S10\_ACH}), (c) $50\,s$ temporal context with majority window length ($V_{C}= \lbrace 1, 3, 5\rbrace$ windows, \texttt{T1000\_V[1,3,5]\_S50\_ACH}).}
  \label{fig:dmae-vsdistance+vsvoting-10s+50s}
\end{figure*}

Fig.~\ref{fig:dmae-vsdistance+vsvoting-10s+50s}.a shows the effect of considering longer temporal contexts. Again, exploiting longer temporal contexts improves the distance estimation results significantly.

To assess the effect of majority voting strategies, Figures~\ref{fig:dmae-vsdistance+vsvoting-10s+50s}.b and~\ref{fig:dmae-vsdistance+vsvoting-10s+50s}.c show the MAE for the $1000\,m$ threshold with 1, 3 or 5 majority voting windows, and using $10s$ or $50s$ time spans, respectively. Contrary to the findings in the classification task, we do observe significant improvements by increasing both the temporal context and the number of windows of the majority voting strategy in this regression task.

\begin{figure}[!t]
  \centering
  \includegraphics[width=.8\columnwidth]{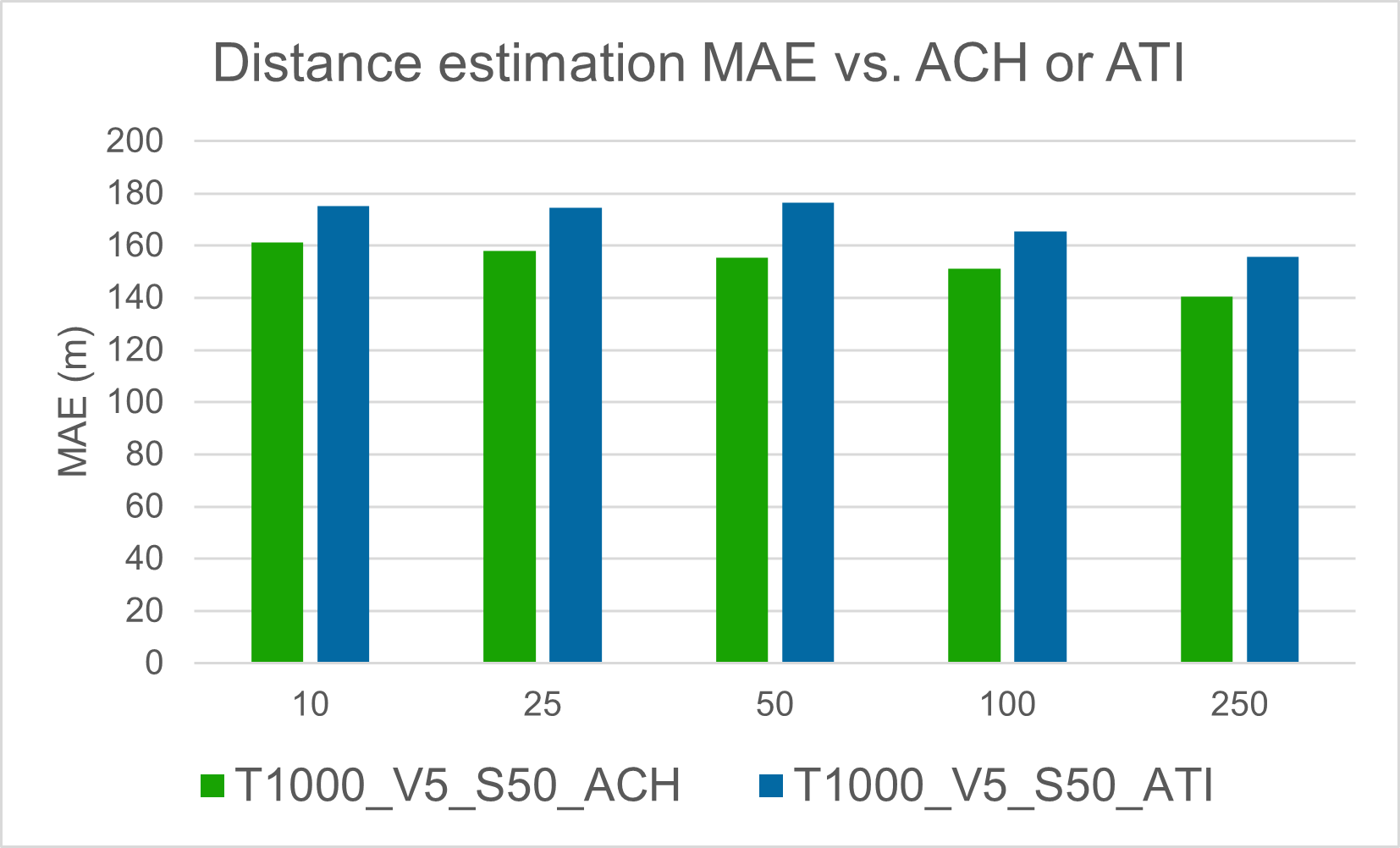}
  \caption{Distance estimation MAE for the $1000\,m$ distance threshold, 50 seconds time span, 5 majority voting windows, evaluating spatial averaging (ACH) vs. spatial+temporal averaging (ATI) (\texttt{T1000\_V5\_S50\_A[CH,TI]}.}
  \label{fig:dmae-1000-V5_S50_ACH+ATI}
\end{figure}

Finally, Fig.~\ref{fig:dmae-1000-V5_S50_ACH+ATI} shows the effect of spatial (\texttt{ACH}) vs spatial+temporal averaging (\texttt{ATI}). Contrary to the findings in the classification task, the inclusion of temporal averaging significantly deteriorates the estimated distances, which can be explained by considering that distance estimation is more severely affected by variations along the temporal span.

\begin{figure}[!t]
  \centering
  \includegraphics[width=.8\columnwidth]{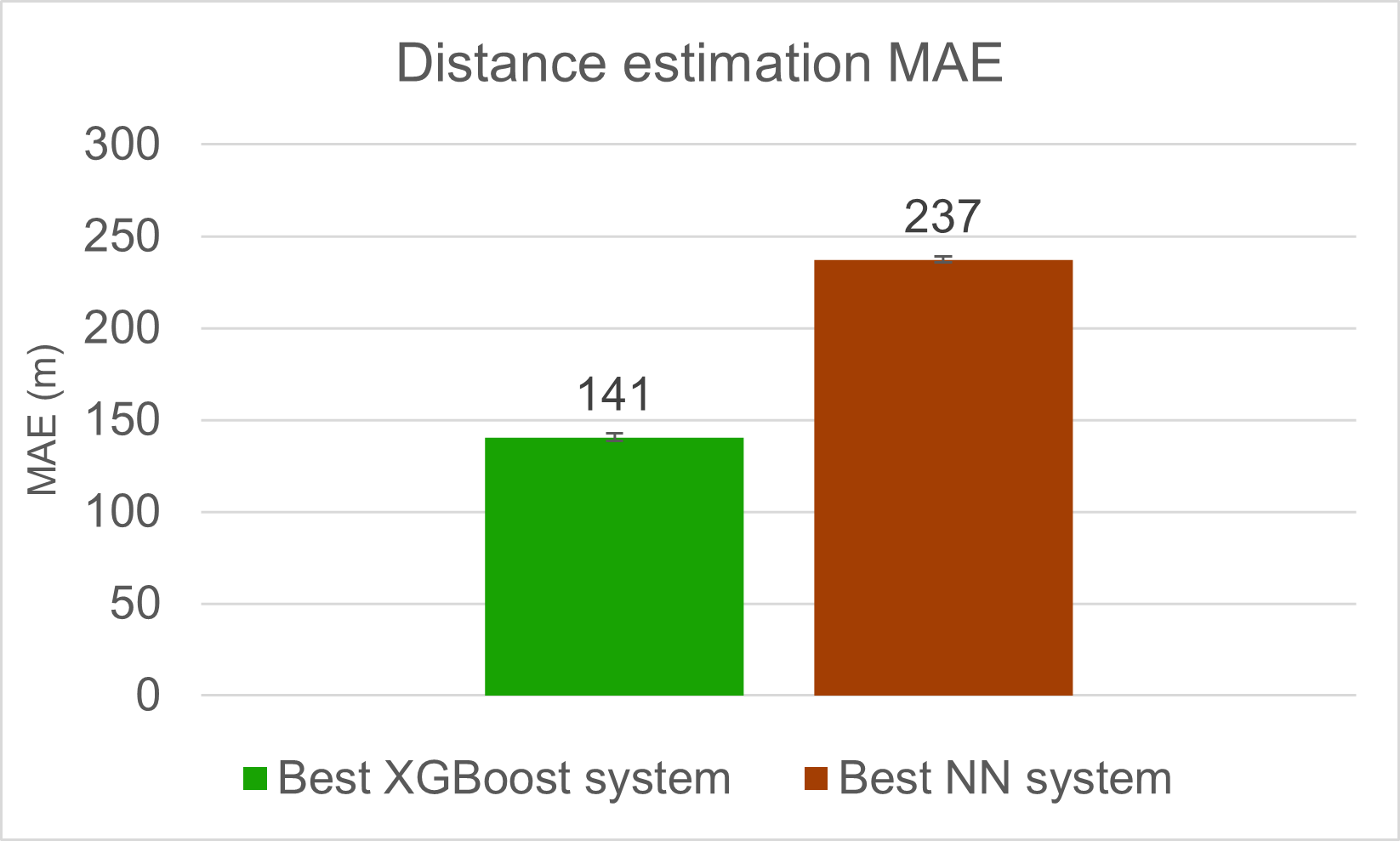}
  \caption{Regression task: Performance comparison between the best XGBoost and NN systems.}
  \label{fig:cmp-regression-best-nn-xgboost}
\end{figure}

Therefore, considering the contributions of the algorithmic variations, the final system we would select for the production vessel distance-estimation task is the one using a distance threshold of $1000\,m$, data from 250 sensors, a 50 second long temporal context, majority voting with 5 windows, and spatial averaging only. We also ran the experiments using the simple NN-based system, with similar observations regarding the effect of the algorithmic modifications. The achieved results are again significantly worse than those obtained with the XGBoost system, as shown in Fig.~\ref{fig:cmp-regression-best-nn-xgboost}.

Videos illustrating the system operating in vessel-detection and distance-estimation modes are provided on the \href{https://geintra-uah.org/psi/index.html\#videos}{supplementary material Web page}, which is also linked from \href{https://github.com/UAH-PSI/das-vessel-detection#supplementary-material}{the project GitHub repository}.

\subsection{Comparison with beamforming methods}
\label{sec:beamforming_comparison}

To compare the proposed ML-based distance-estimation method with a representative conventional DAS localization strategy, we implemented a physics-based beamforming baseline on the same dataset and under the same evaluation protocol used for our ML experiments. A direct comparison with other recently published DAS vessel-monitoring approaches is not currently feasible because no public benchmark dataset, to the best of our knowledge, provides the combination of DAS data, AIS-based labels, task definitions, and evaluation conditions required to train and test different methods under common $\FOneScoreBold$-score and MAE metrics. In particular, DASHip~\cite{Huang_2025} is not directly applicable to our comparison, first because it restricts the bandwidth to 5Hz (excluding relevant higher frequency spectral acoustic components related to vessel behavior~\cite{Karasalo_2017}); and second as the distributed data makes the problem to be formulated as an image-based object detection of ship-passage events over DAS spatio-temporal representations, associated with wake signatures during cable crossings (spanning just $1\,km$ and $1\, minute$ ranges), whereas our work addresses cable-centered vessel detection and continuous distance estimation from spectral DAS features (spanning the full fiber length range and the full time duration of the 10-days recordings).

The closest methodological reference for a conventional physics-based comparison is the work by Paap et al.~\cite{paap2025leveragin}, who proposed a migration-based source-location method for vessel monitoring using submarine DAS data. A direct numerical comparison using the original datasets from~\cite{paap2025leveragin} is also not feasible, since their North Sea dataset cannot be disclosed due to confidentiality restrictions, and the available Oregon dataset does not provide the complete AIS labeling information required to reproduce an equivalent supervised detection and distance-estimation experiment. Moreover, that study evaluates a small number of selected vessel passages, whereas our evaluation is based on a continuous ten-day dataset with a substantially larger number of vessels and operational conditions. For these reasons, we use~\cite{paap2025leveragin} as the methodological reference for a conventional physics-based baseline, and we evaluate an equivalent beamforming formulation directly on our dataset. In particular, we adopted the Steered Response Power (SRP) algorithm~\cite{dibiase2001robust}, which is virtually identical to the implementation of the migration-based method used by Paap (sharing the same propagation assumptions and processing principles), but calculating power in each grid position instead of stacked absolute values of the DAS waveforms. The SRP algorithm was applied to the DAS strain data using the same 250 channels and 10-second temporal windows than in our experimental setup. Because the SRP baseline explicitly depends on propagation delays, the steering model was built using a carefully checked channel-to-position correspondence for the analyzed cable segment.

To optimize the propagation model parameters, we applied Paap's methodology~\cite{paap2025leveragin}, scanning for the optimal sound speed by maximizing the coherency of back-projected waveforms. We tested speeds from $100\,m/s$ to $1900\,m/s$ and determined a maximum response at $1750\,m/s$. As a quantitative sensitivity check on this calibration choice, we also evaluated a reduced set of windows within the $1\,km$ operating range, obtaining the lowest MAE at $1750\,m/s$ ($177.38\,m$), compared with $214.72\,m$ at $1700\,m/s$, $184.95\,m$ at $1800\,m/s$, and $199.46\,m$ at $1850\,m/s$. This value is highly consistent with the $1700\,m/s$ found by Paap~\textit{et~al.} in a comparable North Sea buried-cable environment~\cite{paap2025leveragin}, and can be interpreted as evidence of significant acoustic energy transmission through the seabed top sediment layer, rather than through the water column.

For each time window, the SRP function was evaluated over a two-dimensional spatial grid with $20\,m$ resolution covering the vicinity of the monitored cable (a bounding box defined as $1000\,m$ from the extreme cable positions of the used channels). The algorithm determines the most probable vessel position as the grid point maximizing the steered response power, from which we derived the corresponding distance to the cable. All results presented in this section correspond to vessels located within $1\,km$ of the cable. No temporal continuity constraint, trajectory smoother, or speed-limited tracking model was applied to either the SRP output or the ML regression output. The comparison is therefore performed at the frame level, so that both methods are evaluated under the same distance-estimation protocol rather than as full vessel-tracking systems.

\begin{figure*}[!t]
	\centering
  \includegraphics[width=\linewidth]{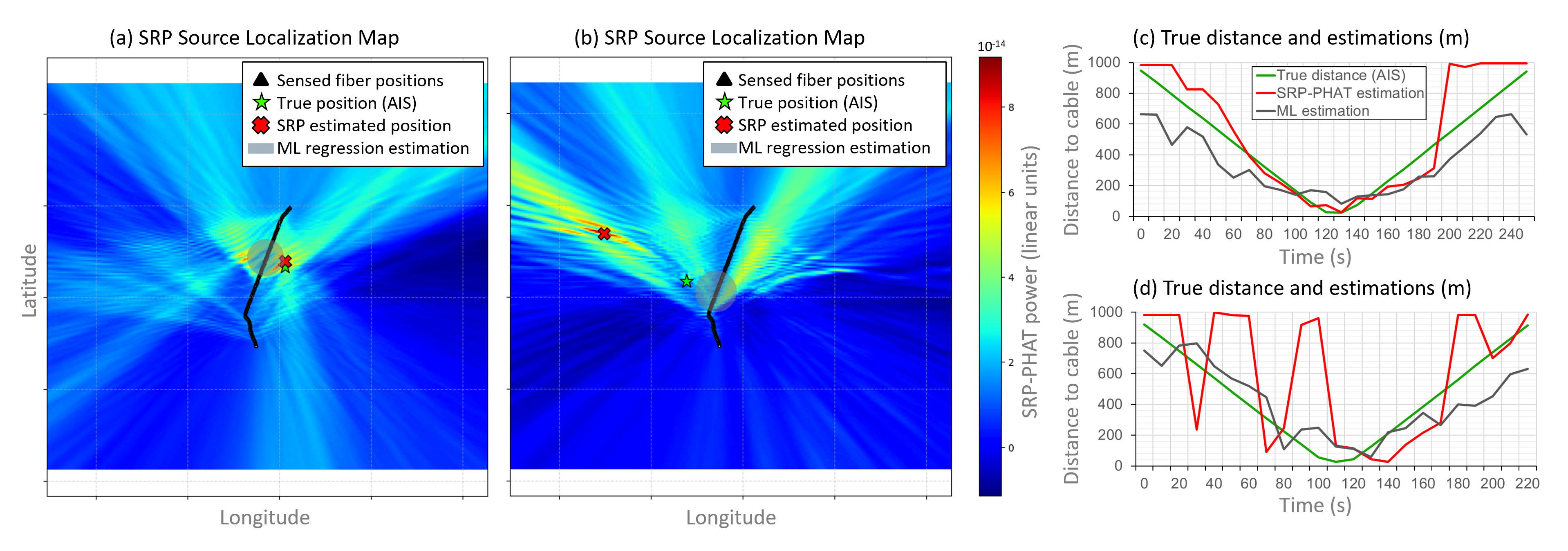}
	\caption{Comparison of SRP beamforming and ML regression methods. (a) SRP map for a case where both methods perform well, particularly as the vessel is close to the cable. (b) SRP map for a vessel that is more difficult to detect; while both methods struggle, the ML regressor's error is lower. (c) A time-series plot of a vessel crossing above the fiber, showing the distance to cable estimations by SRP (\textcolor{red}{red trace}) and the ML regressor (\textcolor{darkgray}{gray trace}) against the ground truth (\textcolor{darkgreen}{green trace}), illustrating a case where both methods track the vessel accurately. (d) A distance to cable estimation time-series plot for a quieter or more challenging vessel, demonstrating a case where the ML regressor performs significantly better than the SRP method.}
	\label{fig:srp_maps}
\end{figure*}
\begin{table}[!h]
    \centering
    \caption{Distance estimation performance of SRP and ML-based regression for vessels within 1~km of the cable.}
    \label{tab:beamforming_vs_ml}
    \begin{tabular}{lcc}
        \toprule
        \textbf{Method} & \textbf{MAE [m]} & \textbf{95\,\% CI [m]} \\
        \midrule
        SRP Beamforming & 225 & [222.17, 227.97] \\
        ML Regression (XGBoost) & 171 & [169.22, 172.83] \\
        \bottomrule
    \end{tabular}
\end{table}
Table~\ref{tab:beamforming_vs_ml} summarizes the MAE results with the 95\,\% confidence intervals (CI) for both methods. 
Fig.~\ref{fig:srp_maps}.a shows a sample SRP map where the steered-response peak accurately identifies the vessel position, a scenario where both the beamformer and the ML regressor perform very well. In contrast, Fig.~\ref{fig:srp_maps}.b shows a more challenging case where the vessel is harder for both methods to detect, though the ML regressor ultimately yields a lower estimation error. The time-series plots in Figs.~\ref{fig:srp_maps}.c and~\ref{fig:srp_maps}.d compare the distance estimations over a time window centered on a vessel crossing, showing the true distance (estimated from AIS), the ML regressor prediction, and the SRP prediction. Fig.~\ref{fig:srp_maps}.c shows an example where both methods track the vessel effectively, whereas Fig.~\ref{fig:srp_maps}.d shows an example with a more challenging vessel where the ML regressor performs much better than the beamforming approach due to the presence of local maxima in the SRP map that do not correspond to the real vessel position. These local maxima may arise from weak vessel signatures, coherent background or interfering acoustic sources, and environmental, bathymetric, or seabed-coupling variability. These ambiguities illustrate the sensitivity of physics-based beamforming to coherent energy patterns in real buried-cable DAS data.

More generally, beamforming methods provide physically interpretable spatial maps, require no training data, and can be adapted to new deployments once the propagation model is specified. However, as discussed above, their performance is highly sensitive to propagation model parameters that are difficult to control and strongly dependent on the deployment and environmental conditions. Additionally, they require dense spatial grids that make them computationally expensive, especially in long-cable deployments.
\begin{figure*}[b]
	\centering
\includegraphics[width=\textwidth]{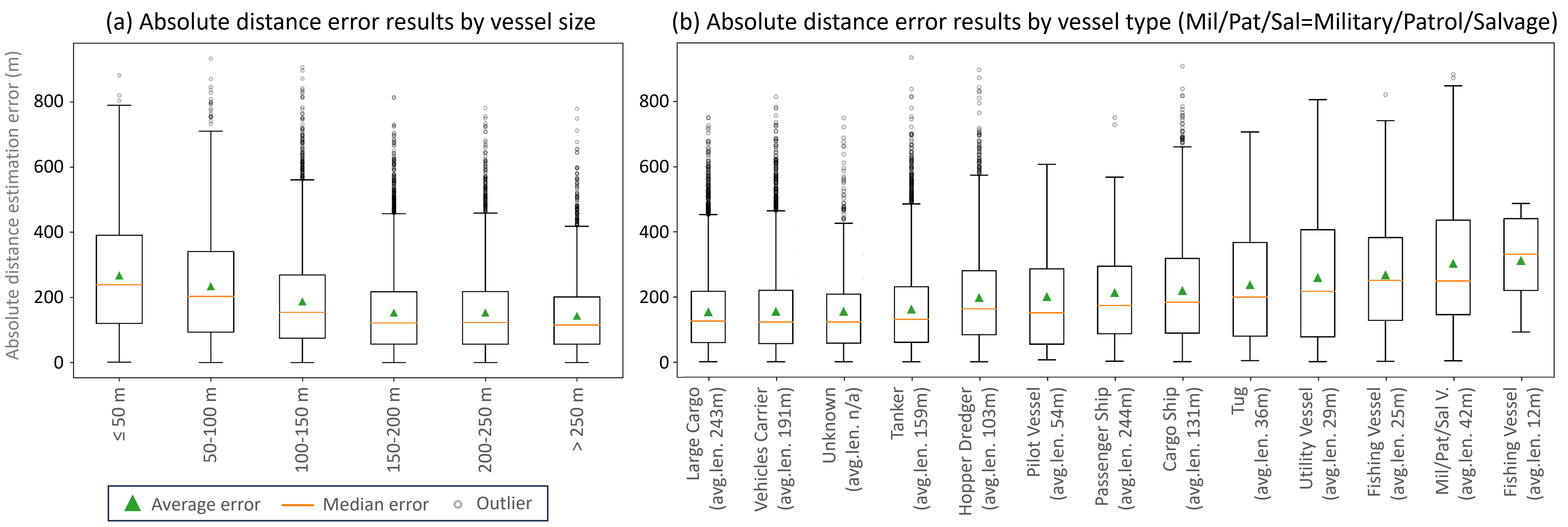}
	\caption{Distribution of absolute distance errors by (a) vessel size range, (b) vessel type (indicating also the average length (avg.len.) for the group type).}
	\label{fig:ae-per-vessel-type-size}
\end{figure*}
In contrast, the proposed ML-based approach avoids strong physical assumptions on the propagation model but depend on representative training data. Once trained, models validated on diverse cable segments can generalize well to new environments, providing faster and more accurate distance estimates by directly mapping signal features to vessel range, effectively exploiting spatio-temporal redundancies across sensors.

\subsection{Dependency on Vessel Size and Type}
\label{sec:depend-vess-size}
To assess the influence of vessel characteristics on localization accuracy, we analyzed the absolute distance error as a function of both vessel size and type. We evaluated the task with the $1000\,m$ distance threshold, 250 channels and 10-second temporal windows (that achieved a MAE of $171\,m$), with the results distribution boxplots shown in Fig.~\ref{fig:ae-per-vessel-type-size}.

In general, larger vessels show the lowest errors, consistent with their expected stronger and more stable acoustic emissions. However, there are cases not fitting in this tendency, such as the Passenger Ships (mainly cruises), achieving higher error than the smaller Pilot Vessels, probably due to their expected lower radiated noise (being focused in passenger comfort).

As a result, vessel size and type clearly influence DAS-based localization performance. A more detailed analysis of this dependency is left for future work, but the current findings already suggest that class-dependent models could further enhance system robustness and enable addressing future vessel-type identification tasks.

\subsection{Computational Requirements and Memory Footprint}

To assess the computational feasibility of the proposal, we measured the execution time required to process one 10-second DAS frame using the selected configuration. The experiments were run on a workstation equipped with an AMD Ryzen 9 5950X processor at 2.2~GHz. Table~\ref{tab:computational_cost} summarizes the mean execution time, standard deviation, and relative contribution of the main processing modules, for both the classification and regression tasks.

\begin{table}[t]
  \centering
  \caption{Computational cost for processing one 10-second DAS frame.}
  \label{tab:computational_cost}
  \begin{tabular}{lcccc}
    \toprule
    Module & \makecell{Mean time \\ (ms)} & \makecell{STD \\ (ms)} & \makecell{\% \\ classif} & \makecell{\% \\ regress} \\
    \hline
    Preprocessing  & $1451.535$ & $36.645$ & $39.65$ & $39.68$ \\
    Feature extraction  & $2206.212$ & $16.920$ & $60.27$ & $60.31$ \\
    Classification &&&&\\
    \quad Model inference & $2.714$ & $5.055$ & $0.07$ &\\
    \quad Result post-processing & $0.001$ & $<0.01$ & $<0.01$ &\\
    Regression &&&&\\
    \quad Model inference  & $0.326$ & $0.145$ &  & $0.01$ \\
    \quad Result post-processing  & $0.015$ & $0.003$ &  & $<0.01$ \\
    \hline
    Total classification & $3660.462$ & -- & $100\%$ &\\
    Total regression     & $3658.088$ & -- &  & $100\%$ \\
    \bottomrule
  \end{tabular}
\end{table}

The total processing time is approximately $3.66\,s$, well below the duration of the 10-second analyzed frame. This indicates that the current non-optimized implementation is compatible with real-time throughput for the evaluated configuration. This execution-time analysis reflects throughput per newly processed 10-second DAS frame, not complete decision latency, which also includes buffering from the 50-second context and 5-window majority-voting scheme. The computational load is dominated by the spectral-feature extraction and preprocessing modules, while model inference and result post-processing are negligible in comparison for both tasks, despite small task-dependent differences in the absolute inference times. Regarding memory footprint, the XGBoost classifier and regressor models occupied $7.59\,MiB$ and $0.46\,MiB$, respectively, implying a very modest storage requirement.

\section{Discussion}
\label{sec:discussion}

The results obtained in this work demonstrate the practical relevance of DAS as a sensing modality for submarine cable protection. First, the vessel-detection results, with an overall $\FOneScore{}$-score above $90\%$, indicate that DAS can reliably identify nearby vessel activity under realistic maritime conditions, even over a continuous multi-day deployment and in the presence of diverse vessel types and operational scenarios. Second, the distance-estimation results, with a mean absolute error of 141\,m, show that DAS can provide meaningful cable-relative proximity information, which is particularly relevant for early-warning and risk-assessment tasks. From an operational perspective, this type of cable-centered information may be more directly useful for infrastructure protection than a generic vessel-monitoring output, since it is explicitely linked to the distance between the vessel and the protected cable. In this sense, the selected $1000\,m$ operating threshold is not only supported by the achieved detection performance, but also provides a practical compromise between practical early-warning requirements, signal detectability, and data balance, making it a suitable working point for cable-protection applications.

Although the present validation is based on a single submarine-cable deployment, the evaluated scenario includes substantial intra-deployment diversity, that goes well beyond proof-of-concept demonstrations. The AIS records comprised 745 unique vessels, 64417 reported positions, and 45 vessel categories (see Table~\ref{tab:vessel-stats-number-per-type} for details). For the day-wise experimental evaluation in the explored sensor range, 74771 processed 10-s DAS feature vectors were generated, with AIS-based labels involving 565 unique vessels, only 11 of which appeared in more than half of the folds. In addition, the meteorological analysis reported in the \href{https://geintra-uah.org/psi/index.html#meteo}{supplementary material Web page} showed statistically significant day-to-day variability, while the rank correlation between the analyzed meteorological indicators and the daily performance scores was negligible. Thus, although cross-deployment generalization remains to be assessed in future work, the present results were obtained under heterogeneous real-world traffic and environmental conditions within the available deployment.


Beyond the numerical performance itself and realistic evaluation conditions, these results are particularly significant because they show that the proposed DAS+ML pipeline can transform distributed submarine-cable measurements into useful, cable-relative information for protection-oriented monitoring of submarine communication and power-cable infrastructure. At the same time, the results should be regarded as a strong indication of feasibility rather than as evidence that DAS can replace existing maritime-monitoring technologies. Instead, they support their role as a complementary monitoring layer, capable of providing continuous, infrastructure-centered awareness in the cable vicinity.


The obtained performance should also be interpreted in light of both the physical characteristics of the DAS acquisition chain and the specific monitoring task being addressed. In this study, the useful spectral information for vessel monitoring is shaped not only by the radiated vessel noise, but also by factors such as the interrogator installation (e.g., induced vibration noise) and the sensing configuration (e.g., the gauge length). The experiments also indicate that vessel detection and distance estimation benefit differently from the same processing strategies (e.g., temporal averaging has a more negative effect on localization than on classification), that suggests that effective DAS-based vessel monitoring depends on the joint design of the sensing, preprocessing, and learning stages, and that the optimal processing strategy should be adapted to the specific operational objective.

The comparison with the beamforming-based baseline helps contextualize the proposed data-driven methodology with respect to classical acoustic localization approaches. Beamforming methods provide physically interpretable spatial maps and do not require training data, which makes them attractive as general-purpose tools. However, in the explored submarine-cable scenario, they are more sensitive to propagation-model assumptions and spatial ambiguities, and their computational cost grows substantially with dense spatial grids. In contrast, the proposed ML-based approach achieves lower localization errors in this study and offers a more scalable alternative for continuous vessel monitoring once suitable training data are available.

Our results should finally be contextualized in relation to the role of AIS in maritime monitoring. AIS and DAS are not fully interchangeable approaches. AIS is a cooperative system that provides vessel identity, trajectory, speed, heading, and georeferenced position over wide areas, making it highly valuable for maritime situational awareness. In contrast, the DAS-based methodology proposed here is intentionally oriented to submarine cable protection, where the key information is whether a vessel is sufficiently near the cable to pose a threat. Therefore, cable-relative proximity is already highly informative for early warning, even without full vessel localization. Rather than replacing AIS, DAS should be understood as a complementary sensing modality for independent physical monitoring near the cable, particularly when AIS data are unavailable, intentionally disabled or unreliable. Extending DAS-based monitoring from distance estimation to full vessel localization relative to the cable geometry remains an important line for future work.

\section{Conclusions and Future Work}
\label{sec:concl-future-work}

Methods aimed at maritime surveillance in submarine cable deployments are crucial to prevent accidental damage or sabotage. This work represents, to the best of our knowledge, the first openly documented systematic study in open waters and realistic conditions to address vessel detection and localization specifically for submarine cable protection, at a much higher scale than previous works in the literature, establishing a foundational framework for leveraging existing underwater fiber infrastructure in security and monitoring applications.

Leveraging a $28\,km$ ocean‐bottom fiber in the Southern Bight of the North Sea and ten days of continuous DAS and AIS data monitoring, we developed and evaluated a systematic, data-driven processing pipeline that includes:
\begin{itemize}
  \item A data-driven \emph{spectral feature extraction} based on logarithmically-spaced frequency‐band energy measurements to capture vessel acoustic signatures.
  \item Two \emph{machine learning} frameworks for (i) vessel detection posed as a binary classification problem (vessel closer of further than a predefined threshold distance) and (ii) vessel distance estimation posed as a regression problem.
  \item A rigorous evaluation over more than $74000$ data frames, adopting a $k$-fold cross-validation strategy that systematically avoids data leakage and increases data variability.
  \item A systematic exploitation of \emph{spatial} (up to 250 channels over $2.5\,km$) and \emph{temporal} redundancies to enhance robustness.
\end{itemize}

For the selected $1000\,m$ threshold distance, our best-performing vessel detection model achieved an overall $\FOneScore{}$-score of over $90\%$, with balanced class-wise performance; and for vessel localization, the selected regression system achieved a mean absolute error of $141\,m$, demonstrating reliable detection and distance estimation under diverse environmental and traffic conditions. We also carried out a comparison with a beamforming-based distance estimation method, showing that the proposed ML approach achieves lower localization errors and offers a more robust and computationally efficient alternative. These results highlight DAS as a practical alternative to conventional maritime surveillance methods, allowing for continuous, all-weather monitoring without reliance on cooperative systems such as AIS. More importantly, they demonstrate the practical relevance of DAS for submarine cable protection, since the combination of high vessel-detection performance and accurate distance estimation under real-world conditions provides actionable information for monitoring vessel activity in the vicinity of critical subsea infrastructure. In this sense, the current work goes beyond a proof-of-concept validation and provides evidence that DAS is a feasible and promising technology for continuous cable-centered maritime surveillance.

The release of the full dataset used in this work will enable other researchers to easily reproduce our machine learning experiments, compare new algorithms, and extend vessel detection efforts in other submarine cable-related contexts.
Our field validation focused on vessel detection and cable-relative distance estimation for submarine cable protection represents, to the best of our knowledge, one of the most extensive DAS-based vessel-monitoring studies specifically aimed at cable-protection applications. Since this validation is based on a single deployment, cross-cable transferability remains to be quantified through additional deployments. Nevertheless, the proposed methodology provides a reproducible framework that can be applied to future deployments.

Our first task will be devoted to applying the methodology to additional submarine cable deployments and emerging public DAS datasets~\cite{ais-data-ieee-dataport-1308-8605-25}, to assess generalization across different bathymetries, seabed couplings, vessel types (proved to have an impact in localization accuracy), traffic profiles, and weather conditions. These extended evaluations will also enable dedicated error-decomposition analyses to quantify how vessel characteristics, operating distance, bathymetry, seabed coupling, traffic profiles, and environmental conditions contribute to distance-estimation accuracy. Beyond binary detection and range estimation, we will also address ML tasks to infer vessel speed, size, and type, as well as multi-target tracking and precise localization, by integrating richer AIS metadata, advanced signal processing and more expressive deep-learning architectures, including CNN-based, recurrent, and transformer-based models. Exploiting meteorological data and bathymetric variables will be considered, either into feature sets or as model conditioning factors, given their significant influence on DAS submarine data acquisition. Finally, for real-world deployments, we will need to adopt efficient, low-latency monitoring pipelines such as the recent proposal in~\cite{biondi2025real}.

\bibliographystyle{IEEEtran-nonote}
\bibliography{paper}

@Article{Malinowski_2002,
  author  = {Malinowski, Stefan and Gloza, Ignacy},
  title   = {{Underwater Noise Characteristics of Small Ships}},
  journal = {Acta Acustica united with Acustica},
  year    = {2002},
  volume  = {88},
  pages   = {718-721},
  OPTmonth   = {09},
  note    = {s29.pdf Typical Sound Frequencies of Ships: Small vessels emit underwater noise with prominent tonal frequencies such as 37.5 Hz (diesel generator), 44 Hz (propeller blade rotation), and 50 Hz (AC power generator). These harmonics dominate narrow-band spectra under certain conditions (e.g., low-speed operation). Propeller cavitation noise typically peaks at frequencies below 100 Hz at higher speeds and has both broadband and narrow-band components. Background on Ship Detection Using ML: Although not directly focused on machine learning, the paper provides foundational acoustic signatures of small vessels, which can be crucial for feature engineering in ML models for ship detection. Machine Learning Applied to Submarine Optical Fiber Signals: The document does not directly address ML or submarine optical fibers but emphasizes narrow-band and wideband spectra analysis, which could inspire feature extraction techniques for ML. Additional Useful Highlights: Radiated noise from small vessels can elevate natural ambient noise by 10–40 dB, depending on the conditions, making them distinguishable even in noisy environments. Machinery noise and propeller cavitation are primary sources of radiated noise, with patterns varying based on vessel speed and operational conditions. Ships with lengths under 60 m experience higher proximity noise due to machinery, an aspect that may influence signal characteristics in nearby optical fibers. Detailed methodologies for noise measurement, such as hydrophone arrays and vibration analysis, are provided and could be useful for benchmarking or validating ML-driven detection systems. Small vessels emit underwater noise with prominent tonal frequencies such as 37.5 Hz (diesel generator), 44 Hz (propeller blade rotation), and 50 Hz (AC power generator). These harmonics dominate narrow-band spectra under certain conditions (e.g., low-speed operation). Propeller cavitation noise typically peaks at frequencies below 100 Hz at higher speeds and has both broadband and narrow-band components. Background on Ship Detection Using ML: Although not directly focused on machine learning, the paper provides foundational acoustic signatures of small vessels, which can be crucial for feature engineering in ML models for ship detection. Machine Learning Applied to Submarine Optical Fiber Signals: The document does not directly address ML or submarine optical fibers but emphasizes narrow-band and wideband spectra analysis, which could inspire feature extraction techniques for ML. Additional Useful Highlights: Radiated noise from small vessels can elevate natural ambient noise by 10–40 dB, depending on the conditions, making them distinguishable even in noisy environments. Machinery noise and propeller cavitation are primary sources of radiated noise, with patterns varying based on vessel speed and operational conditions. Ships with lengths under 60 m experience higher proximity noise due to machinery, an aspect that may influence signal characteristics in nearby optical fibers. Detailed methodologies for noise measurement, such as hydrophone arrays and vibration analysis, are provided and could be useful for benchmarking or validating ML-driven detection systems.},
}

@Article{Karasalo_2017,
  author    = {Karasalo, Ilkka and Östberg, Martin and Sigray, Peter and Jalkanen, Jukka-Pekka and Johansson, Lasse and Liefvendahl, Mattias and Bensow, Rickard},
  title     = {{Estimates of Source Spectra of Ships from Long Term Recordings in the Baltic Sea}},
  journal   = {Frontiers in Marine Science},
  year      = {2017},
  volume    = {4},
  pages     = {164},
  OPTmonth     = jun,
  issn      = {2296-7745},
  OPTnote      = {fmars-04-00164.pdf Typical Sound Frequencies of Different Ship Types: The study estimates underwater radiated noise (URN) source spectra for 943 ships across 2,088 passages in the Baltic Sea. Frequency-dependent source level estimates span 21 1/3-octave bands, ranging from 10 Hz to 1016 Hz. Propeller cavitation and machinery vibrations dominate low-frequency noise below 200 Hz. Background on Ship Detection with ML Using Submarine Fiber Signals: While not directly applied to optical fibers, the methodologies for source level estimation (AIS integration, propagation modeling) align well with signal preprocessing for ML-based detection tasks. Machine Learning Applied to Submarine Optical Fiber Signals with Any Objective: The approach of combining AIS data and sound propagation modeling to estimate ship-specific noise profiles can be adapted for training data preparation in ML-based underwater applications. Other Useful Highlights: The paper presents a cost-effective method for URN monitoring using a single hydrophone, addressing data scarcity for ML model training. Variability in source levels, influenced by ship type and operating conditions, is quantified, with median levels differing by up to 7 dB for major ship categories. Seabed and sound speed profile modeling account for environmental effects on signal transmission, a critical consideration for submarine sensor networks.},
  doi       = {10.3389/fmars.2017.00164},
  publisher = {Frontiers Media SA},
}

@Article{vadov2001long,
  author    = {Vadov, RA},
  title     = {{Long-range sound propagation in the central region of the Baltic Sea}},
  journal   = {Acoustical Physics},
  year      = {2001},
  volume    = {47},
  number    = {2},
  pages     = {150--159},
  note      = {1.1355799.pdf Typical Sound Frequencies of Different Ship Types: The experiments focus on low-frequency sound propagation (<5 kHz), with attenuation coefficients calculated for a range of frequencies. Specific critical frequencies for sound attenuation vary by season (e.g., 100–600 Hz in summer conditions). Background on Sound Propagation: The study emphasizes long-range sound propagation in the Baltic Sea, leveraging the Underwater Sound Channel (USC) as a waveguide for low-frequency signals. Seasonal variations in salinity and temperature significantly influence sound speed profiles and propagation anomalies. Relevance to Ship Detection and ML: While the focus is not directly on ML, the experimental data on sound field decay, attenuation coefficients, and bottom wave characteristics provide a foundation for preprocessing acoustic signals in ML models. Insightful attenuation mechanisms (e.g., internal wave scattering, sediment interaction) could inform ML-based feature engineering for ship noise detection. Other Useful Highlights: The study demonstrates methods for analyzing sound scattering caused by internal waves and seabed interaction, contributing to attenuation at frequencies near 1 kHz. Seasonal sound speed profiles in the Baltic Sea (e.g., spring versus summer) impact sound propagation, with critical implications for training ML models sensitive to environmental variability. Attenuation coefficients derived from experimental data provide a baseline for noise modeling, relevant for creating synthetic datasets in underwater acoustic research.},
  publisher = {Springer},
}

@Article{Rivet_2021,
  author    = {Rivet, Diane and de Cacqueray, Benoit and Sladen, Anthony and Roques, Aurélien and Calbris, Gaëtan},
  title     = {{Preliminary assessment of ship detection and trajectory evaluation using distributed acoustic sensing on an optical fiber telecom cable}},
  journal   = {The Journal of the Acoustical Society of America},
  year      = {2021},
  volume    = {149},
  number    = {4},
  pages     = {2615--2627},
  OPTmonth     = apr,
  issn      = {1520-8524},
  note      = {Abstract Distributed acoustic sensing (DAS) is a recent instrumental approach allowing the conversion of fiber-optic cables into dense arrays of acoustic sensors. This technology is attractive in marine environments where instrumentation is difficult to implement. A promising application is the monitoring of environmental and anthropic noise, leveraging existing telecommunication cables on the seafloor. We assess the ability of DAS to monitor such noise using a 41.5 km-long cable offshore of Toulon, France, focusing on a known and localized source. We analyze the noise emitted by the same tanker cruising above the cable, first 5.8 km offshore in 85 m deep bathymetry, and then 20 km offshore, where the seafloor is at a depth of 2000 m. The spectral analysis, the Doppler shift, and the apparent velocity of the acoustic waves striking the fiber allow us to separate the ship radiated noise from other noise. At 85 m water depth, the signal-to-noise ratio is high, and the trajectory of the boat is recovered with beamforming analysis. At 2000 m water depth, although the acoustic signal of the ship is more attenuated, signals below 50 Hz are detected. These results confirm the potential of DAS applied to seafloor cables for remote monitoring of acoustic noise even at intermediate depth.},
  doi       = {10.1121/10.0004129},
  publisher = {Acoustical Society of America (ASA)},
}

@InProceedings{Drylerakis2024source,
  author       = {Drylerakis, Konstantinos Theofilos and Belal, Mohammad and Mestre, Rafael and Norman, Timothy J and Evers, Christine},
  title        = {{Source detection and tracking for underwater distributed acoustic sensing}},
  booktitle    = {{2024 32nd European Signal Processing Conference (EUSIPCO)}},
  year         = {2024},
  pages        = {1292--1296},
  OPTorganization = {IEEE},
  note         = {Source_Detection_and_Tracking_for_Underwater_Distributed_Acoustic_Sensing.pdf Typical Sound Frequencies of Different Ship Types: Focused on detecting and tracking acoustic sources, including marine vessel motion. The study highlights a 30 Hz eigenfrequency associated with vessel motion, significant for DAS systems. Background on Ship Detection Using Machine Learning: Utilizes PCA (Principal Component Analysis) for feature extraction and dimensionality reduction from DAS data. Implements a Gaussian Mixture Probability Hypothesis Density (GM-PHD) filter for real-time tracking, providing a robust approach to detecting vessel trajectories along submarine optical cables. Machine Learning Applied to Submarine Optical Fiber Signals: Proposes a PCA-based unsupervised framework to identify structured acoustic sources from noisy submarine DAS datasets. Leverages machine learning to preprocess, filter, and track acoustic signals for identifying marine vessels, a direct application for submarine optical fibers. Other Useful Highlights: The study introduces novel unsupervised methodologies to extract features from dense DAS data matrices and use them for tracking vessel movement in real-time. Provides detailed workflows involving band-pass filtering, PCA decomposition, and clustering to separate noise from active source signals in submarine environments. Real-world evaluation conducted using DAS data from a 5.65 km offshore cable in Galway Bay, Ireland. Results show potential for improving marine environmental monitoring and traffic detection with DAS systems.},
  doi={10.23919/EUSIPCO63174.2024.10715378}
}

@InProceedings{Thiem_2023,
  author    = {Thiem, Lukas and Wienecke, Susann and Taweesintananon, Kittinat and Vaupel, Melvin and Landrø, Martin},
  title     = {{Ship noise characterization for marine traffic monitoring using distributed acoustic sensing}},
  booktitle = {{2023 IEEE International Workshop on Metrology for the Sea; Learning to Measure Sea Health Parameters}},
  year      = {2023},
  pages     = {334--339},
  month     = oct,
  publisher = {IEEE},
  note      = {Ship_noise_characterization_for_marine_traffic_monitoring_using_distributed_acoustic_sensing.pdf Typical Sound Frequencies of Different Ship Types: The study uses band-pass filtering from 100–120 Hz to focus on vessel-generated P-wave signals, emphasizing this range for noise reduction and analysis in shallow, noisy environments. Background on Ship Detection Using Machine Learning: While not directly applying ML, the study integrates signal processing and novel methodologies (e.g., persistent homology) to detect and track vessel signals. These approaches provide a foundation for ML-based feature engineering and classification models. Machine Learning Applied to Submarine Optical Fiber Signals: Persistent homology and image processing techniques, such as Gaussian filtering and Otsu’s thresholding, are employed to enhance signal detection. These techniques can be adapted for preprocessing and feature extraction in ML models using DAS data. Other Useful Highlights: Demonstrates the feasibility of DAS for marine traffic monitoring, even with poorly coupled cables in soft sediments. Successfully localizes vessels using direct P-wave arrivals and travel-time inversion, achieving results comparable to GPS tracking from AIS. Highlights the potential of DAS for detecting "dark ships" (vessels without AIS signals), which is significant for maritime security and environmental monitoring. Discusses the challenges and limitations of shallow water environments, such as uncertainties in bathymetry, sound velocity, and cable positioning, which affect localization accuracy. Provides a roadmap for integrating DAS with automatic algorithms and edge computing for future ML applications in signal classification.},
  doi       = {10.1109/metrosea58055.2023.10317227},
}

@Article{Lior_2021,
  author    = {Lior, Itzhak and Sladen, Anthony and Rivet, Diane and Ampuero, Jean‐Paul and Hello, Yann and Becerril, Carlos and Martins, Hugo F. and Lamare, Patrick and Jestin, Camille and Tsagkli, Stavroula and Markou, Christos},
  title     = {{On the Detection Capabilities of Underwater Distributed Acoustic Sensing}},
  journal   = {Journal of Geophysical Research: Solid Earth},
  year      = {2021},
  volume    = {126},
  number    = {3},
  pages     = {e2020JB020925},
  OPTmonth     = mar,
  issn      = {2169-9356},
  note      = {JGR Solid Earth - 2021 - Lior - On the Detection Capabilities of Underwater Distributed Acoustic Sensing.pdf Typical Sound Frequencies of Different Ship Types: While not directly focused on ship types, the study identifies frequencies associated with seismic signals and natural noise sources, including surface gravity waves (0.05–0.3 Hz) and secondary microseisms (0.3–2 Hz), which overlap with some low-frequency ship noises. Background on DAS Applications: Demonstrates the potential of Distributed Acoustic Sensing (DAS) for underwater seismic monitoring, leveraging dark fibers in the Mediterranean Sea. Highlights the capability of DAS to detect transient ground deformations and convert strain-rate measurements into ground motion spectra for earthquakes, analogous to broadband seismometers. Relevance to Machine Learning: The study's robust methods for signal processing, including frequency-wavenumber (f-k) analysis, and phase velocity estimation, provide preprocessing frameworks applicable to ML-based anomaly or event detection in DAS datasets. Insights into the effect of bathymetry and cable coupling on noise and signal quality offer valuable features for ML algorithms. Other Useful Highlights: Explores the influence of bathymetry, sedimentary basins, and cable coupling on DAS signal quality, noting that flatter regions with sediments provide stronger signals, while irregular bathymetry reduces amplitude. Identifies Scholte waves as the dominant mode for DAS signal propagation in underwater environments, with low apparent velocities leading to higher strain amplitudes, favorable for detection. Suggests DAS as a complementary technology to existing seismometers for real-time earthquake early warning (EEW), with implications for broader environmental monitoring. Highlights the challenges of high-frequency instrumental noise and irregular coupling, which are mitigated with advanced interrogators and optimized cable deployment.},
  doi       = {10.1029/2020jb020925},
  publisher = {American Geophysical Union (AGU)},
}

@Article{Mata_Flores_2023,
  author    = {Mata Flores, D and Mercerat, E D and Ampuero, J P and Rivet, D and Sladen, A},
  title     = {{Identification of two vibration regimes of underwater fibre optic cables by distributed acoustic sensing}},
  journal   = {Geophysical Journal International},
  year      = {2023},
  volume    = {234},
  number    = {2},
  pages     = {1389--1400},
  OPTmonth     = mar,
  issn      = {1365-246X},
  note      = {ggad139.pdf Typical Sound Frequencies of Different Ship Types: This paper does not explicitly address ship noise frequencies but identifies oscillation regimes of underwater cables, which could interact with marine acoustic sources. Background on DAS Applications: Highlights two vibration regimes detected via DAS: High-frequency (HF) oscillations (>2 Hz): Triggered by seismic waves and confined to cable sections pinned by seafloor features. Low-frequency (LF) oscillations (<1 Hz): Associated with vortex-induced vibrations (VIV) caused by deep ocean currents in suspended cable sections. Relevance to Machine Learning Applications: While ML is not explicitly applied, the detection and classification of HF and LF oscillations can inform the development of ML models for analyzing mechanical coupling between cables and the seafloor, potentially extending to ship detection scenarios. Other Useful Highlights: Proposes DAS as a tool for monitoring mechanical cable/seafloor coupling over time, providing insights into cable usability for seismic and environmental monitoring. HF oscillations are linked to longitudinal wave propagation along cable segments, with speeds near 4000 m/s, consistent with the properties of steel-armored cables. LF oscillations exhibit quasi-uniform strain rates and are driven by environmental conditions, such as sedimentation and erosion, affecting suspended cable segments. Recommends using LF and HF oscillations as indicators for cable condition, aiding in identifying sections suitable for seismic monitoring or prone to fatigue and damage.},
  doi       = {10.1093/gji/ggad139},
  publisher = {Oxford University Press (OUP)},
}

@Article{Douglass_2023,
  author    = {Douglass, Alexander S. and Abadi, Shima and Lipovsky, Bradley P.},
  title     = {{Distributed acoustic sensing for detecting near surface hydroacoustic signals}},
  journal   = {JASA Express Letters},
  year      = {2023},
  volume    = {3},
  number    = {6},
  OPTmonth     = jun,
  pages     = {066005},
  issn      = {2691-1191},
  note      = {066005_1_10.0019703.pdf Typical Sound Frequencies of Different Ship Types: The study does not focus on ship noise specifically but analyzes DAS's ability to detect acoustic signals from 1 Hz to 700 Hz, including ship-related signals within this range. Background on DAS Applications: Demonstrates DAS as a cost-effective alternative to hydrophones for underwater acoustic monitoring. Compares DAS data with co-located hydrophone recordings in Puget Sound, showcasing DAS's ability to measure acoustic signals in a controlled experiment. Relevance to Machine Learning and Ship Detection: Highlights challenges such as low signal-to-noise ratios (SNR) in DAS data compared to hydrophones, necessitating advanced signal processing (e.g., filtering, channel stacking) to improve SNR for analysis. Suggests potential for ML techniques to enhance signal detection and classification in noisy DAS datasets. Technical Insights: The DAS cable exhibited high sensitivity in shallow waters, detecting acoustic signals from broadband impulsive sources, with clear responses in the 300–500 Hz range. Channel stacking significantly improved SNR, enabling DAS to detect signals up to 700 Hz, albeit with limitations compared to hydrophones. Challenges and Limitations: Variability in acoustic detection due to sediment cover on DAS cables; unburied sections perform better at detecting acoustic signals. Identifies gauge length and channel sensitivity as factors affecting frequency response, suggesting further optimization for broad-spectrum monitoring. Future Directions: Advocates for further studies integrating DAS with intelligent algorithms for noise suppression, signal enhancement, and automatic classification of marine acoustic signals. Recommends extending DAS monitoring to higher frequencies (>700 Hz) and improving calibration techniques using co-located hydrophones.},
  doi       = {10.1121/10.0019703},
  publisher = {Acoustical Society of America (ASA)},
}

@Article{Landro_2022,
  author    = {Landrø, Martin and Bouffaut, Léa and Kriesell, Hannah Joy and Potter, John Robert and Rørstadbotnen, Robin André and Taweesintananon, Kittinat and Johansen, Ståle Emil and Brenne, Jan Kristoffer and Haukanes, Aksel and Schjelderup, Olaf and Storvik, Frode},
  title     = {{Sensing whales, storms, ships and earthquakes using an Arctic fibre optic cable}},
  journal   = {Scientific Reports},
  year      = {2022},
  volume    = {12},
  number    = {1},
  OPTmonth     = nov,
  issn      = {2045-2322},
  OPTnote      = {2022-Sensing whales, storms, ships and earthquakes using an Arctic fibre optic cable-s41598-022-23606-x.pdf Typical Sound Frequencies of Ships: Ship-generated signals were primarily detected in the 35–85 Hz range, leveraging DAS capabilities to monitor acoustic signatures and Doppler shifts for localization and tracking. Innovative DAS Applications: Demonstrates DAS for real-time monitoring of ships, whales, earthquakes, and storms using Arctic fiber optic cables. Highlights the potential of FO cables to transform global monitoring networks, combining DAS data with AIS and satellite data for comprehensive maritime surveillance. Relevance to Machine Learning: Suggests the integration of automated data analysis and classification, including ML, for real-time anomaly detection and tracking of multiple phenomena (e.g., ship traffic, marine mammal movements). Applications for Underwater Monitoring: Successfully detected and tracked the Norbjørn cargo ship, achieving an RMS deviation of ±50 m in localization compared to AIS data. Detected baleen whale calls up to 95 km along the cable, distinguishing between species like blue whales and fin whales based on their vocalization patterns (e.g., 9 Hz for blue whale calls). Technical Highlights: DAS systems achieved unprecedented signal-to-noise ratios with beamforming, enabling both range and bearing determination of acoustic sources. Streamed 250 TB of data over 44 days from Svalbard, showcasing the feasibility of large-scale data collection and analysis. Challenges and Future Directions: Identifies key challenges such as improving SNR, increasing range capabilities, and automating detection and classification. Advocates for broader deployments of DAS for ocean monitoring, offering new insights into phenomena like distant storms, earthquakes, and anthropogenic impacts. Broader Implications: DAS provides a cost-effective, scalable solution for global Earth-Ocean-Atmosphere monitoring, with applications in climate studies, biodiversity conservation, and seismic hazard assessment.},
  doi       = {10.1038/s41598-022-23606-x},
  pages     = {19226},
  publisher = {Springer Science and Business Media LLC},
}

@InProceedings{Malaprade2019,
  author    = {Jacques Malaprade and Ryan Hunt and Gareth Lees},
  title     = {{Toward Detecting Ship Characteristics and Movements using DAS and Machine Learning}},
  booktitle = {Proceedings of the 10th International Conference on Insulated Power Cables (Jicable'19)},
  year      = {2019},
  address   = {Versailles, France},
  month     = {June},
  OPTnote      = {2019-Toward Detecting Ship Characteristics and Movements using DAS and Machine Learning.pdf Typical Sound Frequencies of Ships: Ship crossings were identified using Distributed Acoustic Sensing (DAS) frequency-band energy (FBE) plots, primarily focusing on vibrational energy observed in the low-frequency range (<50 Hz). Application of DAS and Machine Learning: Demonstrates the integration of DAS technology with Machine Learning (ML) to detect and classify ships crossing subsea power cables. Two ML models were tested: Convolutional Neural Networks (CNNs) for image-based feature extraction. Support Vector Machines (SVMs) for simpler datasets with fewer training requirements. Results: SVM performance: Achieved a classification accuracy of 97.1% on the test dataset, with a low false positive rate (0.06). CNN performance: Achieved only 57% accuracy, highlighting the limitations of CNNs with small datasets and overfitting issues. Signal Processing: Used Fast Fourier Transform (FFT) techniques to convert phase signals into frequency-band energy plots. FBE plots were treated as raster images for ML classification, with spatial resolution of 1.28 meters and time resolution of 0.25 seconds. Significance and Future Directions: Demonstrates the feasibility of using DAS for ship detection and tracking in areas where AIS signals are unavailable or intentionally switched off. Highlights the need for larger, more diverse datasets to improve ML model performance, particularly for CNNs. Suggests extending ML applications to detect ship characteristics such as speed, tonnage, and draught, based on DAS signal patterns. Practical Implications: DAS systems can augment maritime navigation and security by providing real-time detection and classification of ships, particularly in regions prone to illegal activities (e.g., smuggling or unregulated fishing).},
  OPTurl       = {https://www.researchgate.net/publication/334090687_Toward_Detecting_Ship_Characteristics_and_Movements_using_DAS_and_Machine_Learning},
  howpublished = {\href{https://www.researchgate.net/publication/334090687_Toward_Detecting_Ship_Characteristics_and_Movements_using_DAS_and_Machine_Learning}{Online, accessed Mar. 2025}}
}

@InProceedings{Brenne2019,
  author       = {O. Brenne and S. Besanger and P. Travers},
  title        = {{Distributed Acoustic Sensing for Submarine Cable Protection}},
  booktitle    = {SubOptic 2019 Conference Proceedings},
  year         = {2019},
  address      = {New Orleans, USA},
  OPTorganization = {SubOptic},
  note         = {2019-Distributed Acoustic Sensing for Submarine Cable Protection - SubOpticConf-OP4-1-BRENNE-ASN.pdf Applications of DAS for Cable Protection: Demonstrates the use of Distributed Acoustic Sensing (DAS) to monitor strain and vibrations along submarine telecom cables in the North Sea. Successfully detects: Large vessels up to 8 km away from the cable. Seabed trawler activity within 2.5 km of the cable. Frequency and Acoustic Wave Modes: Detected frequencies include: P-wave modes (25–90 Hz), associated with water-column vibrations. Scholte waves (dominant at 9 Hz), sensitive to sediment interactions. Love waves (200 m/s propagation speed), excited by seabed contact. Relevance to Maritime Surveillance: DAS can distinguish between vessel types (cargo vessels vs. trawlers) based on wave modes and signal intensity. Provides actionable insights for detecting activities like bottom trawling that could damage submarine cables. Technical Insights: Data were collected using the OptoDAS interrogator, capable of measuring strain vibrations over 130 km. Long cables act as "coherent antennas," amplifying detection ranges for noise-limited signals up to 50 km. Environmental and Operational Benefits: DAS enhances early warning systems against threats like anchors and trawlers, reducing cable damage risks. Ambient noise measurements allow detailed mapping of cable routes, aiding in maintenance and repair planning. Challenges and Future Directions: Seabed variability and cable burial depth affect wave propagation, necessitating site-specific calibrations. Recommends further integration with AIS and advanced processing techniques for enhanced detection accuracy.},
  url          = {https://suboptic.org},
  howpublished = {Accessed june 2026}
}

@Article{Stork_2020,
  author    = {Stork, Anna L. and Baird, Alan F. and Horne, Steve A. and Naldrett, Garth and Lapins, Sacha and Kendall, J.-Michael and Wookey, James and Verdon, James P. and Clarke, Andy and Williams, Anna},
  title     = {{Application of machine learning to microseismic event detection in distributed acoustic sensing data}},
  journal   = {GEOPHYSICS},
  year      = {2020},
  volume    = {85},
  number    = {5},
  pages     = {KS149–KS160},
  OPTmonth     = sep,
  issn      = {1942-2156},
  note      = {10.1190@geo2019-0774.1.pdf Application of Machine Learning with DAS: Demonstrates the first successful use of a convolutional neural network (CNN), specifically YOLOv3, to detect microseismic events in DAS data. Transferability of trained CNN models to datasets from different geological settings and fiber configurations. Advantages of DAS: DAS technology offers dense spatial and temporal sampling, enabling the monitoring of microseismic activity, including induced seismicity at industrial sites. Can capture seismic events with a strain-rate sensitivity that complements or surpasses traditional geophones for certain applications. Synthetic Data for Training: Synthetic microseismic data with real noise added were used to train the CNN, ensuring the training set had a known ground truth. Variations in training data included different event magnitudes, mechanisms, and SNR levels to enhance generalizability. Performance Comparison: YOLOv3 detected approximately 14% more seismic events than traditional filtering techniques and had a false detection rate of 2%. Performance metrics validated its utility for near real-time seismic monitoring, especially in industrial and geophysical contexts. Limitations and Future Directions: Events with low SNR (<3) were often missed, suggesting future training datasets should include such examples. Recommendations include expanding the synthetic dataset with varied configurations and geological settings to improve transferability. Broader Applications: Highlights DAS's potential beyond microseismic event detection, such as its use for geophysical imaging and hazard assessment. Machine Learning in DAS: Demonstrates the application of convolutional neural networks (CNNs), specifically YOLOv3, for microseismic event detection in DAS data. Transferability of the trained model to datasets with different geometries and conditions was validated. Achieves higher event detection rates compared to traditional STA/LTA methods, detecting 14% more events with a low false detection rate (2%). Technical Highlights: Uses a synthetic dataset augmented with real noise for training, allowing robust validation. Processes large DAS datasets efficiently (e.g., 650GB/day for a 2 km cable with 2000Hz sampling). Demonstrates near real-time event detection capability, crucial for industrial applications like hydraulic fracturing monitoring. Comparison of Methods: CNN (YOLOv3) outperforms traditional STA/LTA and f-k filtering methods in terms of detection accuracy and event generalizability. Classical signal processing methods like 2D median and Fourier filtering are also explored as baseline comparisons. Key Contributions: The trained CNN generalizes well across datasets recorded in varying geological settings and by different DAS systems. Provides a framework for applying advanced ML techniques to seismic monitoring and reducing manual inspection workload. Challenges and Recommendations: Data volumes require careful preprocessing and efficient storage solutions. Suggests further training with lower signal-to-noise ratio (SNR) examples to enhance model robustness. Recommends additional synthetic datasets to improve generalization across different DAS configurations. Potential Applications: Microseismic monitoring in hydraulic fracturing and geothermal sites. Real-time event detection for operational decision-making in industrial settings. File Information Filename: 10.1190@geo2019-0774.1.pdf Potential BibTeX Fields: Journal: Geophysics DOI: 10.1190/geo2019-0774.1 Authors: Anna L. Stork, Alan F. Baird, Steve A. Horne, Garth Naldrett, Sacha Lapins, J.-Michael Kendall, James Wookey, James P. Verdon, Andy Clarke, Anna Williams. Year: 2020},
  doi       = {10.1190/geo2019-0774.1},
  publisher = {Society of Exploration Geophysicists},
}

@Article{Chen_2022,
  author    = {Chen, Shaoyi and Zhu, Kun and Han, Jun and Sui, Qi and Li, Zhaohui},
  title     = {{Photonic Integrated Sensing and Communication System Harnessing Submarine Fiber Optic Cables for Coastal Event Monitoring}},
  journal   = {IEEE Communications Magazine},
  year      = {2022},
  volume    = {60},
  number    = {12},
  pages     = {110--116},
  OPTmonth     = dec,
  issn      = {1558-1896},
  note      = {Photonic_Integrated_Sensing_and_Communication_System_Harnessing_Submarine_Fiber_Optic_Cables_for_Coastal_Event_Monitoring.pdf Photonic ISAC Technology: Combines optical communication and sensing using submarine telecommunication fiber optic cables. Utilizes phase-sensitive optical time-domain reflectometry (Phi-OTDR) for detecting acoustic signals. Applications: Ocean Wave Analysis: Studies ocean dynamics using frequency-wavenumber (f-k) domain analysis. Differentiates surface gravity waves (<0.3 Hz) and transoceanic seismic waves (>0.8 Hz). Earthquake Detection: Successfully monitors in-land, coastal, and offshore seismic events (e.g., Mw 4.0). Provides high sensitivity to microseisms (e.g., Mw 1.1) and teleseisms, with frequency ranges of 2–15 Hz. Ferry Monitoring: Detects and tracks vessel activities, including trajectory mapping using acoustic signals from ferries. Advantages of the Approach: Leverages existing submarine fiber optic infrastructure without disrupting communication. Cost-effective and environmentally friendly compared to conventional methods (e.g., ocean-bottom seismometers). Offers real-time monitoring capabilities for coastal geological and marine activities. Technical Insights: Achieves independent operation of communication and sensing channels using wavelength-division multiplexing (WDM). Data from sensing signals are processed with machine learning algorithms to enhance SNR and pattern recognition. Future Directions: Investigating new fiber technologies (e.g., multicore, few-mode fibers) for improved sensing performance. Expanding applications to early warning systems for earthquakes and tsunamis. Challenges: External vibration and acoustic noise from the environment require advanced filtering and processing techniques.},
  doi       = {10.1109/mcom.002.2200191},
  publisher = {Institute of Electrical and Electronics Engineers (IEEE)},
}

@Article{Harati-Mokhtari_Wall_Brooks_Wang_2007AIS,
  author  = {Harati-Mokhtari, Abbas and Wall, Alan and Brooks, Philip and Wang, Jin},
  title   = {{Automatic Identification System (AIS): Data Reliability and Human Error Implications}},
  journal = {Journal of Navigation},
  year    = {2007},
  volume  = {60},
  number  = {3},
  pages   = {373–389},
  note    = {automatic-identification-system-ais-data-reliability-and-human-error-implications.pdf AIS Overview and Importance: Automatic Identification System (AIS) was implemented to enhance navigation safety, efficiency, and environmental protection by providing real-time vessel information. Provides data on vessel identification, location, navigational status, and voyage-related information, useful for collision avoidance and maritime traffic management. AIS Data Challenges: Inaccurate Data: High error rates in AIS fields, including MMSI numbers, vessel types, and voyage-related information such as draught and destination. Errors stem from human input, equipment installation, and lack of standardization. Human Error: Majority of errors are due to forgetfulness or omissions by navigators, particularly in manually entered fields. Training gaps lead to navigational officers misunderstanding or misusing AIS functionality. Systemic Issues and Failures: Design inconsistencies across AIS manufacturers lead to data discrepancies. Lack of interlinking between AIS and other onboard systems (e.g., speed sensors) results in mismatched data fields. The "Swiss Cheese" model is applied to illustrate vulnerabilities in AIS data handling due to human and system errors. Case Studies: Highlighted several collisions (e.g., Hyundai Dominion and Sky Hope) where AIS misuse or errors contributed to accidents. Duplicate MMSI numbers and incorrect vessel status were common issues observed in operational scenarios. Recommendations for Improvement: Standardization: Uniform standards for AIS design and terminology across manufacturers. Integration with other navigation systems for automatic cross-verification of data. Training: Comprehensive training for navigators on AIS operation, data interpretation, and error detection. Regulation and Supervision: Stronger enforcement of data accuracy through regular inspections and penalties for non-compliance. Technological Enhancements: Incorporation of warning systems for detecting inconsistent or conflicting AIS data. Implications for my Research: The study highlights the importance of AIS data accuracy for maritime monitoring, which is crucial when integrating AIS with Distributed Acoustic Sensing (DAS) for ship detection. Provides insights into potential sources of AIS data errors, which can inform preprocessing and validation in ML models using AIS data.},
  doi     = {10.1017/S0373463307004298},
}

@Article{emmens2021promisesAIS,
  author    = {Emmens, Ties and Amrit, Chintan and Abdi, Asad and Ghosh, Mayukh},
  title     = {{The promises and perils of Automatic Identification System data}},
  journal   = {Expert Systems with Applications},
  year      = {2021},
  volume    = {178},
  pages     = {114975},
  note      = {1-s2.0-S0957417421004164-main.pdf AIS Data Usage in Maritime Applications: AIS (Automatic Identification System) is crucial for identifying vessels in navigation, environmental monitoring, and improving navigational safety. Applications include route optimization, collision prevention, and emissions estimation. Challenges with AIS Data: Noise: Significant noise is present in static, dynamic, and voyage-related AIS data. For instance, positional data and Speed over Ground (SOG) often exhibit inconsistencies. Equipment Quality: Variability in equipment performance impacts data accuracy. Issues include unstable Course over Ground and incorrect SOG values. Human Factors: Errors stem from incorrect data entry, equipment misconfiguration, and deliberate signal switch-offs. Applications Highlighted: Safety and Collision Avoidance: AIS supports navigational safety by enabling vessel identification and communication. Environmental Monitoring: AIS can assess adherence to environmental regulations, such as restricted areas or emissions tracking. Supply Chain Optimization: Predictive analytics using AIS enhances port management and facilitates automatic tax collection. Research Insights: Mixed-method design combining quantitative AIS data analysis (from the Port of Amsterdam) with qualitative expert interviews provides validation. Identifies systemic issues like incomplete tracks, data gaps, and the need for better integration of AIS with other systems (e.g., radar, geographic data). Future Directions: Addressing equipment inconsistencies and integrating advanced data filtering to mitigate noise. Enhancing the reliability of static and voyage-related AIS data for broader applications.},
  publisher = {Elsevier},
}

@techreport{ITU_R_M1371_5AIS,
  author       = {{International Telecommunication Union}},
  title        = {{Technical Characteristics for an Automatic Identification System Using Time Division Multiple Access in the VHF Maritime Mobile Band (Recommendation ITU-R M.1371-5)}},
  institution  = {International Telecommunication Union},
  address      = {Geneva, Switzerland},
  year         = {2014},
  type         = {Technical Report}
}

@Article{Wong2019-kfold,
  author   = {Wong, Tzu-Tsung and Yeh, Po-Yang},
  title    = {{Reliable Accuracy Estimates from k-Fold Cross Validation}},
  journal  = {IEEE Transactions on Knowledge and Data Engineering},
  year     = {2020},
  volume   = {32},
  number   = {8},
  pages    = {1586-1594},
  note     = {Reliable_Accuracy_Estimates_from_k-Fold_Cross_Validation.pdf Purpose: Investigates the reliability of accuracy estimates from k-fold cross-validation (CV), especially with repeated iterations. Analyzes dependency relationships between replications of k-fold CV and proposes methods to test these dependencies. Findings: Accuracy estimates from replications of k-fold CV are generally highly correlated, with correlation increasing as the number of folds rises. Dependency among accuracy estimates impacts the calculation of variance, leading to potential underestimation if correlations are ignored. Statistical methods for evaluating dependency strength are proposed. Recommendations: Use a larger number of folds with fewer replications for performance evaluation of classification algorithms to reduce variance. Repeating k-fold CV provides diminishing returns in reducing variance when folds are highly dependent. Implications for Machine Learning: Highlights the importance of variance estimation and reliable performance evaluation using cross-validation, particularly in high-stakes ML model assessments. Provides insights into designing more robust CV strategies tailored to specific dataset characteristics and model requirements. Methodology: Dependency analysis using statistical methods and experiments on 20 datasets. Results emphasize the trade-off between the number of folds and replications. Practical Applications: Performance evaluation in classification tasks where accuracy estimates are critical. Suitable for research comparing multiple ML algorithms using consistent and reliable evaluation metrics.},
  doi      = {10.1109/TKDE.2019.2912815},
}

@Article{zhan2024application,
  author    = {Zhan, Yage and Liu, Lirui and Li, Kehan},
  title     = {{Application of machine learning for signal recognition in distributed fibre optic acoustic sensing technology}},
  journal   = {IET Optoelectronics},
  year      = {2024},
  volume    = {18},
  number    = {4},
  pages     = {81--95},
  note      = {Application_of_machine_learning_for_signal_recogni.pdf Overview: Focuses on Distributed Acoustic Sensing (DAS) based on coherent Rayleigh scattering for real-time vibration and acoustic monitoring. Reviews signal recognition and feature extraction methodologies used in DAS, with comparisons between wavelet-based, Fourier transform-based, and neural network-based approaches. DAS Development Stages: Qualitative Detection: Disturbance detection along optical fibers. Quantitative Detection: Extraction of detailed vibration and acoustic information. Performance Enhancement: Improved spatial resolution (meter to centimeter level) and extended sensing distances (>100 km). Feature Extraction Techniques: Wavelet-Based Methods: Effective for multi-scale, noise-robust analysis of non-stationary signals. Fourier Transform: Suitable for periodic signals, emphasizing frequency spectrum features. Empirical Mode Decomposition (EMD): Handles nonlinear and non-stationary signals, effectively addressing complex time-frequency characteristics. Other Techniques: Includes Wigner-Ville distribution and Spectral Euclidean Distance methods, useful for transient and spectral feature extraction. Pattern Recognition Approaches: Traditional Methods: Support Vector Machine (SVM), Random Forest (RF), and other classifiers provide robust results with appropriate feature engineering. Neural Network Models: Convolutional Neural Networks (CNN): Excels in time-frequency domain feature extraction. Long Short-Term Memory (LSTM): Captures temporal dependencies in vibration signals. Backpropagation Neural Networks (BPNN): Effective for diverse classification tasks with simpler architectures. Applications: Pipeline Monitoring: Leak detection and security. Perimeter Security: Intrusion event recognition. Geological Exploration: Vibration signal analysis for environmental monitoring. Performance Results: Achieved recognition accuracies range from 82% to 100% across various use cases depending on methods and datasets. Multi-feature fusion consistently outperformed single-feature approaches in recognition tasks. Future Directions: Integration of multi-feature fusion algorithms for enhanced accuracy and robustness. Leveraging advanced AI techniques and optimization methods to improve DAS applications in real-world scenarios.},
  doi       = {10.1049/ote2.12120},
  publisher = {Wiley Online Library},
}

@InProceedings{wienecke2023new,
  author       = {Wienecke, Susann and Brenne, Jan Kristoffer},
  title        = {{New advances in fiber optic technology for environmental monitoring, safety, and risk management applications}},
  booktitle    = {2023 IEEE International Workshop on Metrology for the Sea; Learning to Measure Sea Health Parameters},
  year         = {2023},
  pages        = {316--321},
  OPTorganization = {IEEE},
  note         = {New_advances_in_fiber_optic_technology_for_environmental_monitoring_safety_and_risk_management_applications.pdf Distributed Acoustic Sensing (DAS) Overview: Utilizes submarine telecommunication and power cables for real-time monitoring of marine environments. Measures strain along optical fibers via Rayleigh backscatter analysis, detecting acoustic vibrations and other strain-related changes. 2. Applications of DAS: Marine Traffic Monitoring: Detects vessels within 10 km and distinguishes fishing trawls from large vessels based on acoustic modes. Integrates with AIS (Automatic Identification System) for vessel tracking and validation. Seismic and Oceanographic Monitoring: Records seismic events and microseisms (0.01–50 kHz). Monitors ocean-bottom pressure and ambient noise linked to storms. Environmental Conservation: Tracks whale vocalizations and cetacean activities, providing real-time location data to prevent ship strikes. Observes interactions between marine species and human activities. Infrastructure Protection: Identifies threats like anchor drags and fishing activities near submarine cables. Enables early warnings and risk mitigation for submarine cable integrity. 3. Technical Advancements: Extended Range Monitoring: Achieved long-range DAS interrogation up to 130 km with OptoDAS technology. Enhanced signal-to-noise ratio (SNR) and resolution with laboratory validation at 171 km. High-Frequency Sensitivity: Supports very broadband measurements ranging from 0.01 Hz to 50,000 Hz. Live Traffic Integration: DAS coexists with live telecommunication signals in repeatered networks without performance degradation. 4. Field Trials: Svalbard Trials: Monitored cetaceans using a 260-km cable and achieved accurate whale localization over a 9.4-km corridor. Detected microseisms from distant storms and seismic events with high precision. North Sea: DAS used for seismic imaging on a 130-km cable, demonstrating compatibility with active 4D seismic surveys. 5. Challenges and Future Directions: Challenges: Limited DAS interrogation range in repeatered systems due to signal disruption at repeaters. Requires dark fibers or additional infrastructure for long-haul applications. Future Directions: Development of new repeater configurations for uninterrupted sensing. Broad adoption of DAS for global submarine cable networks to unlock underexplored marine ecosystems.},
  doi          = {10.1109/MetroSea58055.2023.10317494},
}

@InProceedings{dias2014,
  author    = {Dias, André Rodrigues and Santos, Nuno Pessanha and Lobo, Victor},
  title     = {{Acoustic Technology for Maritime Surveillance: Insights from Experimental Exercises}},
  booktitle = {OCEANS 2024 - Singapore},
  year      = {2024},
  pages     = {1-7},
  note      = {Acoustic_Technology_for_Maritime_Surveillance_Insights_from_Experimental_Exercises.pdf Motivation: Renewed interest in underwater acoustic surveillance, particularly for anti-submarine warfare and unmanned vehicle detection. Maritime surveillance plays a critical role in national security, especially for countries like Portugal, with extensive maritime domains. Distributed Acoustic Sensing (DAS): DAS transforms fiber-optic cables into a dense network of virtual hydrophones, providing cost-effective and extensive monitoring capabilities. Leveraged for detecting and identifying underwater sound sources, including ships and submarines, through passive acoustic barriers. Technical Overview: DAS relies on Coherent Optical Time Domain Reflectometry (C-OTDR), which detects acoustic vibrations through Rayleigh backscatter. Converts optical fibers into acoustic sensors with spacings of 5–10 meters, offering dense monitoring over long distances (up to 70 km). Applications: Surveillance: Detection and classification of underwater threats such as submarines, illegal vessels, and potential intrusions. Infrastructure Protection: Monitoring underwater power and communication cables, ports, and naval bases. Environmental Monitoring: Captures ocean noise for ecological studies and marine species monitoring. Experimental Exercises: Initial field tests conducted during Robotics Exercise 2022 in Portugal. Tested DAS performance for acoustic data collection at varying distances (100–800 yards) using controlled sound sources. Demonstrated the feasibility of passive acoustic barriers for maritime surveillance. Findings: DAS technology effectively captures and localizes acoustic events, offering significant advantages over traditional hydrophone systems. High sensitivity to both anthropogenic (ship noise) and natural (seismic or marine life) acoustic signals. Burying fiber-optic cables in sand improved signal stability by reducing noise and external interference. Challenges and Limitations: Strong tidal currents affected cable placement during field experiments, highlighting the need for robust cable configurations. Limited to passive detection, requiring complementary systems for range and directional data. Future Directions: Extend testing to a broader range of configurations and geographical locations. Integrate DAS systems with autonomous maritime vehicles for enhanced surveillance capabilities. Further research on long-term durability and real-time data fusion for operational deployment.},
  doi       = {10.1109/OCEANS51537.2024.10682280},
}

@Article{Yu_2021,
  author    = {Yu, Wei and You, Hongjian and Lv, Peng and Hu, Yuxin and Han, Bing},
  title     = {{A Moving Ship Detection and Tracking Method Based on Optical Remote Sensing Images from the Geostationary Satellite}},
  journal   = {Sensors},
  year      = {2021},
  volume    = {21},
  number    = {22},
  pages     = {7547},
  OPTmonth     = nov,
  issn      = {1424-8220},
  note      = {sensors-21-07547-v2.pdf Objective: Proposes a novel method for detecting and tracking moving ships using GF-4 geostationary optical remote sensing images. Focuses on overcoming challenges posed by dim and small ship targets and interference from clouds, islands, and sea clutter. Technological Approach: Image Enhancement: Employs the Adaptive Nonlinear Gray Stretch (ANGS) method to suppress high-brightness cloud clutter and highlight dim ship wakes. Ship Detection: Introduces the Multiscale Dual-Neighbor Difference Contrast Measure (MDDCM) for visual saliency-based target detection. The saliency map is segmented using dynamic thresholds to identify candidate ship locations. Shape-Based Verification: Candidate regions are verified based on the shape and size characteristics of ship wakes to reduce false positives. Tracking: Utilizes the Joint Probability Data Association (JPDA) method for multi-frame data association, enabling ship tracking and trajectory estimation. Applications: Near real-time monitoring of maritime traffic, economic zone surveillance, and fishery safety supervision. Provides ship motion information such as position, heading, speed, and trajectory. Experimental Validation: Tests conducted in the Bohai Sea and East China Sea using GF-4 satellite images. The MDDCM method demonstrated superior performance in detecting small and dim ship wakes compared to other methods like MPCM and PSNR. Performance Metrics: Achieved a recall rate of 93.1%, precision of 98.2%, and a false alarm rate of 1.9% after shape verification and multi-frame data association. Significantly reduced false positives by incorporating wake shape analysis. Challenges and Limitations: Performance declines in scenes with numerous moving clouds or excessive background noise. Requires further refinement for dynamic environments and extended temporal analysis. Future Directions: Application of the method to other low- and medium-resolution satellite systems. Development of enhanced algorithms for tracking in complex scenes with high-density ship traffic.},
  doi       = {10.3390/s21227547},
  publisher = {MDPI AG},
}

@Article{Wawrzyniak_2019,
  author    = {Wawrzyniak, Natalia and Hyla, Tomasz and Popik, Adrian},
  title     = {{Vessel Detection and Tracking Method Based on Video Surveillance}},
  journal   = {Sensors},
  year      = {2019},
  volume    = {19},
  number    = {23},
  pages     = {5230},
  OPTmonth     = nov,
  issn      = {1424-8220},
  note      = {sensors-19-05230.pdf Objective: Introduces a method for detecting and tracking vessels using video streams from monitoring systems installed in ports and rivers. Focuses on creating an efficient solution for varying environmental conditions and vessel types. Challenges Addressed: Environmental variability: Dynamic lighting, sun reflections, and water disturbances like waves. Diverse vessel sizes: From kayaks to large cargo ships. Hardware efficiency: Aims for compatibility with economically feasible systems. Methodology: Detection Zone: Limits analysis to areas where vessel activity is expected. Moving Vessel Detection Algorithm (MVDA): Utilizes background subtraction to isolate moving objects. Applies shape and size filters to distinguish vessels from other objects. Status Update Algorithm (SUA): Tracks vessels over multiple frames, removing artifacts like water wakes and reflections. Water Detection Algorithm: Analyzes edges and contours to differentiate between water disturbances and actual vessels. Experimental Setup: Tested with three datasets captured under varying conditions using different cameras: GoPro Hero 6 (Set A). AXIS IP camera (Set B). Dahua IP camera (Set C). Resolution ranges: Full HD to 4K, with varying frame rates and bitrates. Performance Results: Correct detection rates: Standard settings: ~75% for Sets A and B; ~60% for Set C. High-threshold settings: Improved to 85% for Set C but reduced for Sets A and B due to filtering out smaller vessels. Water-related artifacts (e.g., wakes) remain a challenge, particularly for small vessels. Applications: Enhances real-time vessel traffic services (VTS) and river information services (RIS). Useful for identifying non-conventional vessels lacking AIS transponders. Limitations and Future Work: Limited to daytime operation; struggles with low-light and night-time scenarios. Future enhancements include integrating infrared cameras and optimizing detection for multi-vessel interactions in close proximity.},
  doi       = {10.3390/s19235230},
  publisher = {MDPI AG},
}

@Article{Xie_2020,
  author    = {Xie, Xiaoyang and Li, Bo and Wei, Xingxing},
  title     = {{Ship Detection in Multispectral Satellite Images Under Complex Environment}},
  journal   = {Remote Sensing},
  year      = {2020},
  volume    = {12},
  number    = {5},
  pages     = {792},
  OPTmonth     = mar,
  issn      = {2072-4292},
  note      = {remotesensing-12-00792.pdf Objective: Proposes a novel ship-detection method using spectral reflectance in multispectral satellite images. Aims to address detection challenges in complex environments like shadows, mist, and clouds. Improves detection efficiency and robustness across varying weather and environmental conditions. Methodology: Coarse-to-Fine Detection Framework: Coarse Stage: Uses spectral reflectance gradients to extract ship candidates. Applies Random Forest classifier for initial filtering. Fine Stage: Combines reflectance and synthesized false color images using a lightweight fusion network (LFNet). LFNet refines detection with shape, color, and reflectance features. Feature Construction: Reflectance gradients are calculated to differentiate ships from background objects. Features selected to balance classification precision and recall. Technical Innovations: Spectral Reflectance Gradients: Exploit stable gradient patterns to detect ships in multispectral bands. LFNet Architecture: Combines reflectance and color features using grouped convolutional filters for enhanced efficiency. Multispectral Feature Fusion: Synthesized false color images (NIR–red–green) outperform traditional band combinations. Performance: Tested on 48 images from GaoFen-1, ShiJian-9, ZiYuan-3, and CBERS-04 satellites. Achieved F1-measurement of up to 97.6% across different conditions (clouds, waves, clean). Outperformed comparison methods such as DenseSIFT, SVDNet, and YOLT. Applications: Maritime security, vessel monitoring, and environmental surveillance. Enhances ship detection in challenging scenarios like mist and dense clouds. Comparison with Other Methods: Demonstrated superior precision and recall rates compared to traditional methods using handcrafted features or simpler networks. Achieved robustness against environmental complexities and sensor variations. Challenges and Limitations: Limited precision in coarse detection stage due to overlaps in reflectance gradients. Dependence on specific sensor parameters like resolution and quantization levels. Future Directions: Extend to hyperspectral or multimodal satellite data for improved detection capabilities. Explore advanced learning techniques like semi-supervised or transfer learning for limited-label scenarios. Enhance real-time processing for large-scale monitoring.},
  doi       = {10.3390/rs12050792},
  publisher = {MDPI AG},
}

@article{sladen2019distributed,
  title={{Distributed sensing of earthquakes and ocean-solid Earth interactions on seafloor telecom cables}},
  author={Sladen, Anthony and Rivet, Diane and Ampuero, Jean Paul and De Barros, Louis and Hello, Yann and Calbris, Ga{\"e}tan and Lamare, Patrick},
  journal={Nature communications},
  volume={10},
  number={1},
  pages={5777},
  year={2019},
  publisher={Nature Publishing Group UK London}
}

@misc{OptoDAS,
author = {{Alcatel Submarine Networks}},
OPTtitle = {\href{https://www.asn.com/fiber-sensing/}{OptoDAS interrogator}},
title = {{OptoDAS interrogator}},
  OPThowpublished = {Online, accessed June 2026},
  howpublished = {\href{https://www.asn.com/fiber-sensing/}{Online, accessed june 2026}},
  OPTurl = {https://www.asn.com/fiber-sensing/},
  note         = {}
}

@inproceedings{Chen:2016:XST:2939672.2939785,
author = {Chen, Tianqi and Guestrin, Carlos},
title = {{XGBoost: A Scalable Tree Boosting System}},
year = {2016},
isbn = {9781450342322},
publisher = {Association for Computing Machinery},
address = {New York, NY, USA},
OPTurl = {https://doi.org/10.1145/2939672.2939785},
OPTdoi = {10.1145/2939672.2939785},
booktitle = {Proceedings of the 22nd ACM SIGKDD International Conference on Knowledge Discovery and Data Mining},
pages = {785–794},
numpages = {10},
location = {San Francisco, California, USA},
series = {KDD '16}
}

@article{munoz2022enhancing,
  title={{Enhancing fibre-optic distributed acoustic sensing capabilities with blind near-field array signal processing}},
  author={Mu{\~n}oz, Felipe and Soto, Marcelo A},
  journal={Nature Communications},
  volume={13},
  number={1},
  pages={4019},
  year={2022},
  publisher={Nature Publishing Group UK London}
}

@article{tejedor2016toward,
  title={Toward prevention of pipeline integrity threats using a smart fiber-optic surveillance system},
  author={Tejedor, Javier and Martins, Hugo F and Piote, Daniel and Macias-Guarasa, Javier and Pastor-Graells, Juan and Martin-Lopez, Sonia and Guill{\'e}n, Pedro Corredera and De Smet, Filip and Postvoll, Willy and Gonzalez-Herraez, Miguel},
  journal={Journal of Lightwave Technology},
  volume={34},
  number={19},
  pages={4445--4453},
  year={2016},
  publisher={OSA}
}

@article{tejedor2019contextual,
  title={{A contextual GMM-HMM smart fiber optic surveillance system for pipeline integrity threat detection}},
  author={Tejedor, Javier and Macias-Guarasa, Javier and Martins, Hugo F and Martin-Lopez, Sonia and Gonzalez-Herraez, Miguel},
  journal={Journal of Lightwave Technology},
  volume={37},
  number={18},
  pages={4514--4522},
  year={2019},
  publisher={IEEE}
}

@article{tejedor2017machine,
  title={{Machine learning methods for pipeline surveillance systems based on distributed acoustic sensing: A review}},
  author={Tejedor, Javier and Macias-Guarasa, Javier and Martins, Hugo F and Pastor-Graells, Juan and Corredera, Pedro and Martin-Lopez, Sonia},
  journal={Applied Sciences},
  volume={7},
  number={8},
  pages={841},
  year={2017},
  publisher={MDPI}
}

@article{huynh2025real,
  title={{A real scale application of a novel set of spatial and similarity features for detection and classification of natural seismic sources from distributed acoustic sensing data}},
  author={Huynh, C and Hibert, C and Jestin, C and Malet, J-P and Lanticq, V},
  journal={Geophysical Journal International},
  volume={240},
  number={1},
  pages={462--482},
  OPTmonth={January},
  year={2025},
  publisher={Oxford University Press}
}

@article{paap2025leveragin,
title = {{Leveraging Distributed Acoustic Sensing for monitoring vessels using submarine fiber-optic cables}},
journal = {Applied Ocean Research},
volume = {154},
pages = {104422},
year = {2025},
issn = {0141-1187},
doi = {10.1016/j.apor.2025.104422},
OPTurl = {https://www.sciencedirect.com/science/article/pii/S0141118725000100},
author = {Bob Paap and Vincent Vandeweijer and Jan-Diederik {van Wees} and Dirk Kraaijpoel}
}

@book{efron1994introduction,
  title        = {{An Introduction to the Bootstrap}},
  author       = {Efron, Bradley and Tibshirani, Robert J.},
  year         = {1994},
  publisher    = {Chapman \& Hall/CRC},
  address      = {Boca Raton, FL}
}

@inproceedings{goo2024confidence,
  title={{Confidence Interval Estimation for Machine Learning Models in Forecasting Infectious Diseases}},
  author={Goo, Taewan and Han, Kyulhee and Song, Hanbyul and Park, Jiwon and Liu, Zhe and Oh, Jooha and Jose, Sayooj Aby and Park, Taesung},
  booktitle={2024 IEEE International Conference on Bioinformatics and Biomedicine (BIBM)},
  pages={5914--5919},
  year={2024},
  organization={IEEE}
}

@article{shabbir2024estimation,
  title={{Estimation of Prediction Intervals for Performance Assessment of Building Using Machine Learning}},
  author={Shabbir, Khurram and Umair, Muhammad and Sim, Sung-Han and Ali, Usman and Noureldin, Mohamed},
  journal={Sensors},
  volume={24},
  number={13},
  pages={4218},
  year={2024},
  publisher={MDPI}
}

@data{ais-data-ieee-dataport-1308-8605-25,
doi = {10.21227/1308-8605},
url = {https://dx.doi.org/10.21227/1308-8605},
author = {Wang, Qile},
publisher = {IEEE Dataport},
title = {{DAS data for submarine cable detection}},
year = {2025} }

@misc{navalnews_estlink2_2024,
  title = {{Seabed Cable Damaged in Latest Baltic CUI Incident}},
  author = {{Naval News}},
  year = {2024},
  OPTurl = {https://www.navalnews.com/naval-news/2024/12/seabed-cable-damaged-in-latest-baltic-cui-incident/},
  howpublished = {\href{https://www.navalnews.com/naval-news/2024/12/seabed-cable-damaged-in-latest-baltic-cui-incident/}{Online, accessed june 2026}},
}

@misc{cbsnews_baltic_cables_2024,
  title = {{Undersea cables cut or damaged, leading European nations to ...}},
  author = {{CBS News}},
  year = {2024},
  OPTurl = {https://www.cbsnews.com/news/undersea-cables-cut-europe-finland-germany-hint-russia-sabotage/},
  howpublished = {\href{https://www.cbsnews.com/news/undersea-cables-cut-europe-finland-germany-hint-russia-sabotage/}{Online, accessed june 2026}},
}

@misc{euromaidan_baltic_2025,
  title = {{Finland-Germany submarine cable damaged again in Baltic Sea in possible sabotage act}},
  author = {Yuri Zoria},
  year = {2025},
  OPTurl = {https://euromaidanpress.com/2025/02/21/finland-germany-submarine-cable-damaged-again-in-baltic-sea-in-possible-sabotage-act/},
  howpublished = {\href{https://euromaidanpress.com/2025/02/21/finland-germany-submarine-cable-damaged-again-in-baltic-sea-in-possible-sabotage-act/}{Online, accessed march 2025}},
}

@Article{Nur_DAS_storage2024,
AUTHOR = {Nur, Abdusomad and Muanenda, Yonas},
TITLE = {Design and Evaluation of Real-Time Data Storage and Signal Processing in a Long-Range Distributed Acoustic Sensing (DAS) Using Cloud-Based Services},
JOURNAL = {Sensors},
VOLUME = {24},
YEAR = {2024},
NUMBER = {18},
ARTICLE-NUMBER = {5948},
OPTURL = {https://www.mdpi.com/1424-8220/24/18/5948},
PubMedID = {39338693},
ISSN = {1424-8220},
DOI = {10.3390/s24185948}
}

@Article{Zhan_2020,
  author    = {Zhan, Zhongwen},
  title     = {{Distributed Acoustic Sensing Turns Fiber‐Optic Cables into Sensitive Seismic Antennas}},
  journal   = {Seismological Research Letters},
  year      = {2020},
  volume    = {91},
  number    = {1},
  pages     = {1--15},
  OPTmonth     = dec,
  issn      = {1938-2057},
  doi       = {10.1785/0220190112},
  publisher = {Seismological Society of America (SSA)},
}

@article{gabai2016sensitivity,
  title={{On the sensitivity of distributed acoustic sensing}},
  author={Gabai, Haniel and Eyal, Avishay},
  journal={Optics letters},
  volume={41},
  number={24},
  pages={5648--5651},
  year={2016},
  doi = {10.1364/OL.41.005648},
  publisher={Optical Society of America}
}

@Article{Zou_2015,
  author    = {Zou, Weiwen and Yang, Shuo and Long, Xin and Chen, Jianping},
  title     = {{Optical pulse compression reflectometry: proposal and proof-of-concept experiment}},
  journal   = {Optics Express},
  year      = {2015},
  volume    = {23},
  number    = {1},
  pages     = {512},
  OPTmonth     = jan,
  issn      = {1094-4087},
  doi       = {10.1364/oe.23.000512},
  publisher = {Optica Publishing Group},
}

@inproceedings{dong2020dassa,
  title={{DASSA: Parallel DAS data storage and analysis for subsurface event detection}},
  author={Dong, Bin and Tribaldos, Ver{\'o}nica Rodr{\'\i}guez and Xing, Xin and Byna, Suren and Ajo-Franklin, Jonathan and Wu, Kesheng},
  booktitle={2020 IEEE International Parallel and Distributed Processing Symposium (IPDPS)},
  pages={254--263},
  year={2020},
  organization={IEEE}
}

@ARTICLE{shao2025tracking,
  author={Shao, Jie and Wang, Yibo and Zhang, Yixin and Zhang, Xuping and Zhang, Chi},
  journal={IEEE Geoscience and Remote Sensing Letters},
  title={{Tracking Moving Ships Using Distributed Acoustic Sensing Data}},
  year={2025},
  volume={22},
  number={},
  pages={1-5},
  doi={10.1109/LGRS.2024.3522136}}

@misc{eu_submarine_cables2025,
  author       = {{European Commission}},
  title        = {{Joint Communication: Strengthen Security and Resilience of Submarine Cables}},
  OPThowpublished = {\url{https://digital-strategy.ec.europa.eu/en/factpages/joint-communication-strengthen-security-and-resilience-submarine-cables}},
  OPTnote         = {Accessed: Mar. 5, 2025},
  OPTurl          = {https://digital-strategy.ec.europa.eu/en/library/joint-communication-strengthen-security-and-resilience-submarine-cableshttps://www.researchgate.net/publication/334090687_Toward_Detecting_Ship_Characteristics_and_Movements_using_DAS_and_Machine_Learning},
  howpublished = {\href{https://digital-strategy.ec.europa.eu/en/library/joint-communication-strengthen-security-and-resilience-submarine-cables}{Online, accessed june 2026}},
  year         = {2025}
}

@InProceedings{Waagaard_2021,
  author     = {Waagaard, Ole Henrik and Rønnekleiv, Erlend and Haukanes, Aksel and Stabo-Eeg, Frantz and Thingbø, Dag and Forbord, Stig and Aasen, Svein Erik and Brenne, Jan Kristoffer},
  title      = {{Real-time phase-recording DAS in 171 km low-loss fiber}},
  booktitle  = {Optical Fiber Sensors Conference 2020 Special Edition},
  year       = {2021},
  series     = {OFS},
  pages      = {T2A.3},
  publisher  = {Optica Publishing Group},
  OPTcollection = {OFS},
  doi        = {10.1364/ofs.2020.t2a.3},
}

@article{li2024precise,
  title={{Precise and low-complexity method for underwater Doppler estimation based on acoustic frequency comb waveforms}},
  author={Li, Jie and Qian, ZhiWen and Hong, DeYue and Zhai, JingSheng},
  journal={Frontiers in Marine Science},
  volume={11},
  pages={1365095},
  year={2024},
  publisher={Frontiers Media SA}
}

@Article{Huang_2025,
  author    = {Huang, Wenjin and Chen, Shaoyi and Wu, Yichang and Li, Ruihua and Li, Tianrui and Huang, Yihua and Cao, Xiaochun and Li, Zhaohui},
  title     = {{DAShip: A Large-Scale Annotated Dataset for Ship Detection Using Distributed Acoustic Sensing Technique}},
  journal   = {IEEE Journal of Selected Topics in Applied Earth Observations and Remote Sensing},
  year      = {2025},
  volume    = {18},
  pages     = {4093--4107},
  issn      = {2151-1535},
  doi       = {10.1109/jstars.2024.3525082},
  publisher = {Institute of Electrical and Electronics Engineers (IEEE)},
}

@misc{emodnet2024,
  author = {{EMODnet Bathymetry Consortium}},
  title = {{EMODnet Digital Bathymetry (DTM 2024)}},
  year = {2024},
  url = {https://dx.doi.org/10.12770/cf51df64-56f9-4a99-b1aa-36b8d7b743a1},
  doi = {10.12770/cf51df64-56f9-4a99-b1aa-36b8d7b743a1},
  publisher = {European Marine Observation and Data Network (EMODnet)}
}

@article{bagnall2017great,
  title={{The great time series classification bake off: a review and experimental evaluation of recent algorithmic advances}},
  author={Bagnall, Anthony and Lines, Jason and Bostrom, Aaron and Large, James and Keogh, Eamonn},
  journal={Data mining and knowledge discovery},
  volume={31},
  pages={606--660},
  year={2017},
  publisher={Springer}
}

@book{hartog2017introduction,
  title={{An introduction to distributed optical fibre sensors}},
  author={Hartog, Arthur H},
  year={2017},
  publisher={CRC press}
}

@misc{ramirez2024dasvesseldataset,
  author       = {Erick Eduardo Ramirez-Torres and
                  Javier Macias-Guarasa and
                  Daniel Pizarro-Perez and
                  Javier Tejedor and
                  Sira Elena Palazuelos-Cagigas and
                  Pedro J. Vidal-Moreno and
                  Sonia Martin-Lopez and
                  Miguel Gonzalez-Herraez and
                  Roel Vanthillo},
  title        = {{Marlinks-NS DAS: Dataset for Vessel Detection and Distance Estimation Using Distributed Acoustic Sensing in Submarine Cables}},
  year         = {2026},
  publisher    = {Zenodo},
  doi          = {10.5281/zenodo.15611778},
  url          = {https://doi.org/10.5281/zenodo.15611778}
}

@article{cheng2024photonic,
  title={{Photonic seismology: A new decade of distributed acoustic sensing in geophysics from 2012 to 2023}},
  author={Cheng, Feng},
  journal={Surveys in Geophysics},
  volume={45},
  number={4},
  pages={1205--1243},
  year={2024},
  publisher={Springer}
}

@article{biondi2025real,
    author = {Biondi, Ettore and Tepp, Gabrielle and Yu, Ellen and Saunders, Jessie K. and Yartsev, Victor and Black, Michael and Watkins, Michael and Bhaskaran, Aparna and Bhadha, Rayomand and Zhan, Zhongwen and Husker, Allen L.},
    title = {Real-Time Processing of Distributed Acoustic Sensing Data for Earthquake Monitoring Operations},
    journal = {Seismological Research Letters},
    year = {2026},
    OPTmonth = {03},
    issn = {0895-0695},
    doi = {10.1785/0220250208},
    OPTurl = {https://doi.org/10.1785/0220250208},
    eprint = {https://pubs.geoscienceworld.org/ssa/srl/article-pdf/doi/10.1785/0220250208/7794507/srl-2025208.1.pdf},
}

@article{nasholm2022array,
  title={{Array signal processing on distributed acoustic sensing data: Directivity effects in slowness space}},
  author={N{\"a}sholm, Sven Peter and Iranpour, Kamran and Wuestefeld, Andreas and Dando, Ben DE and Baird, Alan F and Oye, Volker},
  journal={Journal of Geophysical Research: Solid Earth},
  volume={127},
  number={2},
  pages={e2021JB023587},
  year={2022},
  publisher={Wiley Online Library}
}

@incollection{dibiase2001robust,
  title={{Robust localization in reverberant rooms}},
  author={DiBiase, Joseph H and Silverman, Harvey F and Brandstein, Michael S},
  booktitle={{Microphone arrays: signal processing techniques and applications}},
  pages={157--180},
  year={2001},
  publisher={Springer}
}

@article{guo2021improved,
  title={{Improved kinematic interpolation for AIS trajectory reconstruction}},
  author={Guo, Shaoqing and Mou, Junmin and Chen, Linying and Chen, Pengfei},
  journal={Ocean Engineering},
  volume={234},
  pages={109256},
  year={2021},
  publisher={Elsevier}
}

@article{Martins2026ModelingSurface,
  author = {Martins, Pedro and van Golde, Ilmer and Silva, Susana and Frazão, Orlando and Sousa, Ricardo},
  title = {{Modeling of surface vessels using distributed acoustic sensing data and physics-based optimization}},
  journal = {Journal of the European Optical Society-Rapid Publications},
  year = {2026},
  volume = {22},
  number = {1},
  pages = {13},
  publisher = {EDP Sciences},
  doi = {10.1051/jeos/2026008},
  url = {https://doi.org/10.1051/jeos/2026008}
}

@inproceedings{Oers2025EnabledDetection,
  author = {van Oers, Alexander M. and Dees, Roos C. H. M.},
  title = {{AI-enabled detection of vessels in distributed acoustic sensing (DAS) data using submarine fiber-optic cables}},
  booktitle = {{Proceedings of SPIE}},
  year = {2025},
  volume = {13679},
  pages = {136791D},
  publisher = {SPIE},
  doi = {10.1117/12.3069866},
  url = {https://doi.org/10.1117/12.3069866}
}

@inproceedings{Anhaus2025TowardsShip,
  author = {P. Anhaus and A. Bueno Rodriguez and J. Schmidt and M. Stephan and E. Peters},
  title = {{Towards DAS ship detection using signal-based features}},
  volume = {13678},
  booktitle = {Emerging Technologies and Materials for Security and Defence 2025},
  editor = {Chantal Andraud and Roberto Zamboni and Andrea Camposeo and Luana Persano and Martin Laurenzis and Gerald S. Buller and Robert A. Lamb},
  organization = {International Society for Optics and Photonics},
  publisher = {SPIE},
  pages = {136780P},
  year = {2025},
  doi = {10.1117/12.3069711},
  URL = {https://doi.org/10.1117/12.3069711}
}

@article{pedersen2025feasibility,
  title={{A Feasibility Study of Automated Detection and Classification of Signals in Distributed Acoustic Sensing}},
  author={Pedersen, Hasse B and Heiselberg, Peder and Heiselberg, Henning and Simonsen, Arnhold and S{\o}rensen, Kristian Aalling},
  journal={Sensors},
  volume={25},
  number={17},
  pages={5445},
  year={2025},
  publisher={MDPI}
}

@misc{zhang2026deeplearningframeworkmarine,
      title={{A deep learning framework for marine acoustic and seismic monitoring with distributed acoustic sensing}},
      author={Chun Zhang and Weiqiang Zhu and Barbara A. Romanowicz and Richard M Allen and Kenichi Soga and Yuxin Wu},
      year={2026},
      eprint={2603.14844},
      archivePrefix={arXiv},
      primaryClass={physics.geo-ph},
      url={https://arxiv.org/abs/2603.14844},
}

@inproceedings{shan2026distributed,
  title={{A distributed acoustic sensing technology method for ship feature extraction based on multitask learning}},
  author={Shan, Chun and Lu, Yankai and Zeng, Yong},
  booktitle={{Second Distributed Optical Fiber Sensing Technology and Applications Conference (DOFS 2025)}},
  volume={14112},
  pages={404--417},
  year={2026},
  organization={SPIE}
}






\begin{IEEEbiographynophoto}{Erick E. Ramirez-Torres}
 received the Engineering degree in Automation and the M.Sc. degree in Biomedical Engineering from Universidad de Oriente (UO), Santiago de Cuba, Cuba, in 2003 and 2017, respectively. From 2013 to 2022, he was a Professor and Researcher at UO. He later worked as a Researcher with the University of Alcal\'a (UAH), Madrid, Spain. Since March 2026, he is an Experienced Consultant of AI and Generative AI Solutions with EPAM NEORIS. He is currently pursuing the Ph.D. degree at UAH, with research focused on intelligent methods for monitoring submarine critical infrastructure. 
\end{IEEEbiographynophoto}

\begin{IEEEbiographynophoto}{Javier Macias-Guarasa}
  received the Ph.D. degree from Universidad Politécnica de Madrid (UPM), Madrid, Spain, in 2001. From 1990 to 2007 he held different teaching positions at UPM and he is currently an Associate Professor with the Department of Electronics and a member of the GEINTRA research group at UAH. In 2003, he was a visiting scientist at the International Computer Science Institute (ICSI). He has co-authored more than 140 refereed journal and conference papers. 
  His current research interests include the use of machine learning in the context of health-related applications and distributed acoustic sensing systems for fiber-optic cable monitoring.
\end{IEEEbiographynophoto}

\begin{IEEEbiographynophoto}{Daniel Pizarro}
  received the Ph.D. degree from UAH in 2008. From 2004 to 2012, he was an Assistant Lecturer at UAH, and from 2012 to 2014, he was an Associate Professor at the Université de Clermont-Auvergne, France, and a member of the EnCoV research group. In 2014, he joined the permanent staff at UAH as an Associate Professor, and he is a Full Professor since 2021. He is currently a member of the GEINTRA Group at UAH. He has led several publicly funded projects and has co-authored more than 100 refereed papers in the field of computer vision, including image registration, deformable reconstruction, and their applications to minimally invasive surgery.
\end{IEEEbiographynophoto}

\begin{IEEEbiographynophoto}{Javier Tejedor}
  received the Ph.D. degree in Computer and Telecommunication Engineering from Universidad Autónoma de Madrid (UAM) in 2009. From 2003 to 2013, he was with HCTLab research group at UAM as Assistant Professor in the Institute of Technology. He has been with GEINTRA research group in Universidad de Alcalá as an Associate Researcher since 2014 and is currently associate professor at the Institute of Technology in Universidad San Pablo CEU in Madrid. His main interests are speech technology, and pattern recognition applied to pipeline monitoring.
\end{IEEEbiographynophoto}

\begin{IEEEbiographynophoto}{Sira Elena Palazuelos-Cagigas}
  received the Ph.D. degree in Telecommunications Engineering from UPM, in 2001. Since 2003 she is an Associate Professor with the Department of Electronics at UAH. She has participated in over 80 research projects, including more than 15 as Principal Investigator, and has co-directed several Ph.D. theses. She has co-authored numerous publications in international journals and conferences, as well as patents and utility models. Her research interests include machine learning for signal processing, computer vision, and multisensory systems for human activity analysis and functional assessment.
\end{IEEEbiographynophoto}

\begin{IEEEbiographynophoto}{Pedro J. Vidal-Moreno}
  received the Ph.D. in applied photonics from UAH in 2025. His research has focused on distributed optical fiber sensing, with emphasis on extended dynamic range and long-term stability techniques for Rayleigh distributed dynamic strain sensing. During his doctoral studies, he completed a research stay at the Optics and Photonics Lab, Tel Aviv University, Israel. He has contributed to both theoretical and experimental advancements in signal-to-noise ratio optimization and pulse coding for distributed sensing. He has collaborated with the Spanish National Research Council (CSIC) and several companies in the field of distributed sensing.
\end{IEEEbiographynophoto}

\begin{IEEEbiographynophoto}{Sonia Martin-Lopez}
  received the Ph.D. degree from Universidad Complutense de Madrid, Spain, in 2006. She completed a predoctoral research stay with the Nanophotonics and Metrology Laboratory (NAM), École Polytechnique Fédérale de Lausanne (EPFL), Switzerland. She was a Postdoc Researcher with the Institute of Applied Physics and the Institute of Optics, Spanish National Research Council (CSIC), Spain, and subsequently held a five-year Ramón y Cajal contract at UAH, where she became an Associate Professor in 2019. She is currently a Scientific Researcher with the Institute of Optics, CSIC. She has authored or co-authored more than 200 papers in international refereed journals and conference proceedings. Her research interests include nonlinear fiber optics and distributed optical fiber sensors.
\end{IEEEbiographynophoto}

\begin{IEEEbiographynophoto}{Miguel Gonzalez-Herraez}
  received the Ph.D. degree from UPM in 2004. He was a Research Assistant and a Postdoc Fellow at the Applied Physics Institute at CSIC, carrying out several long stays at the NAM Laboratory, EPFL. In 2006, he became Associate Professor with the Department of Electronics at UAH, being promoted to Full Professor in 2017. Since 2025, he is a Full Professor with the Institute of Optics, CSIC. He has authored or co-authored over 130 papers in international refereed journals, more than 150 conference contributions, and over 30 invited or plenary talks. His research interests cover the wide field of nonlinear interactions in optical fibers. Prof. González-Herráez has received several research awards, including the European Research Council Starting Grant, the ``Miguel Catalán'' prize, and the ``Agustín de Betancourt'' prize.
\end{IEEEbiographynophoto}

\begin{IEEEbiographynophoto}{Roel Vanthillo}
  received the M.Sc. degree in Electromechanical Engineering from KaHo St Lieven/KU Leuven in 2007 and the MBA degree from Vlerick Business School in 2014. He is co-founder of Marlinks, a company specializing in advanced monitoring solutions for subsea power cables, focused on reducing failure rates and maintaining transmission efficiency. He has actively contributed to publications in the area of distributed fiber-optic sensing and submarine-cable monitoring.
\end{IEEEbiographynophoto}



\end{document}